Summer 7-10-2020

# Adaptive Encoding for Constrained Video Delivery in HEVC, VP9, AV1 and VVC Compression Standards and Adaptation to Video Content

Gangadharan Esakki
*University of New Mexico - Main Campus*

Gangadharan Esakki

*Candidate*

Electrical and Computer Engineering

*Department*

This dissertation is approved, and it is acceptable in quality and form for publication:

*Approved by the Dissertation Committee:*

Marios Pattichis , Chairperson

Manuel Martinez-Ramon

Ramiro Jordan

Sylvia Celedon-Pattichis

Andreas Panayides

# Adaptive Encoding for Constrained Video Delivery in HEVC, VP9, AV1 and VVC Compression Standards and Adaptation to Video Content

by

## Gangadharan Esakki

B.E., Anna University, India, 2009
M.S., University of New Mexico, 2014

DISSERTATION

Submitted in Partial Fulfillment of the
Requirements for the Degree of

Doctor of Philosophy

Engineering

The University of New Mexico

Albuquerque, New Mexico

July 2020

# Dedication

இந்த விளக்கவுரை என் தாய் திருமதி.சுப்புலட்சுமி, தந்தை திரு.இசக்கி சங்கரபாண்டியன், என் அன்பு தங்கை பிரபாவதி இவர்களின் அளவற்ற அன்பிற்கும், தூய உள்ளத்திற்கும், என்னுடைய அறிவு தேடலுக்கு உறுதுணையாக எப்போதும் என்னோடு நடந்ததற்கு சமர்ப்பணம்

இ.கங்காதரன்

*To Mom Subbulakshmi, Dad Esakki Sankarapandian*
*and my loving Sister Prabhavathy,*
*for their pure, ingenuous heart of love and sacrifice,*
*supporting me in this quest for seeking knowledge and wisdom.*

# Acknowledgments

I would like to express my gratitude to my advisor, Dr. Marios S. Pattichis, for his motivation, advice and encouragement throughout my years of research in graduate school. I am indebted to him for all the patience and time he spent on me as many of the research ideas own large contributions to him.

I would also like to thank my dissertation committee chairs Dr. Manel Martinez-Ramon, Dr. Ramiro Jordan, Dr. Sylvia Celedon-Pattichis for their time and support in reviewing and checking this work. Special thanks to Dr. Andreas Panayides who has been crucial in providing critical insights, valuable feedback and helped me overcome obstacles in my research endeavors

I would also like to thank my friends and lab mates at IVPCL: Venkatesh Jatla, Wenjing, Sravani whose advice and encouragement have been crucial and have become my family far from home. Thanks to Krishna Poddar for helping me with code automation, Miguel for always being a supportive friend. Thanks to Luis Tapia Sanchez for all the jokes and music, Zhen Yu and Phuong Tran for board games and crazy dance parties. Thanks to staff and friends in the Department of Electrical and Computer Engineering and Department of Spanish & Portuguese language for supporting me early in my years of grad school and throughout my on-campus jobs and scholarships.

Thanks to the internships at **_Intel PerC lab_** where I learned and developed new algorithms for camera depth sensing and performance metrics.

I would also like to express my gratitude to Dr. Sylvia Celedon-Pattichis from College of Education, Elsa Maria Castillo from School of Engineering, Maria Wolfe from UNM Alumni office for their continued financial assistance. I would also like to thank the Tamil families here in ABQ who made me feel at home to the likes of Namasivayam and Suresh to celebrate Indian festivals and for their moral support. Besides, my favorite Winning Coffee shop and Frontier restaurant for amazing coffee and best burritos. The last word of acknowledgment I have reserved for my greatest passion is *"yoga"*. All through my PhD, there were many occasions yoga kept me calm, with innate breath control and mental fortitude to persevere and keep moving.

# Adaptive Encoding for Constrained Video Delivery in HEVC, VP9, AV1 and VVC Compression Standards and Adaptation to Video Content

by

## Gangadharan Esakki

B.E., Anna University, India, 2009

M.S., University of New Mexico, 2014

Ph.D., Engineering, University of New Mexico, 2020

## Abstract

The dissertation proposes the use of a multi-objective optimization framework for designing and selecting among enhanced GOP configurations in video compression standards. The proposed methods achieve fine optimization over a set of general modes that include: (i) maximum video quality, (ii) minimum bitrate, (iii) maximum encoding rate (previously minimum encoding time mode) and (iv) can be shown to improve upon the YouTube/Netflix default encoder mode settings over a set of opposing constraints to guarantee satisfactory performance. The dissertation describes the implementation of a codec-agnostic approach using different video coding standards (x265, VP9, AV1) on a wide range of videos derived from different video datasets. The results demonstrate that the optimal encoding parameters obtained from the Pareto front space can provide significant bandwidth savings without sacrificing video quality. This is achieved by the use of effective regression models that allow for the selection of video encoding settings that are jointly optimal in the encoding time, bitrate, and video quality space. The dissertation applies the proposed methods to x265, VP9, AV1 and using new GOP configurations in x265, delivering over 40% of the optimal encodings in two standard reference videos. Then, the proposed encoding method is extended to use video content to determine constraints on video quality during real-time encoding. The content-based approach is demonstrated on identifying camera motions like panning, stationary and zooming in the video. Overall, the content-based approach gave bitrate savings of 35 % on the zooming & panning motion from Shields video, and 51.5 % on stationary & panning motion from Parkrun video. Additionally, the dissertation develops a segment-based encoding approach that delivers bitrate savings over YouTube's recommended bitrates. Using BD-PSNR and BD-VMAF, a comparison is made of x265, VP9, AV1 against the emerging VVC encoding standard. The new VVC-VTM encoder is found to outperform all rival video codecs. Based on subjective video quality assessment study, AV1 was found to provide higher quality than x265 and VP9.

# Contents

# List of Figures

# List of Tables

# Chapter 1

# Introduction

## 1.1   Video Streaming Industry & CODEC Wars

The deployment of effective video coding standards in 5G networks aims to address the rapid growth of network traffic and bandwidth-hungry applications. Additionally, video streaming dominates the delivery of video content. Video-On-Demand (VOD) and Video streaming applications worldwide are experiencing an exponential growth with applications such as video based learning, adaptive medical video communications [5–7], mobile gaming and AR/VR. According to Cisco [8], global IP video traffic will account for 82% of the internet traffic in 2020 which is significantly higher than 70% back in 2015. Our everyday life is surrounded by devices connected to the internet and with so many apps, we are increasing the internet traffic with videos. Over the years, YouTube has become the major source of video traffic accounting for a significant portion of the Internet traffic followed by the streaming providers (e.g., Amazon, Disney+, Hulu and Netflix).

Limiting the pre-encoded formats to a fixed set of combinations often may or may not provide the best quality for users, since user constraints keep changing all the time. With the evolution of new codec standards and the encoder configuration options growing exponentially, the challenge for streaming video providers is not only to come up with the optimal set of pre-encoding configurations that best suit user profiles, but also to choose the "one" optimal video encoded format that gives the best possible quality for a specific user (compute power, bandwidth, display resolution, network delay). With 4K becoming standard and now 8K and even higher resolutions on the horizon, the increase in higher quality video, along with the need for such a real-time capable, resource optimizing video control system is growing more than ever before. This optimized video delivery for best quality problem applies to enterprises, consumers and government users alike.

Apart from higher consumption of videos on a daily basis due to regular streaming, video content providers also promise that their videos are always of higher quality. For example, YouTube [4] encodes 480p video @2.5Mbit/s, 720p @5Mbit/s and 1080p @8Mbit/s which are rather high values. Netflix, Facebook and Apple use their own conservative encoding-bitrate ladder settings. Even though streaming providers like Amazon Prime or Netflix have their own streaming techniques that aim to deliver higher video quality, we as users/consumers have always noted that they lack significant drop in visual quality. The quality of experience really matters when it comes to streaming VOD where the drop in bandwidth is felt directly by the consumer. So, an efficient system would be to offer higher or an acceptable quality even when there is a sudden drop in bandwidth and to sustain the quality throughout the video. Such systems can be built only after a thorough understanding of the video encoding pipeline with better optimization and control. To understand the bitrate constraints on the problem, typical recommendations by Netflix are given in Fig 1.1.

H.264/AVC has been widely adopted as the default video encoding standard as MP4 container format is the most widely used extension. On the other hand, Google's VP9 codec is deployed in YouTube and the new AV1 codec from Alliance for Open Media (AOM) is used in YouTube and other streaming platforms very recently. Adaptive streaming technology standards such as Adobe HDS (HTTP Dynamic Streaming), Google's webRTC, and Apple's HLS and (MPEG-DASH) have become very popular and cover a range of codec services and support. For example, Apple's HLS only supported H.264 even after H.265/HEVC was released in 2013 but still used the former standard in all its devices owing to patent and royalty issues. However, it is not only a matter of royalty issues as there are other bottlenecks in video streaming like video buffering, frames frozen/stalled, and latency issues because of insufficient bandwidth which can happen during peak traffic hours and emergency scenarios [9, 10]. Currently, at the time of this thesis we have COVID-19 [11] and there is a huge number of people working from home, students taking classes online and much more. All these situations have led to a crisis where the streamed videos will have frequent buffering, stalled frames or rendered with pixellations which is very visual and results in a direct impact on the user's overall satisfaction of the video quality being delivered. Amazon Prime/Netflix/YouTube are definitely not going to be happy when the video is of lower quality with artifacts. A possible solution is to adapt video encoding based on content and/or user provided constraints.

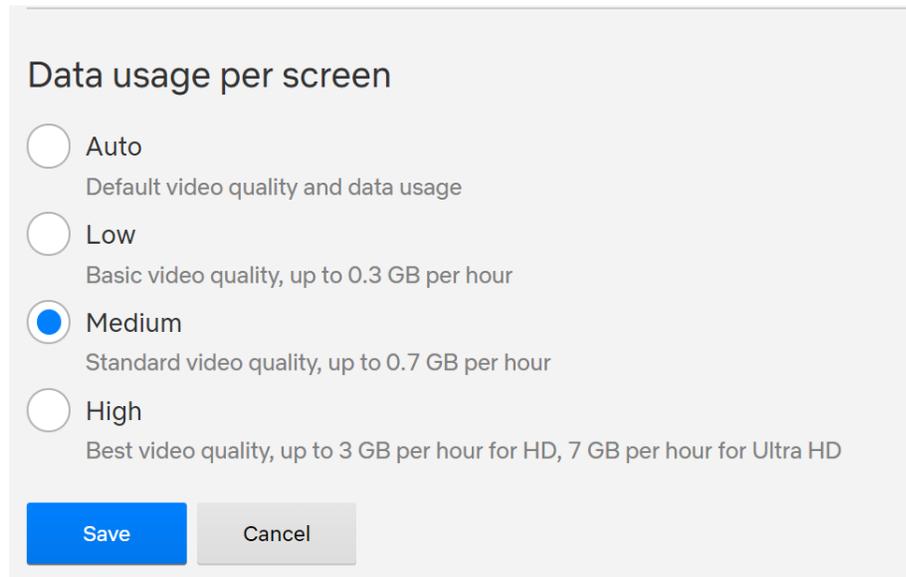

Figure 1.1: Netflix Video Quality Settings

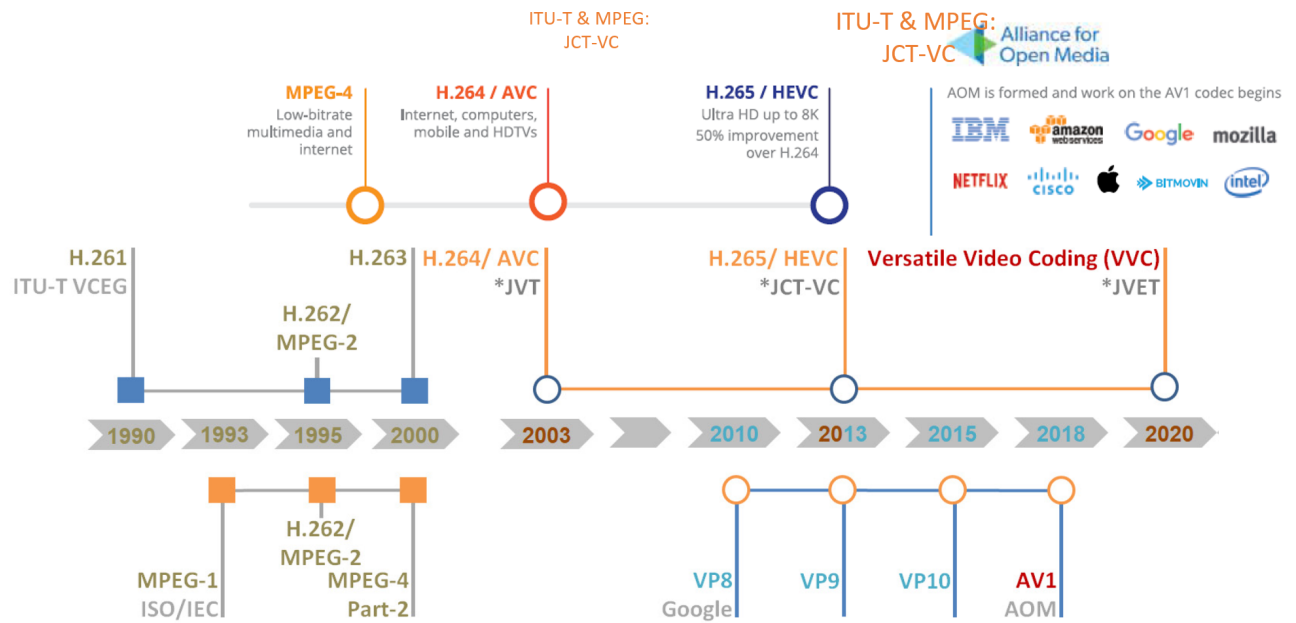

Figure 1.2: Video Codec Standards showing both MPEG ITU-T and AOM Codecs [12]

But before jumping into the solution, we need to understand video codecs and their pivotal role in this dynamically changing ecosystem. ITU-based codec standards have evolved from H.264 to H.265 (HEVC or High Efficiency Video Coding), and Google, along with the Alliance for Open Media (AOM), has been pushing their standards from VP9 to AV1. ITU-T standard based encoding systems come with royalties for commercial deployment whereas, the open media alliance codecs (VP8/VP9/AV1) are royalty free. Performance-wise, the older video encoders (VP8 and VP9) do not offer as much bitrate savings as HEVC (which provides the same quality as H.264 at half the bit rate). A general description of all the video coding standards from both MPEG ITU-T and AOM is shown in Fig 1.2.

The biggest challenge with HEVC is that it never got fully adopted because of royalty and patent issues and hence was never widely deployed as its predecessor H.264. Established in 2015, the AV1 codec was founded by Alliance for Open Media (AOM), originally the advanced VP10 (the next upgrade to Google's VP9 [13] Codec), eventually merged with other open source Codecs like Daala from Mozilla and Thor from Cisco incorporating different video codec tools into AV1. AOM/AV1 [14] was originally created to be the future codec, open sourced and royalty free available for video streaming and to cater to the web and for delivery on browsers supporting multiple device platforms. AOM/AV1 or the libaom codec (introduced in 2017) is already available in Google Chrome and Mozilla Firefox browsers and even YouTube started streaming its videos in AV1 while it still has a lot of room for improvement. Libaom/AV1 was the first codec AOM created and there are several ongoing implementations. AOM provided SVT-AV1 (Scalable Video Technology - AV1) from Intel which is built for VoD and live streaming applications. Netflix [15] very recently adopted SVT-AV1 [16] to stream all of its content in the new open-sourced codec and has been jointly working with Intel to optimize them for their platform. The wide spread adoption of VP9/libaom in YouTube and SVT-AV1 in Netflix has created a war between open sourced codecs versus the MPEG's next upgrade H.266 or Versatile Video Coding (VVC). The AOM has been supported by a large array of software and hardware companies, the majority of them in Silicon Valley, harboring the means to bring AOM/SVT-AV1 to be supported in all device platforms. For example, Twitch which is an online gaming platform, heavily uses VP9 for its online streaming which is an FPGA based high performance VP9 encoding. Thus, while VVC promises big bitrate savings over open source encoders, the challenge is for VVC to get adopted as quickly as possible to avoid being overrun by open source solutions.

## 1.2 Motivation

The primary motivation of this dissertation is to develop a segment based encoding approach which can be applied to any encoding standard with Video-On-Demand (VOD) and streaming as applications. Prior work with DRASTIC [17] was done with a focus on specialized hardware architectures developed specifically for MJPEG, H.264/AVC, H.265/HEVC standards and a software based approach was extended to HEVC Intra coding. In [18], a joint optimization methodology was taken to study the CU depths and Intra coding together to achieve precise control and modeling at a frame level using RDO budget constraints.

## 1.3 Thesis statement

The thesis of the dissertation is that a multi-objective based approach can provide optimal video encodings for video delivery applications. This dissertation has developed methods to build models that can provide optimal encoding parameters across standards and can support adaptive encoding based on dynamic constraints. This research heavily focuses on the development of Group of Picture (GOP) level control with newer GOP structures for x265/HEVC, VP9, SVT-AV1 Video Coding Standards with applications in optimal encoding, adaptive encoding using Camera activity classification and GOP level adaptation for VP9 and AV1 encoding standards. The approach uses Pareto based segment modeling and predicts the optimal encoding parameters for the next segment within a video subject to dynamic constraints.

## 1.4 Contributions

This dissertation demonstrates the use of Segment based encoding with efficient use of encoding parameterization and joint-optimization of rate, quality and encoding rate on software configurations available to the codecs. All of these contributions came through a DRASTIC optimization framework.

A summary of the contributions includes:

- **DRASTIC Framework with new GOPs for x265/H.265 standard:**
  Newly introduced GOP structures for x265 encoder are tested on the UT-LIVE Video Dataset. The results show that the new GOPs have improved performance across a range of videos.

- **Video Content adaptation based on Camera activity Classification**
  Motion vectors are used as feature vectors input to a classifier to demonstrate adaptive encoding based on different camera motions.

- **Segment based encoding with x265 and Local Pareto Models**
  The Pareto front is used to build a regression model and uses it within video segments to predict encoding parameters along the Pareto front. This approach eliminates the need for re-encoding. VMAF based model fitting was done from the Pareto front.

- **Open-Source Video Coding Standards: Google VP9 & AOM/SVT-AV1 Codec**
  The Pareto front is used to build a regression model and use it within video segments to predict encoding parameters for VP9, AOM/SVT-AV1.

- **Subjective Video quality assessment**
  This dissertation provides both subjective and objective Video quality assessment for x265, VP9, SVT-AV1 encoders. Thirty two human subjects were shown different videos encoded with different video quality levels and asked to score them. From the tests, VMAF metric proved AV1 as the winner in the perceptual quality test.

- **VVC Encoding Standard & BD-PSNR and BD-VMAF measurements done on wide video datasets.**
  The emerging VVC standard was studied and was used within BD-PSNR and BD-VMAF measurements with x265, VP9, SVT-AV1 encoders. VVC gave the best results followed by SVT-AV1 and VP9.

## 1.5    Organization

We have organized the chapters in the following order:

- **Chapter 1: Video Streaming & CODEC Wars**

  The first chapter provides motivation, a thesis statement and a description of the primary research contributions.

- **Chapter 2: Optimal GOP Configurations for x265 HEVC Encoder in DRASTIC Framework**

  This chapter describes the new GOP structures introduced with the x265 encoder and evaluates their performance on different videos.

- **Chapter 3: Adaptive video encoding based on Camera activity Classification**

  This chapter covers the use of motion vectors for adaptive video encoding with x265 and SSIM.

- **Chapter 4: Segment based x265 encoding with adaptive Local Pareto models for Video On Demand(VoD)**

  This chapter covers segment-based encoding for x265, describes how to build Pareto models, and summarizes how to predict optimal encodings using VMAF.

- **Chapter 5: Analysis of the libVPx Codec and Segment based VP9 encoding at GOP level optimization**

  This chapter covers VP9 and the implementation of segment-based encoding.

- **Chapter 6: Overview of AOM Video Coding Standard with SVT-AV1 Codec in Multi-objective optimization**

  This chapter covers the new AOM SVT-AV1 codec and its new tools, GOP structures and implementation of segment based encoding.

- **Chapter 7: Emerging VVC encoding standard with VMAF metric evaluation**

  This chapter briefly explains the emerging VVC standard and its tools, BD-PSNR and BD-VMAF rate curves with coding standards HEVC, VP9, AV1. This chapter also provides subjective video quality assessments for x265, VP9, SVT-AV1 Codecs for Spatio-Temporal datasets.

- **Chapter 8 provides a conclusion and suggestions for future work.**

# Chapter 2

# Optimal GOP Configurations for x265 HEVC Encoder in DRASTIC Framework

## 2.1　Introduction

The recent emergence of HEVC software implementations provides several different encoding options that can simultaneously affect video quality, bitrate, and encoding time. Unfortunately, there is no established approach for selecting optimal encoding configurations. The current chapter recommends the use of a multi-objective optimization framework for selecting optimal encodings that can be subsequently used for solving constrained optimization problems that are functions of quality, bitrate, and encoding time. The proposed optimization framework is used to select optimal configurations from 3,600 possibilities based on GOP configurations, the quantization parameter, deblocking filtering, sample adaptive offset, and software presets that control the coding tree unit size (CTU size), prediction sizes, and the transform unit sizes.

We implement our approach using the x265 encoder and demonstrate on an example from the UT LIVE video quality database [1, 19, 20], and a second standard 2K video example from [3]. The results demonstrate the success of the proposed approach by selecting optimal configurations and eliminating sub-optimal encodings.

The recent introduction of x265 open source HEVC encoder with several presets associated with different encoding times motivates the study of a unifying approach that can consider all of the presets together [21–24]. Beyond the standard use of rate-distortion theoretic methods, this chapter introduces a unifying approach that considers the multi-objective optimization of encoding time, video quality, and bitrate for selecting and extending x265 HEVC presets.

To formally define the multi-objective optimization framework, let $\mathtt{Q}$ denote a metric of video quality, $\mathtt{BPS}$ denote the number of bits per second, and $\mathtt{T}$ denote the required encoding time. An optimal video encoding configuration needs to simultaneously maximize image quality, minimize the required bitrate and also minimize encoding time. More compactly, in vector form, the multi-objective optimization framework requires that we solve as follows:

$$\min_{\mathtt{EP}} \left( -\mathtt{Q}(\mathtt{EP}), \mathtt{BPS}(\mathtt{EP}), \mathtt{T}(\mathtt{EP}) \right) \tag{2.1}$$

for the optimal encoding parameters $\mathtt{EP}$. Here, we note that the negative sign for video quality comes from the fact that maximizing the video quality is equivalent to minimizing the negative of video quality. Furthermore, in what follows, we will drop the $\mathtt{EP}$ argument from the objectives. In other words, we write $\mathtt{Q}, \mathtt{BPS}, \mathtt{T}$ with the understanding that they depend on the encoding parameters $\mathtt{EP}$.

The solution of the vector optimization problem given in (2.1) defines a Pareto front. The Pareto front is defined by the set of configurations for which no other configuration can be found that improves on all of the objectives ($\mathtt{Q}, \mathtt{BPS}, \mathtt{T}$) at the same time. Thus, a configuration $EP_{opt}$ is optimal if there is no way to find another configuration $EP$ that gives better image quality, lower bitrate, and requires less encoding time. Here, we need not consider the very unlikely case that another configuration can have the same objectives as $EP_{opt}$.

In order to select an optimal configuration, we then define optimal communications modes as (also see related work in [25–27]). Here, the goal is to find optimal solutions subject to realistic constraints on encoding ($\mathtt{T} \leq \mathtt{T_{max}}$), bitrate ($\mathtt{BPS} \leq \mathtt{BPS_{max}}$), and image quality ($\mathtt{Q} \geq \mathtt{Q_{min}}$). We are then primarily interested in optimal modes defined as [26]: (i) minimum encoding time mode, (ii) minimum bitrate mode, and (iii) maximum video quality mode, subject to opposing constraints from the two remaining objectives.

There are several challenges associated with the application of the multi-objective framework to HEVC encoding. First, we note that the Pareto-front will significantly vary from video to video, and even from GOP to GOP within each video. In [25, 26], the authors considered a bottom up approach that allowed the variation of DCT hardware cores and the quantization parameter (QP) for each image. In [27], in another bottom-up approach, the authors considered a multi-objective optimization approach that was applied to HEVC intra-coding.

Here, we take a top down approach where we consider the development of a unifying approach for all HEVC modes. Second, it is important to acknowledge that the current x265 encoder for HEVC [21, 28, 29] provides a very sparse sampling of the space of encoding time - video quality - bitrate. Unfortunately, such sparsity imposes fundamental limits on the usefulness of the proposed, multi-objective optimization framework [30]. Thus, to address this problem, the current chapter introduces extended HEVC presets in x265 that include new GOP configurations. This combination of new GOP configurations with the variation of QP, De-blocking filtering, and other parameters produces a large number of optimal configurations that allows for significantly better sampling of the multi-objective space. Third, the use of extended HEVC configurations requires the compression of each video under each one and can thus impose significant storage requirements. To address this issue, we introduce an offline approach that only stores the optimal configuration parameters (without the compressed videos) associated with the Pareto front. Then, the optimal configuration is selected by solving the optimization problem associated with each optimization mode. The optimally compressed video is then reproduced by running the x265 encoder with the optimal parameters.

In terms of related work, we also mention earlier research focused on the use of multiple objectives in hardware implementations, unrelated to video compression. We have the use of parallel cores for single-pixel processors in [31], the development of one-dimensional filtering in [32], and two-dimensional filter bank approaches in [33]. The current chapter differs significantly from these previous hardware approaches applied to digital filtering by focusing on a top-down approach.

The rest of the chapter is organized into four sections. In section 2.2, we summarize the methodology. We provide the results in 2.3 and give concluding remarks in 2.4.

## 2.2 Methodology

We summarize the proposed method in Figure 2.1. For each given video, we present the computation of the Pareto front based on the GOP configurations, the HEVC profiles, and related parameters.

As stated earlier, the resulting Pareto front is simply expressed in terms of a mapping from each optimal GOP configuration, HEVC profile, and related parameters to the three objective functions (video quality, encoding time, and bitrate requirements). For any given optimization mode, we select and apply the optimal encoding configuration as shown in Fig. 2.1.

As stated earlier, efficient implementation of the optimization modes requires an **function OptEnc**(`V`, `Vc`, `ParetoFront`, `OptPars`)
▷ **Input:** video `V`, Pareto front in `ParetoFront`,
▷          optimization mode specified in `OptPars`.
▷ **Output:** compressed video in `Vc`.

    `ParetoEntry` ← **Find** an optimal solution specified
          by `OptPars` that lies on `ParetoFront`.
    **if** (valid `ParetoEntry` has been found) **then**
       `Vc` ← **Compress** `V` using configuration
          (P, GOPconfig, ParVec)
          extracted from `ParetoEntry`.
    **else**
       `ParetoEntry` ← **Search** `ParetoFront`
          for an entry that violates the constraints by the
          least amount.
       `Vc` ← **Compress** `V` using configuration
          (P, GOPconfig, ParVec)
          extracted from `ParetoEntry`.
    **end if**
**end function**

Figure 2.1: Optimal mode encoding using the Pareto front.

extension of the standard GOP configurations. We present a diagram with some of the new GOP configurations in Fig. 2.2. We provide a detailed summary of the proposed GOP configurations in Table 2.1.

From the Pareto front, we can extract the following optimal modes:

- ***Minimum encoding time mode:***

$$\min_{\text{EP}} \text{T} \quad \text{subject to} \quad (\text{Q} \geq Q_{min}) \text{ and } (\text{BPS} \leq \text{BPS}_{\text{max}}) \tag{2.2}$$

  In this mode, the goal is to minimize encoding time provided that the video can be communicated within the given bitrate and it is of sufficiently good quality.

- ***Minimum bitrate mode:***

$$\min_{\text{EP}} \text{BPS} \quad \text{subject to} \quad (\text{Q} \geq Q_{min}) \text{ and } (\text{T} \leq \text{T}_{\text{max}}). \tag{2.3}$$

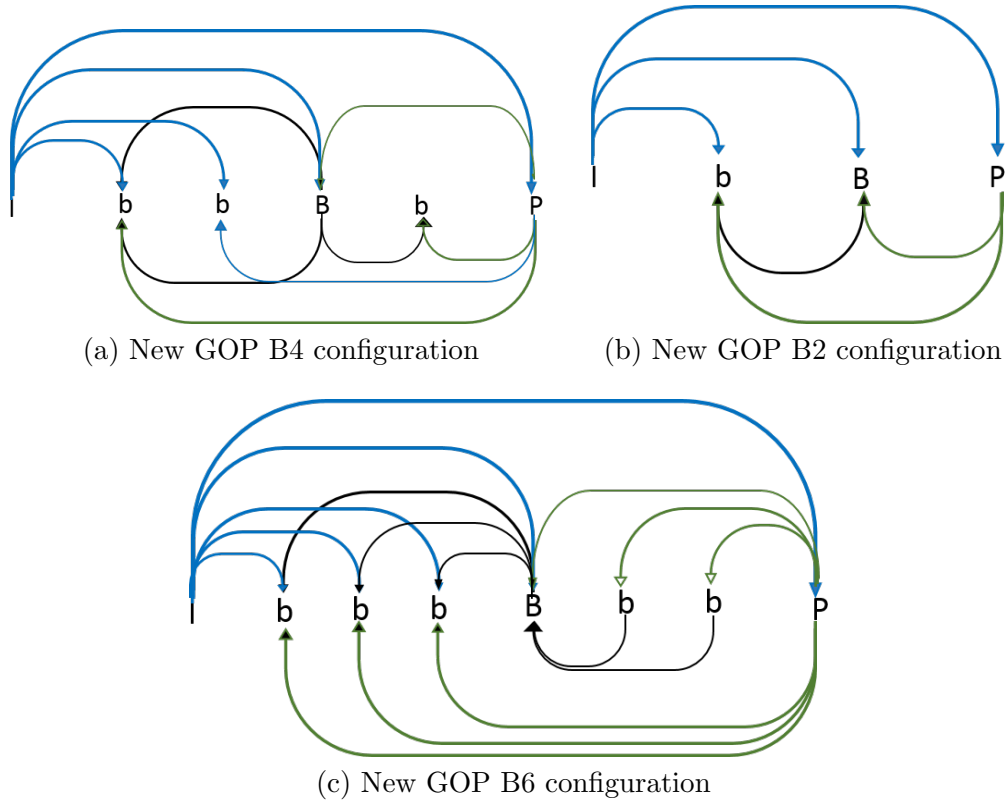

(a) New GOP B4 configuration    (b) New GOP B2 configuration

(c) New GOP B6 configuration

Figure 2.2: New GOP configurations. (a) Extended GOP configuration by removing a **b** frame. (b) Extended GOP configuration by adding a **b** frame.

In this mode, the goal is to minimize bandwidth requirements provided that the video is of sufficient quality and we do not spend a large amount of time encoding it.

- ***Maximum video quality mode:***

$$\max_{\texttt{EP}} \texttt{Q} \quad \text{subject to} \quad (\texttt{BPS} < \texttt{BPS}_{\texttt{max}}) \text{ and } (\texttt{T} < \texttt{T}_{\texttt{max}}). \tag{2.4}$$

Here, the goal is to reconstruct the video with the highest possible video quality that does not require more bandwidth that is available and within reasonable encoding time.

Table 2.1: Encoder GOP configuration setup broken into two groups. Group A presets are extensions of GOP B4 into new GOP B2, B6 and consist of: ultra fast (U), super fast (S), very fast (V), faster (Fr), fast (F), medium (M), and slow (S). Group B profiles are extensions of default GOP B8 into new GOP B6, B10 and consist: slower (Sl), very slow (VS) and Placebo (P). There are a total of 3600 possible configurations.

| Parameter | Profile Group A | Profile Group B |
|---|---|---|
| Presets | U, S, V, Fr, F, M, S | Sl, Vs, P |
| GOP | AI, **B2**, B4, **B6**, ZL | AI, **B6**, B8, **B10**, ZL |
| GOP Str | Open/Close | Open/Close |
| QP | 22, 27, 32, 37, 42 | 22, 27, 32, 37, 42 |
| SAO | On/Off | On/Off |
| DBF | On/Off | On/Off |
| Tuning | PSNR, ZL, FD | PSNR, ZL, FD |
| Configs. | 360 per profile | 360 per profile |

## 2.3 Results

For testing our approach, we consider optimal encoding for videos as shown in Figs. 2.3(a), 2.3(b), and 2.3 (c) [1, 3, 19, 20]. For measuring the encoding time, we run the x265 ver 1.4 reference software [24] on a Windows 8 64-bit platform with 64GB RAM using an Intel(R) Xeon(R) CPU E5-2630v3 microprocessor with 8 cores (16 threads) running at 2.40 GHz. Overall, as we document in Fig. 2.3, we find that we can generate relatively dense Pareto fronts provided that we have predictable, translational motions. Furthermore, we note that the new GOP configurations contributed (i) 40.64 % of the optimal 438 configurations for the Jockey video from Tampere Dataset, (ii) 40.97 % of the optimal 881 configurations for the Pedestrian video, Refer to Table 2.2 for more details.

The relatively dense Pareto fronts for the Jockey and Pedestrian videos allow us to investigate optimization modes as given in equations (2.2), (2.3), and (2.4). We present three DRASTIC mode optimization examples in Table 2.3. For the examples, all of the constraints have been met. Also, as expected, the optimal mode

Table 2.2: Optimal GOP configurations. The new GOP configurations are shown in bold.

| GOP conf. | Optimal Configurations (%) | | |
|---|---|---|---|
| | **Jockey** | **Pa** | **Rb** |
| AI | 63 (14.38%) | 101 (11.46%) | 34 (59.64%) |
| **B2** | 145 (33.1%) | 208 (23.6%) | 5 (8.77%) |
| B4 | 25 (5.7%) | 179 (20.31%) | 3 (5.2%) |
| **B6** | 33 (7.53%) | 136 (15.43%) | 1 (1.75%) |
| B8 | 6 (1.36%) | 27 (3.06%) | 0 (%) |
| **B10** | 0 (0%) | 17 (1.92%) | 0 (%) |
| ZL | 166 (37.89%) | 213 (24.17%) | 14 (24.56%) |
| Pareto | **438** (100%) | **881** (100%) | **57** (100%) |

Table 2.3: Mode Optimization. We measure bitrate in bits per second, PSNR in dB, and time in seconds. We use BR for bitrate, Q for image quality, and T for encoding time. In each case, we present the quantity that is optimized in bold. Refer to Table 2.1 for abbreviations. Refer to (2.2), (2.3), and (2.4) for definitions of the modes and the constraints. Note that all of the constraints have been met in these examples.

2KJockey 1920x1080 @30 FPS, 150 frames

| Mode | GOP | Profile | Time | Bitrate | PSNR |
|---|---|---|---|---|---|
| Max Q | B2 | SF | 4.8 | 4167.3 | **42.8** |
| Constraints | | | 5.0 | 5000.0 | |
| Min T | B2 | M | **6.9** | 1049.2 | 39.1 |
| Constraints | | | | 1300.0 | 39.0 |

Pedestrian 768x432 @25 FPS, 250 frames

| Mode | GOP | Profile | Time | Bitrate | PSNR |
|---|---|---|---|---|---|
| Min BR | ZL | Fr | 2.3 | **147.0** | 31.9 |
| Constraints | | | 3.0 | | 31.0 |

result from finding solutions that are close to the bounds required by at least one of the constraints. To see this, we consider the maximum quality mode in Table 2.3 that requires $T_{max} < 5$ seconds and $BPS_{max} < 5000$ bps. Then, the maximum quality mode requires 4.8 seconds of total encoding time that is close to the upper bound of 5 seconds. On the other hand, we note that there was a lot more bitrate that could have been used. Yet, an outstanding image quality of 42.8 dB with less bitrate is achieved.

## 2.4 Conclusion

In this chapter, we have presented a unifying framework that allows us to jointly optimize for encoding time, bitrate, and image quality. We introduced new GOP configurations that allow for fine optimization control. The system has been demonstrated to work well with videos characterized by translational motions.

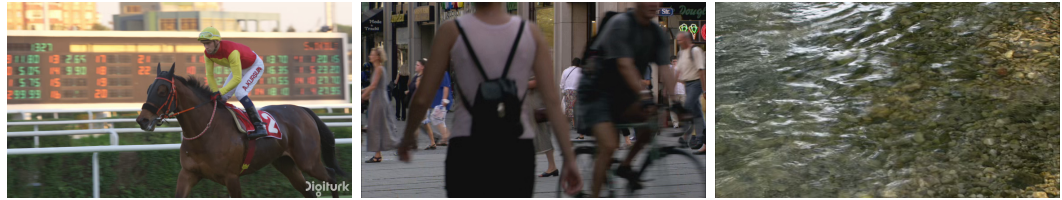

(a) Jockey [3].    (b) Pedestrian [1, 19].    (c) Riverbed [1, 19].

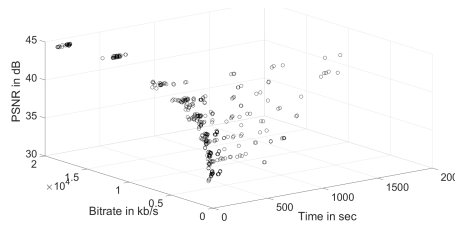

(d) Pareto front for UHD video: Jockey (1920x1080, 30 fps, 150 frames).

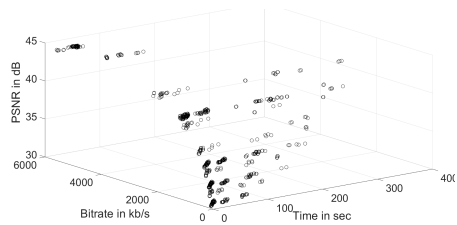

(e) Pareto Front for Pedestrian video (768x432, 25 fps, 250 frames).

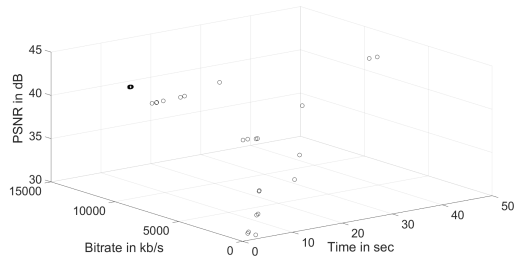

(f) Pareto Front for Riverbed video 768x432

Figure 2.3: Test videos and resulting Pareto fronts. (a) UHD video with strong predictable, translational motions. (b) Pedestrian video with multiple, yet predictable, translational motions. (c) Riverbed video with very complicated motions created by the flowing water. (d) Pareto front for UHD video demonstrating a relatively dense front. (e) Pareto front for Pedestrian video demonstrating a relatively dense front. (f) Pareto front for Riverbed video with fewer optimal points on pareto front.

# Chapter 3

# Adaptive Video Encoding based on Camera activity Classification

## 3.1   Introduction

We present a framework for adaptive video encoding based on video content. The basic idea is to analyze the video to determine camera activity (tracking, stationary, or zooming) and then associate each activity with adaptive video quality constraints. We demonstrate our approach on the UT LIVE video quality assessment database. We show that effective camera activity detection and classification is possible based on the motion vectors and the number of prediction units used in the HEVC standard. In our results, by applying leave-one-out validation, we get a 79% correct classification rate. We also present two examples for real-time, high-quality video encoding achieving bitrate savings of 35% and 51.5%.

The current chapter considers an adaptive encoding framework for effective video communications. Our goal is to automatically detect different video activities and associate quality constraints based on a specific task. Thus, we effectively compress the video for specific tasks that can be adjusted by the users or the owners of the video content.

To begin with, we note that video quality assessment is an area of active research as discussed in [34–37]. In our case, we consider a simple and fast method for assessing image quality based on SSIM as discussed in [38]. Furthermore, our approach is motivated by the well-known fact that visual attention is task dependent as documented in early research reported in [39] and also more recently in [40].

While viewers can have very different tasks that they are interested in, many times, it is possible to identify the goal of the photographer by analyzing the video content itself. In our approach, we identify video segments where the camera is moving, zooming, or held stationary and adaptively encode the video based on the content of each segment. For example, we interpret a camera zooming operation as an obvious attempt by the photographer to draw attention to his or her subject. As a result, we associate camera zooming with the need to encode the video at a higher video quality level. On the other hand, camera motions can be more difficult to interpret. If we associate camera motions as a search operation for obvious targets, then video quality can be lower than level used during zooming. On the other hand, if the camera motion is used to draw attention to the activity, we would expect higher video quality to visualize what is happening (e.g., in sports events). Thus, our focus is to provide a flexible framework that allows the users to adaptively encode the video based on different camera activities. We will next demonstrate our approach using two video examples.

We present an example that demonstrates camera tracking, zooming, and then held stationary in Fig. 3.1. Originally, the camera is following a presenter while he is pointing at different images of shields (see Fig. 3.1(a)). Once a particular shield of interest has been found, the camera motion ends, and the camera remains stationary on the target (see Fig. 3.1(b)). Then, the camera zooms on the target shield as shown in Fig. 3.1(c). For this example, we would require higher video quality during zooming and when the camera is held stationary over the found target.

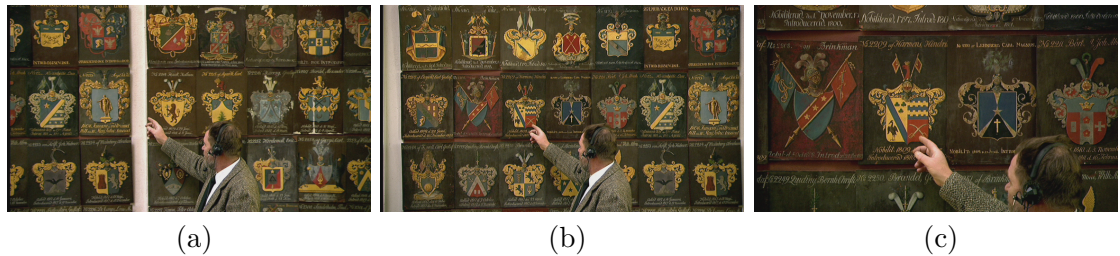

Figure 3.1:  Test video Shields from UT LIVE Video Quality Database [1].  (a) Camera moving as the man is pointing his finger at different shield images.  (b) Camera remains stationary over the target shield image. (c) Camera zooming in the particular shield that is of interest.

A second example that demonstrates different priorities is shown in Fig. 3.2. Here, as shown in Fig. 3.2(a), the camera is tracking the man as he runs. Then, the camera stops tracking as the man stands still for the remaining of the video (see Fig. 3.2(b)). Clearly, if we are interested in identifying the region where the man stops, we would require higher quality during the stationary phase of the video.

The current research is an extension of earlier, related work on selecting optimal

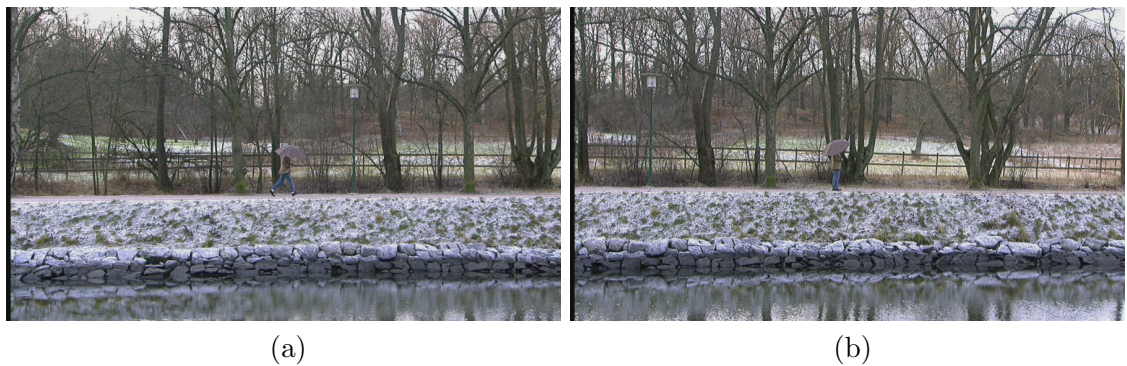

Figure 3.2:  Test video Parkrun from UT LIVE Video Quality Database [1].  (a) Camera moving and tracking the man during a running activity.  (b) Camera remains stationary when the man stops running.

HEVC encodings based on multi-objective optimization as reported in [41]. In [41], our focus was to select optimal video encoding for entire video sequences. The current paper represents a significant extension over [41] by developing an adaptive encoding paradigm.

The rest of the chapter is organized as follows. In section 3.2, we describe the underlying methodology and provide results in section 3.3. Concluding remarks are given in section 3.4.

## 3.2   Methodology

In order to implement the proposed adaptive video encoding approach, we will first need to develop a video activity classification system. Here, we develop a camera activity classification system based on HEVC features so as to minimize the computational complexity of our approach. We present a system diagram that summarizes the components of the adaptive video encoder in Fig. 3.3

We begin with a description of the camera activity classification system. Initially, we encode the video using B2 GOP since this basic prediction mode is subset to more advanced GOPs [41]. The bidirectional motion vectors (MV) and the number of prediction units (PU) are extracted from the encoded video to be used in the classification process. For feature vectors, we compute the magnitude and orientation histograms of the motion vectors using 25 bins and use them to provide estimates of the corresponding cumulative distribution functions (CDFs). We perform a nonparametric test to select histogram bins that can differentiate between the camera activities. Furthermore, using the selected features, we consider the use of a fast K nearest neighbor classification.

For dynamic adaptation, we rely on the use of the percentage change in the number of prediction units to detect camera activity changes. To understand how this works, consider a change from a moving camera tracking the object to a stationary position. Due to the complexity of the camera motion, tracking would be expected to include a substantial number of prediction units. On the other hand, when the moving camera becomes stationary, the number of prediction units will be substantially reduced as the complexity of the motions is also substantially reduced. Similarly, there will be a substantial increase in the number of prediction units when going from stationary camera to zooming. Furthermore, note that a tracking (moving) camera will normally stop moving before zooming.

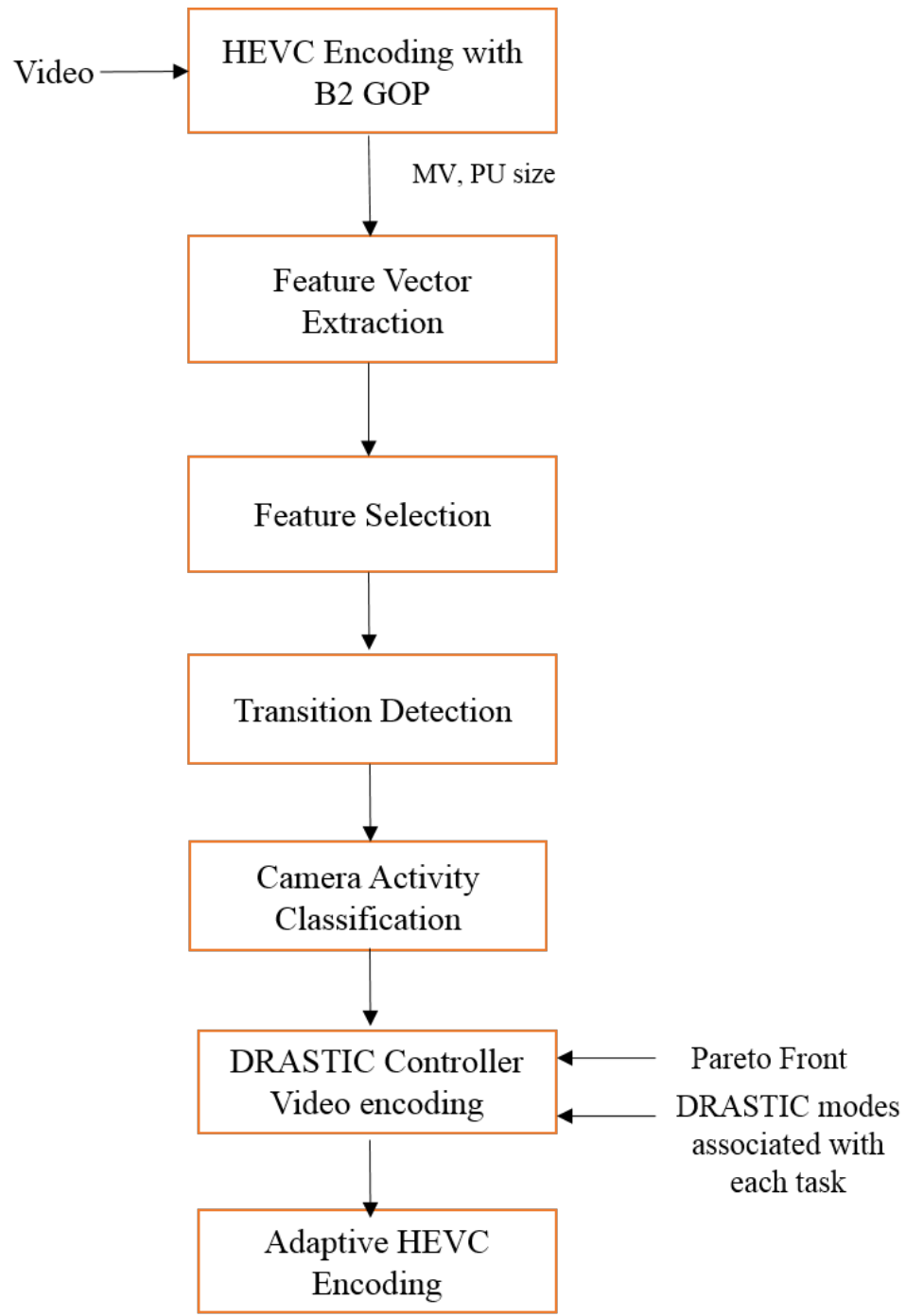

Figure 3.3: Block Diagram of Video activity detection with Classifier.

Once the camera activities have been successfully classified, we associate different video quality constraints for each task. For efficient video encoding, we consider the implementation of the minimum bitrate mode associated with the DRASTIC mode described in [41]. Here, we compute optimal QP and GOP encodings by solving:

$$\min_{\mathtt{EP}} \quad \mathtt{BPS} \quad \text{subject to} \quad (\mathtt{Q} \geq \mathtt{Q_{min}}) \text{ and } (\mathtt{T} \leq \mathtt{T_{max}}). \tag{3.1}$$

where $\mathtt{EP}$ denotes the encoding profile, $\mathtt{BPS}$ refers to the bits per sample, $\mathtt{Q_{min}}$ refers to the minimum acceptable video quality, $\mathtt{Q}$ refers to the achieved video quality, $\mathtt{T}$ refers to the encoding time, and $\mathtt{T_{max}}$ refers to the maximum allowable encoding time. Thus, in (3.1), we can achieve real-time encodings by controlling $\mathtt{T_{max}}$ and control encoding video quality by adjusting $\mathtt{Q_{min}}$. To solve (3.1), we can use the Pareto-front of optimal encodings as discussed in [41].

## 3.3 Results

We begin with a summary of camera activity classification. We then present results for adaptive video encoding for the video examples described in Figs. 3.1 and 3.2 in the introduction.

For camera activity classification, we establish ground truth by manually segmenting the UT LIVE video quality databases into tracking, stationary, and zooming

Table 3.1: Adaptive video quality and encoding time constraints based on camera activity classification. For all cases, we consider the minimum bitrate modes. For real-time encodings, we require that the total encoding time is less than 10 seconds for encoding the 500 frames (50 frames per second). For comparison, we consider leaving the same required video quality level (SSIM) over the entire video and the specific video region of interest. The bitrate savings result from the use of lower video quality constraints over video regions that are not of interest.

Shield video

| Mode | Frames | Activity | Constraints |
|------|--------|----------|-------------|
| Min bitrate | $1 - 500$ | NA | SSIM $\geq 0.94$ |
| | | | TIME $\leq 10$ |
| Min bitrate | $1 - 272$ | Track | SSIM $\geq 0.88$ |
| | | | TIME $\leq 10$ |
| Min bitrate | $273 - 364$ | Stationary | SSIM $\geq 0.94$ |
| | | | TIME $\leq 10$ |
| | $365 - 500$ | Zoom | SSIM $\geq 0.94$ |
| | | | TIME $\leq 10$ |

Park run video

| Mode | Frames | Activity | Constraints |
|------|--------|----------|-------------|
| Min bitrate | $1 - 500$ | NA | SSIM $\geq 0.94$ |
| | | | TIME $\leq 10$ |
| Min bitrate | $1 - 400$ | Track | SSIM $\geq 0.85$ |
| | | | TIME $\leq 10$ |
| Min bitrate | $401 - 500$ | Stationary | SSIM $\geq 0.95$ |
| | | | TIME $\leq 10$ |

Table 3.2: Camera activity classification results for three binary classifiers used to detect camera motion (tracking), stationary camera, and zooming.

| Classifier | Tracking | Stationary | Zoom |
|------------|----------|------------|------|
| Tracking vs | 5 | 0 | - |
| Stationary | 1 | 4 | - |
| Zoom vs | - | 4 | 0 |
| Stationary | - | 2 | 4 |
| Tracking vs | 4 | - | 1 |
| Zoom | 0 | - | 4 |

activities [1]. We ended up with 14 distinct camera video activity segments. Then,

Table 3.3:  Camera activity classification results for all video activities based on the binary classifiers of Table 3.2.  For the results, we use the UT LIVE video quality database [1] with $N = 10$ original videos segmented into 14 actual camera activities.

| **Classification** | Tracking | Stationary | Zoom |
|---|---|---|---|
| Tracking | 4 | 0 | 1 |
| Stationary | 0 | 4 | 0 |
| Zoom | 1 | 1 | 3 |

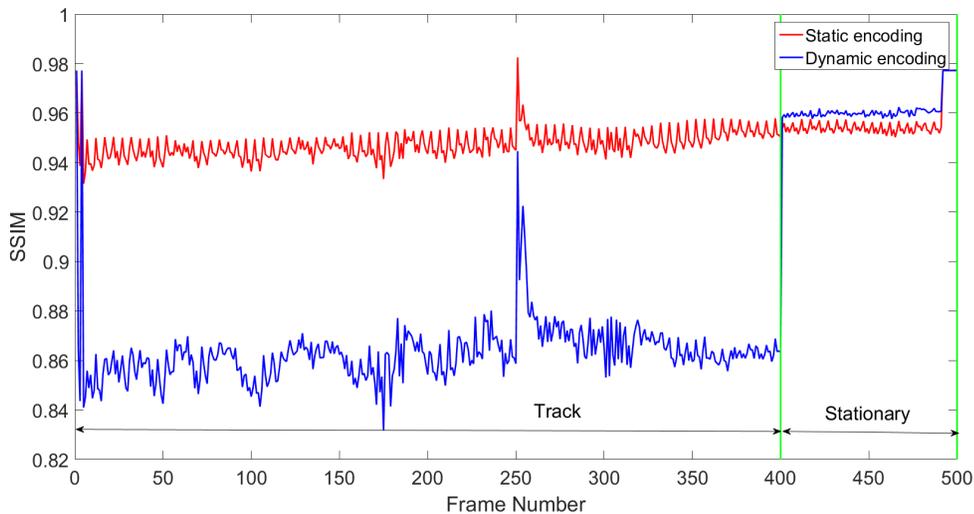

Figure 3.4:  Adaptive video encoding example for the Parkrun video from the UT LIVE Video Quality Database [1].  Refer to Table 3.1 for the bitrate constraints. Bitrate savings results from reducing the SSIM video quality constraint over the stationary portion of the video.

to differentiate among the activities, we design three binary classifiers as summarized in Table 3.2.  Furthermore, for each incoming video segment, we run all three binary classifiers and we use the number of activity wins to classify it.  Thus, for example, if the tracking classification wins in the two corresponding binary classifiers, the activity is classified as tracking.  We present the full confusion matrix in Table 3.3. Classification results were computed using leave-one-out cross validation.

From the results, it is clear that we can correctly classify camera activity from the HEVC features.  The impact of misclassification is minimized when we consider high-quality encodings as we do in our adaptive video encoding examples. Overall, we had a 79% correct classification rate.

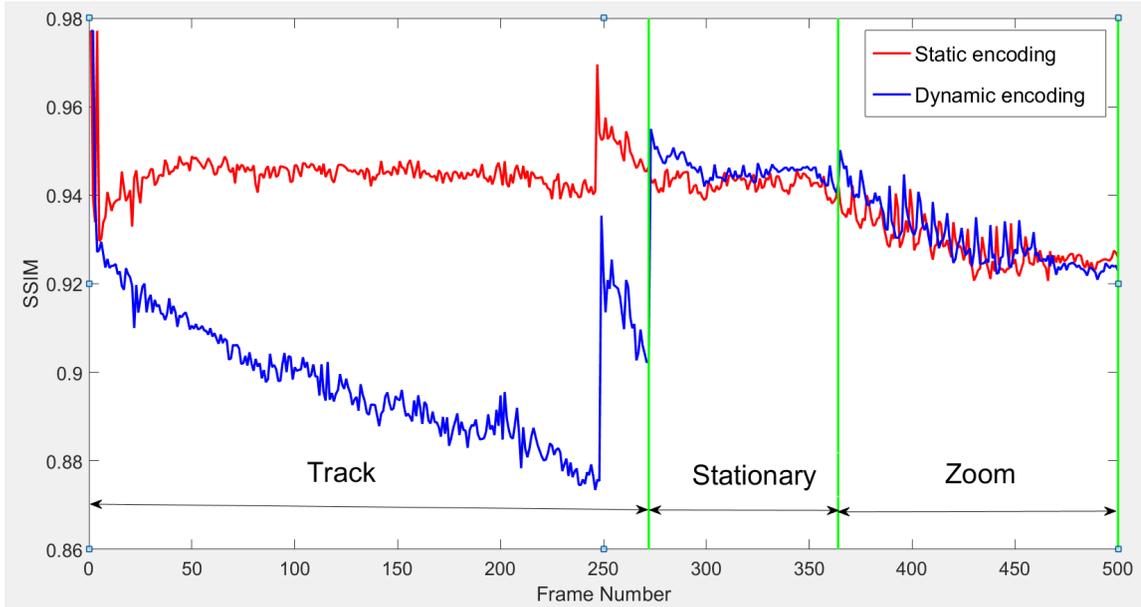

Figure 3.5:  Adaptive video encoding example for the Parkrun video from the UT LIVE Video Quality Database [1]. Refer to Table 3.1 for the bitrate constraints. Bitrate savings results from reducing the SSIM video quality constraint over the stationary portion of the video.

We next present adaptive video encoding results for the shield and park run videos considered in the introduction. For all video segments, we maintain high video quality requirements by requiring that SSIM remains above 0.85 (see [34]). For all cases, we require real-time encoding performance using the x265 software [24]. The basic idea is to maintain high video quality requirements during video regions of interest and reduce the requirements over the remaining video regions. For both video examples, our approach selected the correct encoding modes associated with each assigned task. Refer to Table 3.1 for the full description of the adaptive constraints that were selected.

For the shield video example, we have a reduced video quality requirement over the long tracking portion of the video as described in Table 3.1. On the other hand, we maintain high quality over the stationary and zooming portions of the video. As a result, we have substantial bitrate savings of 35%. Bitrate requirements were reduced from 996.22 kbps to 640.86 kbps.

We also present results for the park run video in Fig. 3.2. For this example, we increase video quality requirements at the end when the man stops. Recall that the goal here is to identify the location where the man stopped running. In this example, we have a 51.5% reduction in bitrate requirements from 5595 kbps to 2711.62 kbps.

## 3.4    Conclusion

This chapter presented an adaptive encoding method that uses video content to determine constraints on video quality for real-time encoding. The basic approach was demonstrated on identifying camera motions but could be extended to cover other types of video content. Overall, the approach shows that substantial bitrate savings can be attained depending on the length of the activity of interest.

# Chapter 4

# Segment-based x265 Encoding with Adaptive Local Pareto Models for Video On Demand (VoD)

## 4.1 Introduction to Segment-based encoding

Video streaming requires significant computing power, bandwidth, and memory so as
to deliver high-quality video under significant constraints. Streaming video technolo-
gies generally are resource (compute power, bandwidth, memory buffer and delay)
hungry, especially since end-users always desire high quality video, in spite of their
resource constraints. The main challenge that streaming video providers face is to
maximize the quality of experience the user desires subject to a wide variety of user
resource constraints. To address this challenge, we have to deal with hundreds of
encoding formats and associated storage requirements, in order to optimize quality of
content delivery for video on demand or live (real-time) services. Popular providers such as YouTube, Netflix and Amazon solve this real-time streaming quality problem by storing a couple of hundred pre-encoded container formats and deliver them based on user needs. For instance, YouTube uses a neural [42] net to deliver adaptive bitrate (ABR) streaming on the web. With millions of videos watched everyday [43], YouTube uses multi-pass video encodings targeting different bitrates [4] for each ABR segment, without requiring multi-pass encoding techniques to enable several millions of videos to their users. The neural-net model learns based from the video content and updates its model parameters using simple features taken from the video segments.

Netflix applies a brute-force approach of encoding each title/film category into 120 codec and bitrate combinations [44]. Each of these streaming platforms has its own encoding ladder, meaning that it targets specific bitrate per resolution such that the streams are encoded without significant artifacts. But this "One-size-fits-all" bitrate ladder, even though it achieves good quality encodings for certain bitrates, the methods cannot adapt to high camera motion or complex scenes. Given the diversity of Netflix movies/titles, this static encoding might store and encode video titles with best quality but not necessarily the optimal one because the static solution might store more bits than the allocated budget to achieve the same perceptual video quality.

Hence, we have the development of Per-Title encoding [45] which use machine learning techniques to select a couple of hundred encodings from a much larger set of possibilities. The selected bitrate-resolution combination tends to be efficient, in the sense that the encoded video is of high quality for the target bitrate. Netflix introduced the Video Multimethod Assessment Fusion (VMAF) [46] video quality metric to measure quality at different Constant Rate Factor (CRF) levels and bitrate-resolution pairs. Netflix uses the VMAF scores to identify the best quality resolution at each applicable data rate. The method only works for a smaller video dataset and it requires extensive computing resources to run hundreds of encoding combinations for each title/movie. In contrast, Per-Chunk [47] pushed boundaries to deliver videos at low bitrates especially using VP9 [13] and H.264/AVC [48]. Per-Chunk encoding fundamentally takes into small chunks of videos in minutes based on estimate of encode chunk complexity (in terms of motion, detail, film grain, texture) and with more encoding parameters produces mobile encodes that have same average bitrate for each chunk in a title with high video quality. Further tuning the methodology, Netflix transitioned into Per-Shot [49] encoding optimization with a Dynamic Optimizer [50] (DO) framework which essentially uses Spatio-Temporal characteristics of the video and builds an encoding ladder based on actual shot complexity. Optimal encoding parameters are chosen from the Convex hull so that it will satisfy the constraints and saves bit per shot. Although Per-Shot optimization does reduce bandwidth, its disadvantages come from its limited ability to adapt to video content, the use of an exhaustive number of combinations of bitrate-resolution pairs, and the lack of estimation of CRF levels or QPs from the encodings. In contrast, the proposed DRASTIC approach [17, 18, 41] allows for proper multi-objective optimization that infers the encoding parameters using predictive models that can also adapt to time-varying constraints.

This chapter presents a novel methodology to adaptively encode video with different content and camera motions. The basic idea here is to fit a Pareto surface using regression models and dynamically adapt them as the video is transmitted one GOP (Group of Pictures) at a time. The current chapter considers an adaptive encoding approach on a GOP level for effective video communications using an x265 encoder. For this, a versatile set of video databases with varied spatial and temporal motions were chosen with different resolutions as the input dataset to the x265 encoder. An offline approach is used to create a large number of encoding configurations for each individual video from the database with different GOP structures and other parameters to create the Pareto surface. Our main objective is to come up with a local model that starts with any GOP structure and switches adaptively depending on the DRASTIC mode constraints. Our goal is to make this system work with minimal computational requirements and model the pareto fronts per segment without strong bounds on computational complexity as duly noted in Per-Shot, Per-Chunk and Per-Title encoding approaches.

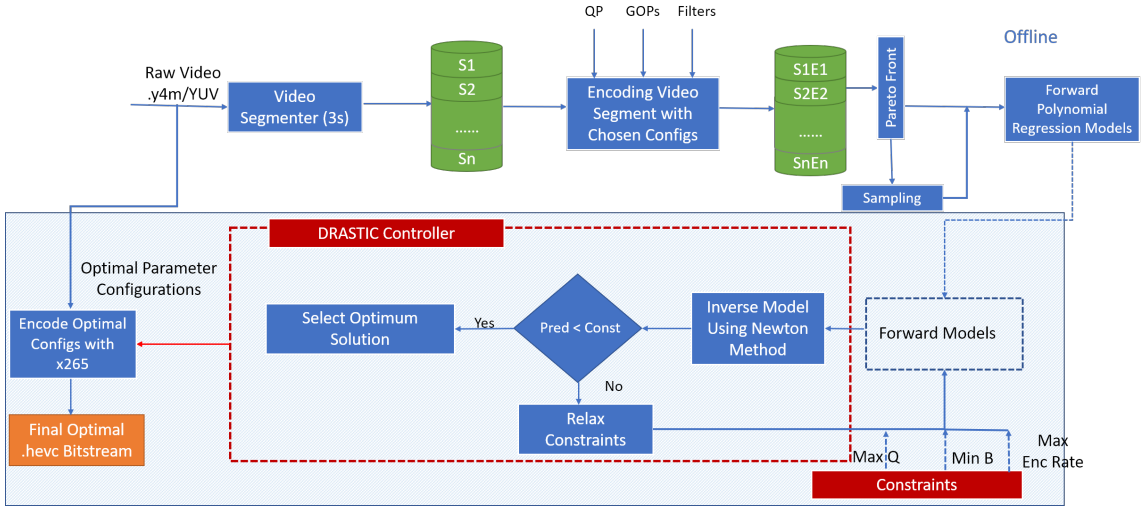

Figure 4.1: Block Diagram of Segment based Local Pareto Models with DRASTIC Control Modes.

HEVC provides new encoding configurations that allow users to compress videos using different presets that offer internal trade-offs with the encoding tools and provide a variety of mode decisions in rate-control to effectively encode videos for a given target bitrate or constant quality. The default preset in x265 is set to *medium* performs with good quality encodes without considerably overclocking or overusing the CPU resources as this implementation of HEVC encoding standard is known for its highly parallelized, multi-threaded operations which enables fewer options in the rate control so that the encoding is processed at real-time in a more efficient way. Compared that to the placebo mode, which is the last preset enables all the major encoding mode decisions for rate control and can produce the highest video quality but at the expense of enormous computational cycles and slower encoding times. So these presets each of them has a selected amount of mode decisions and as we go higher in terms of speed, x265 performs a faster encode but the compression efficiency is not at the best compression ratio which provides the means to define other presets and so on. Similar, to the encoding mode presets of x264 from Ultrafast to Placebo x265 has been implemented to provide a wide variety of encoding decisions to obtain the best bitrate compression ratio. These presets combined with different rate-distortion optimization (RDO), mode decisions, tuning parameters and with GOP structured can achieve the optimal quality without spending too much on the bits. As described in Section 2.2, we have introduced 3 more GOPs structures (B2, B4, B6) in x265 encoding configurations. Furthermore, there are strong variations in the performance of each video preset based on video content.

The current research uses a multi-objective optimization for designing and selecting among enhanced GOP structures for encoding. The basic approach relies on the use of the joint optimization of encoding time, bitrate requirements, and video quality to select the optimal Pareto point from the pareto surface which is fit to a regression model. These models vary for different GOPs and the content of the video determines the shape of the surface. Complex motions in the video force the Pareto surface model to use higher order polynomials (Quadratic, Cubic) while low motion videos use linear models. We implement our approach using the x265 [24] encoder for UT LIVE [1] (VQA) video quality database and HEVC [2] Standard test sequences.

In this chapter, we will develop methods to build forward regression models and inverse Newton's equations which will be adapted according to the DRASTIC operating modes. At the times of this thesis writing, the world is facing [11] COVID19 crisis where severe bandwidth limitation has occurred and all streaming platforms have reduced their bandwidths. DRASTIC provides very fine tuned solutions with higher video quality at low bandwidth scenarios and adapts with acceptable video quality well within the recommended bitrate ladders. The rest of this chapter is organized as follows. In section 4.2, we describe a brief account on video quality metrics, the underlying methodology and provide results in section 4.3 and concluding remarks

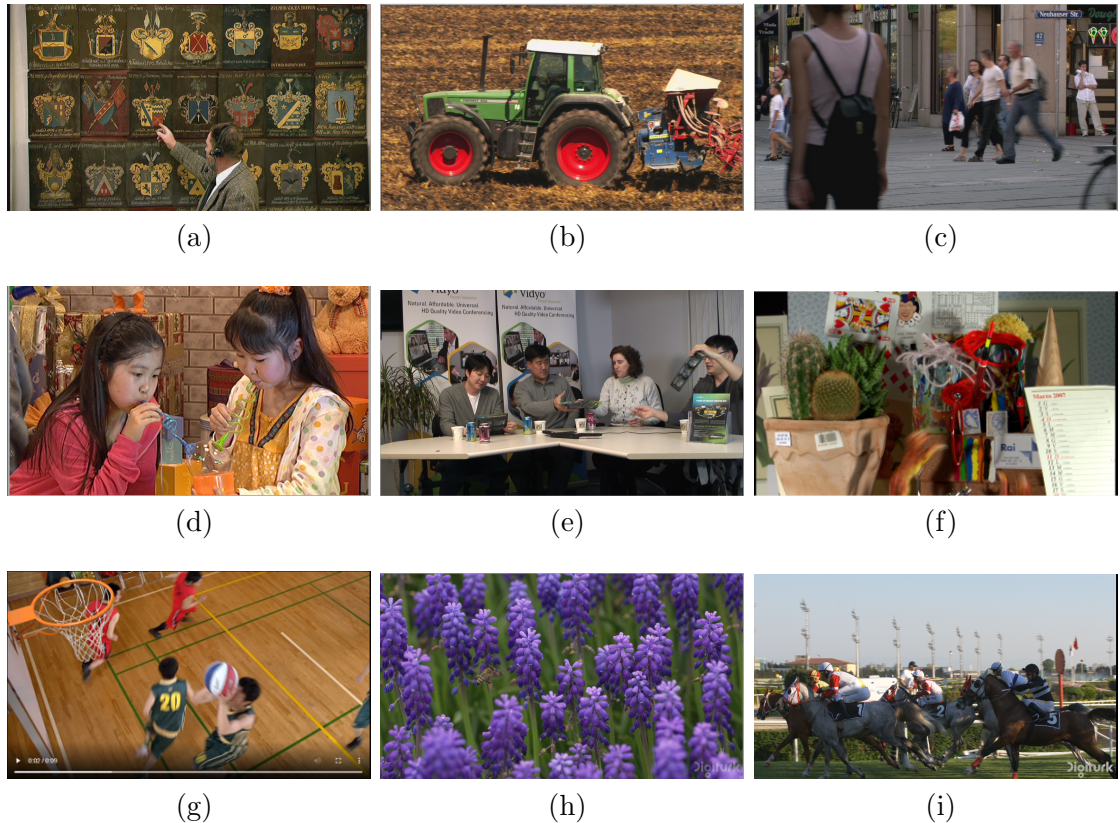

Figure 4.2: Test video Shields from UT LIVE Video Quality Database [1] (a), (b), (c) Shields, Tractor, Pedestrian video with resolution 768x432 of 50, 25, 25 fps respectively from UT LIVE Video Quality Database. (d) Blowing Bubbles of 480x240 from Class D with 50fps, (e) Four People with resolution 1280x720 from Class E with 60fps from HEVC Standard Test video sequences. (f), (g) Cactus, Basketball Drill video with resolution 1920x1080,50 fps and 832x480, 50fps respectively from HEVC [2] Video Test sequences. (h), (i) HoneyBee, ReadysetGo videos with resolution 1920x1080, 60 fps publicly available from Ultra Video group, Tampere [3] University.

are given in section 4.4.

## 4.2    DRASTIC x265 Segment-Based Encoding

### 4.2.1    VMAF - Video Multimethod Assessment Fusion

Peak-Signal-Noise-Ratio (PSNR) is used primarily as an objective video quality metric employed by all major encoding systems , which measures the intensity of the image to the average noise and more of a quality measure from a objective point of view and do not correspond very well perceptually. For example, a very high quality image with PSNR 44dB can still have visually noticeable artifacts even though the PSNR measurement says otherwise, and do represent how the video represents subjectively. As humans we are visually perceptive to intensity or other words brightness of the an image and this is well exploited in video compression and not necessarily represent perceptual video quality.

Video Multimethod Assessment Fusion (VMAF) [51–53] co-invented by Netflix combines human vision modeling and machine learning to measure the viewer's perception of streaming video content. VMAF measures multiple metrics on a frame level like spatial Index (SI) and Temporal Index (TI) and spatial feature extraction done from the pixel neighborhood. When videos are compressed and sent as a streaming content, they are bound to compressing artifacts like blocking, ringing and mosquito noise which cause poor video quality at user side who's viewing on their devices. To accurately measure human perception of video quality which is consistent across the video content, we need to evaluate video content by visual validation in addition to the PSNR, SSIM metrics. Typically, a VMAF score ranges from 0-100 which is mapped from the ACR scale category (20-Worse, 40-Bad, 60-Fair, 80-Good, and any score $\geq$ 90 - Excellent) as it has been trained using encoders ranging from CRF 22 1080p (highest quality) to CRF 28 240p (lowest quality). The former is mapped to score 100 and the latter is mapped to score 20. Also, in order to have a noticeable difference in visual quality, a VMAF score difference of at-least 6 should be established. In video quality research conducted by Netflix, visual quality degrades primarily due to two types of artifacts:

- ***Artifacts due to lossy Compression*** and

- ***Artifacts due to scaling*** (low bitrates, rebuffering [54], Lower bandwidth scenarios ,rebuffering [55]).

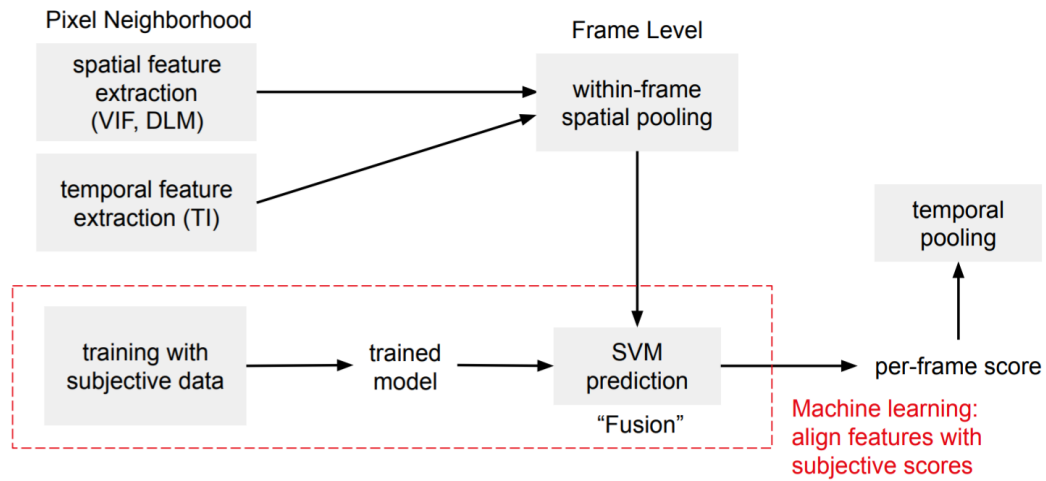

Figure 4.3: Block Diagram of VMAF Framework.
[56]

VMAF [57] was formulated to address the two aforementioned artifacts, which will outperform the objective video quality metric by giving an accurate prediction of how a human would have perceived. VMAF scores reflect subjective video quality assessment by combining multiple metrics using fusion techniques machine learning procedures. SVM regressors are deployed by fusing the elementary metrics as features with weights into final metrics which conserves all the intensities of the individual elementary metrics and presents the final subjective score. To obtain the machine-learning model shown in Figure 4.3, training and testing was done on Differential

Mean (DMOS) Opinion Scores obtained through the subjective experiment on Netflix dataset [58]. The elementary metrics used in VMAF framework consider both the Spatio-Temporal characteristics of a video content by taking into account the below features:

- *Visual Information Fidelity* (VIF) [59] A Full-reference image quality assessment metric that is built upon natural scene statistics of an image and correlates well with human visual system. VMAF uses a modified version inside the framework and governs the image quality of each video frame.

- *Detail Loss Metric (DLM)* [60] measures the loss of detailed information due to compression artifacts and textures of objects which severely impact the subjective quality. Both VIF and DLM represent the spatial feature representation of the video frames.

- *Motion*. The last feature is a temporal feature information of the video. This is achieved by calculating the temporal differences at pixel level of the luma component $Y$ between adjacent frames. By calculating this temporal feature, we obtain the motion characteristics of the video.

### 4.2.2   Video Encoding Configurations

We present a top to bottom approach in describing the proposed method in Figure 4.1. As described in Figure 4.4, we start by splitting the video into 3-second segments and encode them with different GOP, QP, filter combinations as a function of encoding configuration and for each objective video quality, bitrate, and encoding rate/time respectively we obtain their corresponding GOP models to be operated under their respective DRASTIC modes.

The inputs to the forward models were encoding configurations that include Closed GOPs, Quantization parameter (QP), Deblocking-SAO switching filter, Ultra-fast preset, Reference frames 1,3 CTU size 64 and threads enabled with a maximum of 8. We consider different GOPs (B2, B3, B4, B6, ZL) with an I Instantaneous Decoder Refresh (IDR) frame inserted every 150th frame or 3 secs.

We chose segment duration of 3s as the encoding representations since it is directly related to the VMAF model training of a few frames to calculate the subjective score for the particular segment. Also, this segment length is used in streaming based delivery like DASH [61, 62] which encodes up to 20 different combinations, meaning a variety of encoders, resolutions, target bitrates each with different segment lengths (1s, 2s, 4s, 6s, 10s, 15s) respectively. Additionally, this 3s segment length comprising of 150 frames in all our source video sequences will be optimal for the model to capture the entire encoding representation for that particular segment. For longer videos, we might accommodate different segment durations and it is totally suitable for our system to be adapted to include different segment lengths. We simply used GOP configurations, QP and filters to model the 3s segment and in the future, we will add more encoding decisions like MVs, RDO modes and test it on longer video sequences.

We considered different GOP structures B2, B4, B6 and ZL which stands for Zero Latency mode comprising of 'I' and 'P' frames and along with the default GOP B3 adds 5 different GOP representations for encoding. Further, with the fastest preset of x265 'Ultrafast' and with different encoding options using the HEVC x265 encoder as shown in Table 4.1 would add up to 200 encoding configurations per segment. With only these few parameters, we come up with a model that fits the Pareto front and predicts encoding parameters based on the constraints.

We measure Encoding Rate in frames per second (FPS), Bitrate in kilobits per second (kbps) and Video Quality using PSNR & VMAF. We first build sample space from different encoding configurations, fit the Pareto points and estimate the coefficients for local model individually per segment for each of the videos in the dataset. The local model predicts the objectives based on the constraints and, depending upon the DRASTIC [26] mode, can provide estimates for the next 150 video frames. Different encoding combinations were considered before we finalized configurations that directly impact the encoding visually and compression ratio. The model building process was kept simple by choosing quantization parameter (QP) which plays a huge role in the rate-distortion optimization (RDO) and hence the range of QPs were chosen from 16 to 45 (in steps of 3). Regarding the GOPs, we weighed upon both open and closed GOPs with individual structures (B2, B3, B4, B6, ZL) and additionally we added the Deblocking and SAO filters turned ON/OFF as these parameters directly impact the video quality for single pass encoding.

We used VMAF [63] SDK tool inside the encoding system to measure VMAF per segment and will be incorporated into the model building along with other objectives. The default VMAF model (model/vmaf_v0.6.1.pkl) is trained to predict the quality of videos displayed on a 1080p HDTV in a living-room-like environment. All the subjective data were collected in such a way that the distorted videos (with native resolutions of 1080p, 720p, 480p etc.) get rescaled to 1080 resolution and shown on the 1080p display with a viewing distance of three times the screen height (3H). Note that 3H is the critical distance for a viewer to appreciate 1080p resolution sharpness.

```
 1: function ADAPTIVE VIDEO ENCODING()
 2: ▷ Input: Video encoding parameters
 3: ▷ This procedure adaptively encodes the Video stream.
 4:
 5:    while  (more video GOP segments to encode) do
 6:        Allocate Constraints and choose Optimization mode
 7:        Compute all available configs Cfg1, Cfg2
 8:            and QP ranges QP_i and QP_n for different GOPs
 9:        Combine Configs and QP ranges into
10:            candidate sets C_all and QP_all
11:        Compute predicted objective values:
12:            PSNR_all, VMAF_all, FPS_all, Bits_all
13:            by applying the Forward Regression models to
14:            candidate sets C_all and QP_all.
15:        Compute Pareto-front by eliminating
16:            Points whose objectives are not
17:            Pareto-optimal
18:        Find the Optimal encoding Parameters
19:            selecting the C_Opt, QP_Opt by Newton's method that
20:            produce points that lie on the Pareto-front
21:            candidate sets C_all and QP_all.
22:        Robust parameter estimation and optimization for next segment
23:            Apply QP_all and C_all based on the current model.
24:            Solve optimization problem using local search.
25:        if either QP_all or C_all is out of range then
26:            Update constraints and fix encodings
27:                new estimates of QP and Cfg
28:            Constrain QP to be within ±4 of
29:                neighboring QP ranges.
30:            Enforce QP and Cfg within valid ranges.
31:            Use Previous Forward model with new estimates of QP_all & C_all
32:        end if
33:
34:        Encode the video using C_Opt and QP_Opt
35:        Compute PSNR_Opt, VMAF_Opt, FPS_Opt, Bits_Opt
36:            for current GOP segment
37:        Save by applying the regression models to
38:            candidate sets C_all and QP_all.
39:    end while
40: end function
```

Figure 4.4: Overview of DRASTIC Segment based encoding framework.

The need for an adaptive and dynamic video encoding implies that current systems use constant quality mode or constant rate factor as recommended by the encoding ladders [4], and there is no guarantee that these static systems provide efficient bitrate savings or render the video with a higher quality. All of these static systems employ one set of encoding parameters for all the videos not taking into consideration the varied motion content, textures and frame rates. The proposed dynamic system framework encodes videos by breaking them into small segments and then encodes them with different encoding combinations with various GOP structures. Also, this method is applicable to videos with varying spatio-temporal characteristics and different camera motions that occur in the video. For effective usage of bits, we employ the QP and for overall image quality we utilize both the filters. More encoding parameters can be added but we wanted to demonstrate the effectiveness of a simple segment based encoding system that offers greater flexibility in choosing the encoder parameters. Though it is an exhaustive encoding system but only for the first segment and then the forward model adapts as the video progresses. The Pareto modeling follows the constraints and does an efficient job of predicting the encoding parameters for the next segment without re-encoding. We present a system diagram that summarizes the components of the Segment based Local Pareto Models with DRASTIC modes in Table 4.1.

Table 4.1: x265 HEVC Encoder Configurations for Ultrafast Preset

| Parameter | Value | Parameter | Value |
|-----------|-------|-----------|-------|
| Presets | Ultrafast | Frame Threads | 8 |
| GOP Structures | B2,B3,B4,B6,ZL | SAO filter | On/Off |
| GOP Type | Open/Close | Deblocking filter | On/Off |
| QP | 16-45 | Tune | PSNR |
| Key-Interval | 25,30,50 | CTU | 64 |
| Total encoding combinations per segment | 200 | | |

### 4.2.3 Build Forward Regression Models based on x265 Configurations

We build the forward models with the encoding configurations as shown in the Table Though it is an exhaustive encoding system but only for the first segment and then the forward model adapts as the video progresses. The pareto modeling follows the constraints and does an efficient job of predicting the encoding parameters for the next segment without re-encoding the next segment. Our proposed method build forward models from the encoding configurations as an off-line system to build the modelsfor the first 3 seconds and then deploy the models to predict the encoding parameters instead of plain encoding the whole video segments. The model predicts the encoding objectives, filter settings and the quantization parameter and gives to the encoder resulting in an optimal way of encoding.Initially, several linear regression methods were explored and studied carefully with statistical package Python [64] different models fitting the Pareto points.

The model building is a cumulative process since we have to exhaustively combine so many different encoding configurations and then obtain the resulting objectives along with its parameter setting and store them as tables. For each GOP structure encoded we obtain the pareto points which is used in the model building with various encoding combinations and the resulting optimal models are saved to be used for the next segment. For all of these model fittings, the order of the model equations are varied from linear, quadratic and cubic order and also this varies depending on the video content.While constructing a model, we considered many parameters for the equation like Open/Closed GOP structure, different presets, tuning settings but it only made the modeling complicated. Hence, we tries Step-wise regression to find which parameters had a significant impact on the response variables. We simplified the model equation which started from 6 predictor variables to 2 which are QP and SAO and DBSA filters. Qp, being an integer has significant impact on the quality of the video as it controls the step size of the quantizer inside any codec and directly affects the rate-control mechanism. Deblocking and SAO filters on the other enhance the frame quality during reconstruction inside the codec buffer. So both, these variables have significant effect on the quality of the video and also they are simple two variable equations.

$$\ln(\text{PSNR})_i = \alpha_0 + \beta_1 \cdot \text{QP}_i + \beta_2 \cdot \text{QP}^2{}_i + \beta_3 \cdot \text{QP}^3{}_i$$

$$\ln(\text{VMAF})_i = \alpha_1 + \beta_{11} \cdot \text{QP}_i + \beta_{12} \cdot \text{QP}^2{}_i + \beta_{13} \cdot \text{QP}^3{}_i$$

$$\ln(\text{Bits})_i = \alpha_2 + \beta_{21} \cdot \text{QP}_i + \beta_{22} \cdot \text{QP}^2{}_i + \beta_{23} \cdot \text{QP}^3{}_i$$

$$\ln(\text{FPS})_i = \alpha_3 + \beta_{31} \cdot \text{QP}_i + \beta_{32} \cdot \text{QP}^2{}_i + \beta_{33} \cdot \text{QP}^3{}_i$$

where $\beta_1, \beta_{i,1}, \beta_{i,2}, \beta_{i,3}$ represent QP coefficients and, $\alpha_0, \alpha_1, \alpha_2, \alpha_3$ denote the constants of the polynomial regression equation.

We spent a lot of time on regression analysis to generate different model equations that can accurately describe the statistical relationship between QPs and the objectives PSNR, VMAF, Bitrates and FPS respectively. In the case of VMAF, which was later added to our DRASTIC Segment based encoding system, it directly corresponded to the QP variable when it was assessed. Hence we built the model equations that can measure both subjective (VMAF) and objective (PSNR) video quality together. The other objectives bitrate had a similar correspondence to QP and was not hard. The only objective that was harder and perhaps sophisticated was encoding rate or FPS which was quite difficult to do the model fit as the Adjusted R squared value often falls below 0.7 as for all model fitting we generally keep a higher threshold of 0.9 to satisfy the model criteria.

Another factor that we analyzed with our model equations is how well the predicted variable in our case QP statistically related to the response variables (in our case the objective VMAF, PSNR, Bitrate, FPS) is given by the p-value which ranges from 0 to 1. Smaller p-value of ($\leq 0.05$) [64] indicate that any changes in the QP will have a significant impact on the responses (objectives) while larger values do not have any changes that impacts the objectives. The other important thing is to notice the estimated coefficients of filters are significantly lower than QP. Other factors include were the complexity of the video content as high motion videos often end up with Quadratic or a Cubic model fit.

### 4.2.4   Estimating Inverse Models by Newton's Method

Following the forward model for each GOP built, we solve for the optimal encoding parameters using the inverse Newton's method depending on the DRASTIC mode, and the corresponding encoding constraints. For example, in maximum video quality mode, we obtain bitrate and encoding rate constraints as inputs to the model building. The Pareto based system then finds the suitable forward model equation, and, an inverse prediction method uses the forward model equation from encoding rate and bitrate and solves for a QP that maximizes the quality of the video. We apply the Newton method starting with QP=27, which is the default QP for x265 encoder and terminate the search for an optimal QP when the estimated QP remains unchanged.

The QP values generated by the prediction is a floating point value and we approximated to the real-integer as the encoders accept only integer based QP value. By far, there might be prediction errors from the system accounting to forward modeling process so we allow soft violations say 10% for bitrates and encoding frame rates and finally 3-5% for video quality respectively. By this, we generate multiple solutions for QP which in our case is the dominant predictive variable and any error might significantly affect the objectives. So, we carefully determine the QP values generated by the Newton method by estimating whether they can obey the constraints and if in case of a failure it will do a local search around the QP neighborhood which is in the case $(QP + 4 , QP - 4)$ and then repeat the prediction process again until the constraints are satisfied. Let's start with maximum video quality mode, where the bitrate constraint is dominant factor compared to the FPS. Now using the constraints and from the fitted forward models, we deploy the Newton's inverse equation to predict a AP value that satisfy the constraints within the threshold/violations. Thus, we get maximum four QP values from both the bitrate and encoding rate models without any violations in the constraints. By combining multiple QP solutions generated from the inverse model and then applying the constraints we obtain the optimal encoding parameters for that particular segment and then encode them. The resulting objectives VMAF, PSNR, Bitrates, FPS calculated for that segment is within the constraint bounds and if there is a violation, the system executes a local search and recalculates the encoding parameters and then encode the segment. By this mechanism we can always have a constrained optimal solution that is always within the constraint and we can solve any constrained optimization problem provided we relax the violations.

## 4.3 Results and Discussions

We begin with a summary of DRASTIC modes of operation for adaptive video encoding.

- **Maximum video quality mode:**

$$\max_{\text{EP}} \text{Q} \quad \text{subject to} \quad (\text{BPS} \leq \text{BPS}_{\text{max}}) \text{ and } (\text{FPS} \geq \text{FPS}_{\text{min}}). \tag{4.1}$$

    Here, the goal is to reconstruct the video with the highest possible video quality than does not require more bandwidth that is available and within reasonable encoding time.

- **Minimum bitrate mode:**

$$\min_{\text{EP}} \text{BPS} \quad \text{subject to} \quad (\text{Q} \geq Q_{min}) \text{ and } (\text{FPS} \leq \text{FPS}_{\text{min}}). \tag{4.2}$$

In this mode, the goal is to minimize bandwidth requirements provided that the video is of sufficient quality and we do not spend a large amount of time encoding it.

- ***Maximum encoding Rate mode:***

$$\min_{\mathtt{EP}} \mathtt{T} \quad \text{subject to} \quad (\mathtt{Q} \geq Q_{min}) \text{ and } (\mathtt{BPS} \leq \mathtt{BPS_{max}}) \tag{4.3}$$

In this mode, the goal is to maximize the frame rate provided that the video can be communicated within the given bitrate and it is of sufficiently good quality.

All of the Segment-based encoding was implemented using the x265 open source software run on a Windows 10 Dell Precision Tower 7910 Server 64-bit platform with Intel(R) Xeon(R) Processor E5-2630 v3 (8 cores, 2.4GHz, Turbo, HT, 20M, 85W). In what follows, we summarize the benefits of considering different encoding configurations in the proposed adaptive framework, describe the resulting prediction models, highlight the significance of using Pareto optimal solutions, and demonstrate adaptive video encoding efficiency compared to YouTube recommended standard bitrates per resolution and we apply per each segment.

## 4.3.1   Maximum Video Quality Mode
### Basketball Drive Video HEVC 1080p Dataset

In this optimization mode, the objective is to maximize the video quality while conforming to bandwidth constraints in terms of typical upload data rates as recommended by YouTube [4]. We demonstrate this using Basketball Drive Video from Class-B HEVC test Sequence [2] where a bunch of players passing around the ball in the basketball court with a duration of 10s and a frame count of 501. Note this video involves a lot of motions as all the players are continuously moving on all the frames. As per YouTube suggestions for 1080p video, the recommended bitrate is 12000kbps. In our demonstration we have two encoding settings. The Default mode is where we use a QP value that approaches/achieves the recommended bitrate for each segment and then we do an average across the whole video to obtain the PSNR, VMAF, Bitrates and FPS, respectively. In the DRASTIC mode, we give the overall average bitrate and encoding rate (FPS) as the constraint to model the objectives in each segment. We will next provide a summary of the Basketball Drive video.

| Seg ID | CQP | Fil | GOP | Bitrate (kbps) | PSNR (dB) | VMAF |
|--------|-----|-----|-----|----------------|-----------|-------|
| Seg0 | 28 | On | B3 | 10825.29 | 38.73 | 96.98 |
| Seg1 | 28 | On | B3 | 11676.14 | 38.34 | 95.86 |
| Seg2 | 28 | On | B3 | 11037.35 | 38.424 | 96.21 |
| Seg3 | 28 | On | B3 | 11441.38 | 38.113 | 95.45 |
| ***Avg*** | | | | **11205.77** | **38.45** | **96.26** |

Table 4.2: Default Mode - YouTube Recommended Bitrate achieved by CQP.

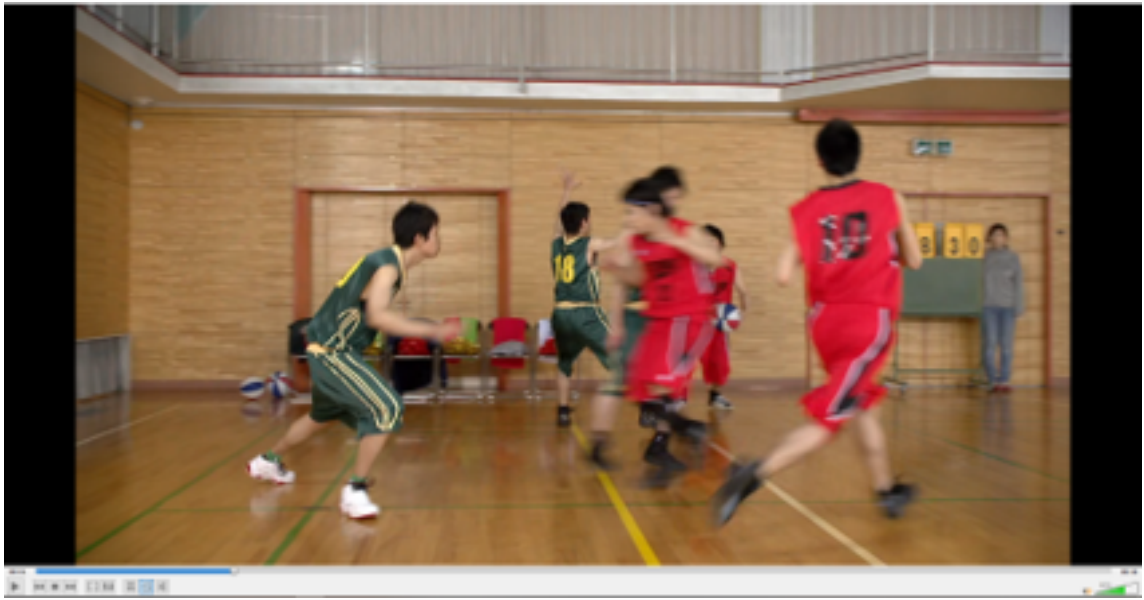

Figure 4.5: BasketballDrive from HEVC [2] Video Sequence,1920x1080, 50fps.

Here we break the video into 3s segments which gives a total 4 segments with three 3 second segments and one 1s segment. We then encode each segment with the following settings to achieve what YouTube recommended as a bitrate for that resolution which is summarized in Table 4.2. Using a QP value of 28, with default GOP B3 and both the filters Deblocking and SAO turned ON, the default mode achieves an overall average bitrate of 11205.77 kbps, PSNR 38.45dB and VMAF 96.26, respectively.

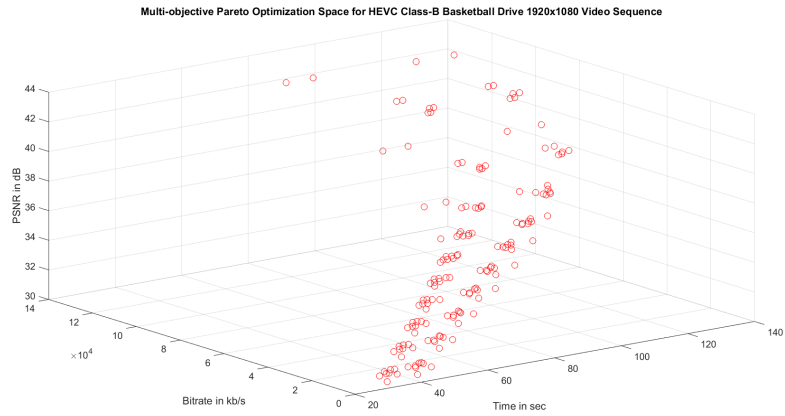

Figure 4.6: Pareto Space for BasketballDrive 1920x1080, 50 fps.

## Recommended video bitrates for SDR uploads

To view new 4K uploads in 4K, use a browser or device that supports VP9.

| Type | Video Bitrate, Standard Frame Rate (24, 25, 30) | Video Bitrate, High Frame Rate (48, 50, 60) |
|---|---|---|
| 2160p (4k) | 35-45 Mbps | 53-68 Mbps |
| 1440p (2k) | 16 Mbps | 24 Mbps |
| 1080p | 8 Mbps | 12 Mbps |
| 720p | 5 Mbps | 7.5 Mbps |
| 480p | 2.5 Mbps | 4 Mbps |
| 360p | 1 Mbps | 1.5 Mbps |

Figure 4.7: YouTube Recommended Bitrates for different resolutions [4].

We then take the Default's average PSNR, Bitrate and FPS as constraints to the maximum video quality mode. The B6 GOP is the optimal GOP picked from the quadratic model based on the constraints with the coefficients reported in Table 4.3. All the objectives PSNR, VMAF, Bitrate and encoding rate for B6 GOP have a higher adjusted R square values 0.99, 0.99, 0.99 and 0.99, respectively. This model equation is employed to predict optimal encoding parameters for the next segments even for the minimum bitrate mode as well. In this section, we will present videos of different resolutions with maximum video quality and minimum bitrate modes.

In the Basketball Drive video, for the first segment with bitrate constraint as 11205.77 and encoding rate (FPS) constraint greater than 25, the inverse prediction methodology described in Section 4.2.4 obtains an optimal encoding configuration with GOP B6 with both Filters ON and encodes the first segment with a bitrate of 10639.83 kbps with a PSNR value of 38.771 dB and VMAF 97.008 and FPS as 44.12 meeting all the constraints. On close examination from Figure 4.8, we can see that the default has a bitrate of 10825.29 kbps and obtains a PSNR 38.73 dB with VMAF 96.98, whereas our maximum quality mode uses 10639.83 kbps and 38.77 dB and VMAF of 97.008 slightly higher in the quality at a bitrate lower than default mode. From the model equation of B6 GOP, we predict the optimal encoding parameters for the second segment. As evident from table 4.4, the second segment has a higher bitrate requirement since there is a high motion involved between the players so the default uses up to 11676.14 kbps, and gives 38.34 dB, respectively.

DRASTIC gives higher values of PSNR which is 38.41 dB, (with 0.07dB) at a lower bitrate 11124.35 kbps, obeying the constraints. For the third and fourth segments, DRASTIC achieves a significant increase in PSNR video quality of 38.477 dB and 38.23 dB with significant increase in the video quality. In maximum quality mode, DRASTIC achieves an overall higher quality while saving bitrates in each segment. The proposed framework adjusts to this change by considering finer improvements in quality per segment by employing a QP of 28 while using the same encoding structure B6. Real-time encoding performance is also maintained. Here, there is a mild violation of 10 % in terms of bandwidth demands and 10 % for encoding rate which is, however, within the acceptable limits. Overall, we save up to 2.40 % in bitrate and a PSNR improvement of 0.07 dB and a corresponding improvement

| *Coefficients* | $\beta_0$ | $\beta_1$ | $\beta_2$ | GOP.Str | Model Order | Adjusted $R^2$ |
|---|---|---|---|---|---|---|
| log(PSNR) | 3.866 | -0.005 | -6.521e-05 | B6 | Quadratic | 0.99 |
| log(VMAF) | 3.965 | 0.058 | -0.001298 | B6 | Quadratic | 0.99 |
| log(Bits) | 15.946 | -0.304 | 0.0024092 | B6 | Quadratic | 0.99 |
| log(EncRate) | 1.872 | 0.095 | 0.0098901 | B6 | Quadratic | 0.99 |

Table 4.3: Model Equations for Maximum Video Quality Mode

of 0.08 in VMAF shown in Table 4.5.

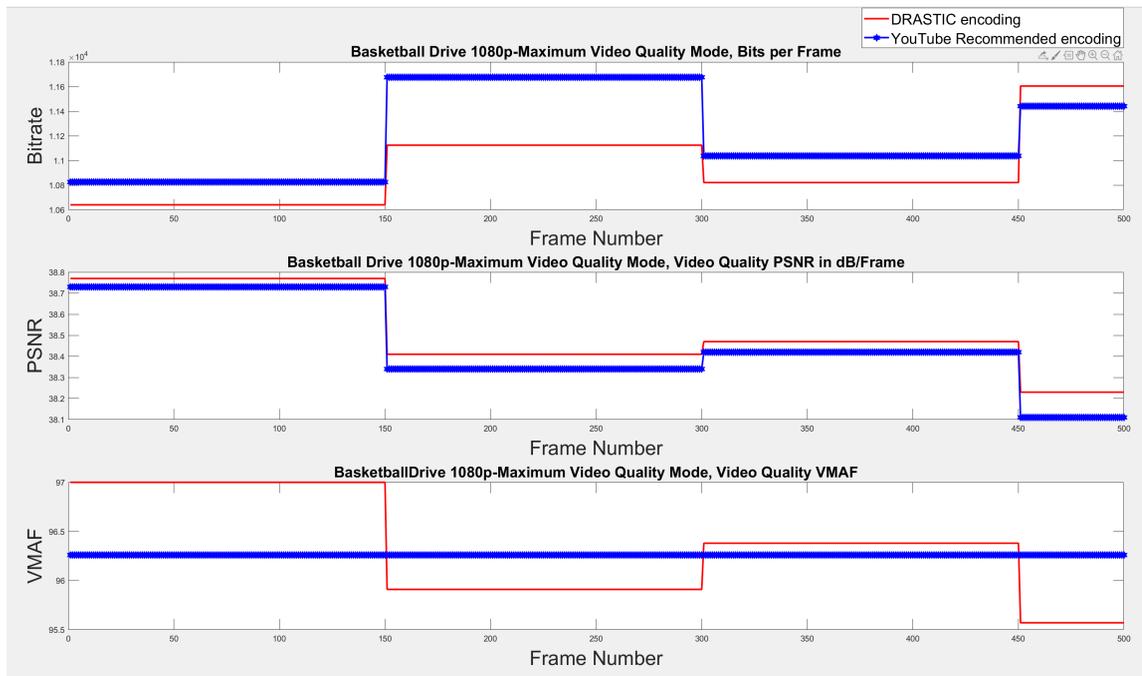

Figure 4.8: HEVC Test sequence, 1920x1080, Basketball Drive maximum Quality Mode.

| *Seg ID* | *QP* | *Fil* | *GOP* | *Bitrate (kbps)* | *PSNR (dB)* | *FPS* | *VMAF* |
|----------|------|-------|-------|------------------|-------------|-------|--------|
| Seg0 | 28 | On | B6 | **10639.83** | 38.771 | **44.12** | 97.008 |
| | | | | <=11205.77 | | >=25 | |
| Seg1 | 28 | On | B6 | **11124.35** | 38.41 | **44.84** | 95.91 |
| | | | | <=11205.77 | | >=25 | |
| Seg2 | 28 | On | B6 | **10820.78** | 38.477 | **46.3** | 96.38 |
| | | | | <=11205.77 | | >=25 | |
| Seg3 | 28 | On | B6 | **11604.94** | 38.232 | **41.25** | 95.57 |
| | | | | <=11205.77 | | >=25 | |
| *Avg* | | | | **10935.982** | **38.52** | **44.7** | **96.34** |

Table 4.4: DRASTIC Maximum Video Quality Mode for Basketball Drive 1920x1080, 50 *fps*.

| *Overall Bitrate Gain* | *Overall PSNR* | *Overall VMAF* |
|------------------------|----------------|----------------|
| 2.40 % | 0.07 dB | 0.08 |

Table 4.5: Overall DRASTIC Gains from Maximum Quality Mode.

## 4.3.2    Minimum Bitrate Mode
## Basketball Drive Video HEVC 1080p Dataset

In the minimum bitrate demands mode, the goal is to minimize bandwidth require-
ments while maintaining acceptable video quality and real-time performance. Such
scenarios are likely to occur in disaster incidents like COVID19 [11] with many people
accessing the network in a crowded area and also in developing countries where wire-
less networks resources are unstable and shared by many users. Here, there is a mild
violation of 5 % in terms of quality demands and 10 % for encoding rate. We use the
default's average PSNR 38.45 dB as an acceptable video quality while maintaining a
minimum FPS above 25 as constraints per segment. In Table 4.6, DRASTIC for the
first segment achieves a PSNR of 38.52 dB at 9477.13 kbps and maintains a higher
FPS of 42.04 and 95.58 for VMAF score. For the next two segments, DRASTIC
obtains a PSNR of 38.21dB and 38.45dB which is 0.24 dB & 0.15 dB less than the
default mode while maintaining bitrates of 9986.59 and 9591.56 kbps, respectively.
For the last segment, DRASTIC convincingly wins with a PSNR higher than 38.45
dB.

In Figure 4.9, DRASTIC reaches above the minimum acceptable PSNR in the
first and fourth segments but overall, the minimum bitrate mode saves 13.41 %
while losing around 0.06 dB in video quality. For a human, this video will still be
perceived as high quality even though the objective video quality metric PSNR has
lower values in the second and third segments. Overall, a PSNR difference of -0.06
dB and subjective video quality VMAF scores an overall difference of only -0.72
which cannot be distinguished from the default video which has a 96.26 as VMAF
score. DRASTIC here has provided finer optimization with encoding parameters and
the model prediction can significantly reduce the bitrate demands while still produce
videos at a higher quality which overall it saves 13.41 % in bitrate gains.

| Seg ID | QP | Fil | GOP | Bitrate (kbps) | PSNR (dB) | FPS | VMAF |
|--------|----|----|-----|----------------|-----------|-----|------|
| Seg0 | 29 | On | B6 | 9477.13 | 38.523 | 42.04 | 95.58 |
| | | | | | >=38.45 | >=25 | |
| Seg1 | 29 | On | B6 | 9986.59 | 38.21 | 42.18 | 94.5 |
| | | | | | >=38.45 | >=25 | |
| Seg2 | 29 | On | B6 | 9591.56 | 38.29 | 43.52 | 95.11 |
| | | | | | >=38.45 | >=25 | |
| Seg3 | 29 | On | B6 | 9853.99 | 38.883 | 41.39 | 94.09 |
| | | | | | >=38.45 | >=25 | |
| Avg | | | | 9701.98 | 38.39 | 43.23 | 95.541 |

Table 4.6: DRASTIC Minimum Bitrate Mode for Basketball Drive 1920x1080, 50 fps.

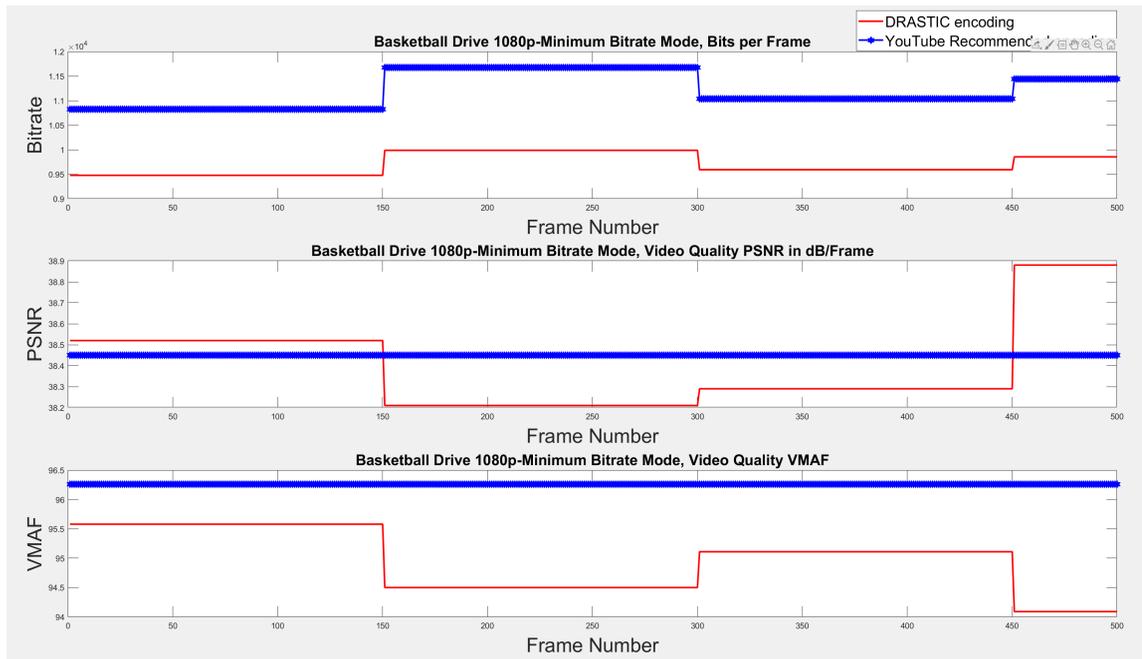

Figure 4.9: HEVC Test sequence, 1920x1080, Basketball Drive minimum Bitrate Mode.

| *Overall Bitrate Gain* | *Overall PSNR* | *Overall VMAF* |
|---|---|---|
| 13.41 % | -0.06 dB | -0.72 |

Table 4.7: Overall DRASTIC Gains from Minimum Bitrate Mode

### 4.3.3 Maximum Video Quality Mode
### Cactus Video HEVC Dataset

In the second example, we take Cactus video of 1920x1080 resolution and 500 frames from Class-B HEVC test sequence [2]. Cactus video has a toy moving in circular direction, faces on poker cards in the background rotating and a Cactus plant revolving around with the distinct spines on its surface. All of these features make this a harder video to encode with added complexity of multiple objects with different textures and hard to encode. Especially, the spines on the cactus are hard to capture during the encoding because the revolving the spines have complex textures. Here, there is a switching of GOP occurring that effectively captures all of these motions and the complexity of the textures. As per YouTube's recommended bitrate, for the default mode we encoded each of the segments and then we did an average for all the frames and obtained PSNR, bitrate, VMAF and FPS, respectively. We will now describe the maximum video quality mode using this Cactus video.

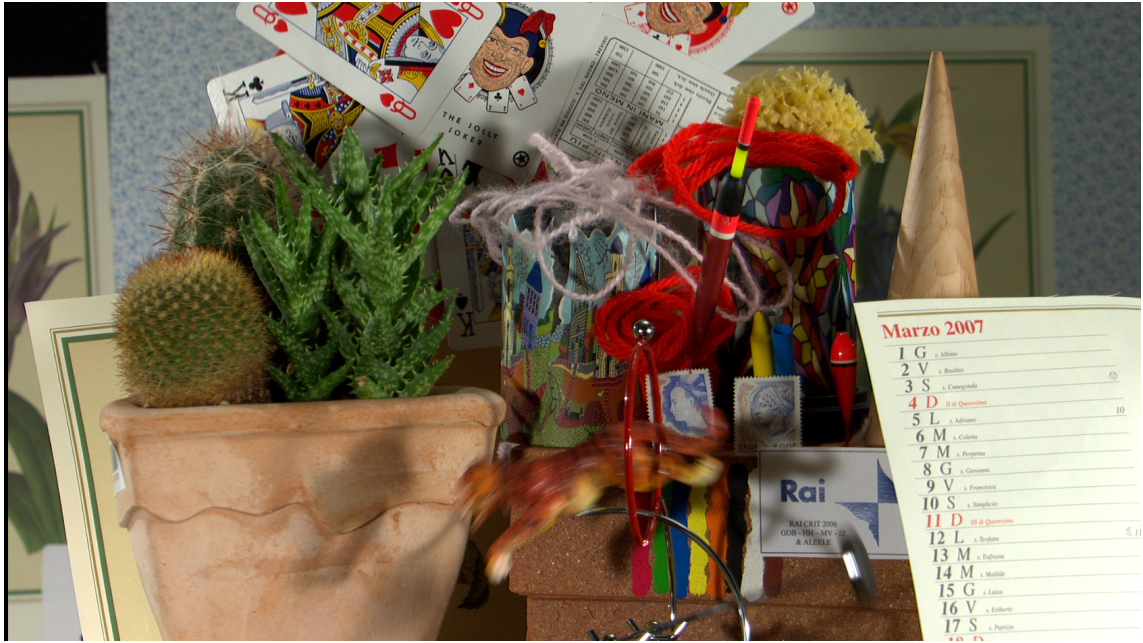

Figure 4.10: Cactus from HEVC [2] Video Sequence,1920x1080, 50fps.

| Seg ID | CQP | Fil | GOP | Bitrate (kbps) | PSNR (dB) | VMAF |
|--------|-----|-----|-----|----------------|-----------|-------|
| Seg0 | 28 | On | B3 | 11129.81 | 37.12 | 92.48 |
| Seg1 | 28 | On | B3 | 10148.02 | 37.23 | 92.85 |
| Seg2 | 28 | On | B3 | 11313.50 | 37.09 | 92.42 |
| Seg3 | 28 | On | B3 | 10347.74 | 37.30 | 92.95 |
| *Avg* | | | | **10812.173** | **37.16** | **92.62** |

Table 4.8: Default Mode - YouTube Recommended Bitrate achieved by CQP.

| Seg ID | QP | Fil | GOP | Bitrate (kbps) | PSNR (dB) | FPS | VMAF |
|--------|-----|------|------|----------------|-----------|---------|--------|
| Seg0 | 28 | On | B3 | **10093.05** | 37.399 | **52.93** | 92.1 |
| | | | | <=10812.173 | | >=25 | |
| Seg1 | 28 | On | B3 | **9191.61** | 37.514 | **51.8** | 92.4 |
| | | | | <=10812.173 | | >=25 | |
| Seg2 | 28 | On | B2 | **10260.5** | 37.382 | **52.91** | 92.06 |
| | | | | <=10812.173 | | >=25 | |
| Seg3 | 28 | On | B2 | **10377.76** | 37.662 | **46.99** | 92.58 |
| | | | | <=10812.173 | | >=25 | |
| **Avg** | | | | **9901.32** | **37.45** | **51.99** | **92.22** |

Table 4.9: DRASTIC Maximum Video Quality Mode for Cactus 1920x1080, 50 *fps*.

In the Cactus video, for the first segment with 10812.173 kbps as bitrate constraint and encoding rate (FPS) constraint greater than 25, the inverse equation predicts the optimal encoding parameters as follows: GOP B3, both filters ON. This results in the first segment being encoded with a bitrate of 10093.05 kbps, PSNR 37.399 dB, VMAF of 92.10 and achieving 52.93 fps, respectively. The second segment is encoded with B3 GOP with a bitrate of 9191.61 kbps, PSNR 37.514 dB, 51.8 fps and VMAF 92.40 still within the constraints. The model equations for the corresponding segments for each GOP is given in Tables 4.10 and 4.11 where both the GOP model orders were quadratic which correlates to complex motions occurring in the video.

For the third segment, there is a GOP switch to B2 which encodes with a bitrate of 10260.5 kbps obtaining a PSNR of 52.91 dB and VMAF 92.06 as in this segment the faces on the poker card and the spines make slightly prominent movement and in the last segment, B2 GOP is able to manage with bitrate of 10377.76 kbps, PSNR 37.662 dB and VMAF of 92.22 with a very high quality as shown in Figure 4.11, achieving an overall gain of 8.1 % with 0.29 improvement in PSNR and very negligible loss in VMAF -0.4 given in Table 4.12.

| *Coefficients* | $\beta_0$ | $\beta_1$ | $\beta_2$ | GOP.Str | Model Order | Adjusted $R^2$ |
|---|---|---|---|---|---|---|
| log(PSNR) | 3.86 | -0.00661 | -7.489899e-05 | B3 | Quadratic | 0.99 |
| log(VMAF) | 3.80 | 0.069566 | -0.0015476 | B3 | Quadratic | 0.99 |
| log(Bits) | 16.65 | -0.319803 | 0.002179 | B3 | Quadratic | 0.99 |
| log(EncRate) | 0.706 | 0.153 | -0.001585 | B3 | Quadratic | 0.96 |

Table 4.10: B3 GOP Model Equations for Maximum Video Quality Mode.

| *Coefficients* | $\beta_0$ | $\beta_1$ | $\beta_2$ | GOP.Str | Model Order | Adjusted $R^2$ |
|---|---|---|---|---|---|---|
| log(PSNR) | 3.89 | -0.00854 | -4.36249e-05 | B2 | Quadratic | 0.99 |
| log(VMAF) | 3.84 | 0.066985 | -0.00149858 | B2 | Quadratic | 0.98 |
| log(Bits) | 16.97 | -0.337398 | 0.00244438 | B2 | Quadratic | 0.99 |
| log(EncRate) | 0.69 | 0.1580597 | -0.001727 | B2 | Quadratic | 0.97 |

Table 4.11: B2 GOP Model Equations for Maximum Video Quality Mode.

| *Overall Bitrate Gain* | *Overall PSNR* | *Overall VMAF* |
|---|---|---|
| 8.4 % | 0.29 | -0.4 |

Table 4.12: Overall DRASTIC Gains from Maximum Video Quality Mode.

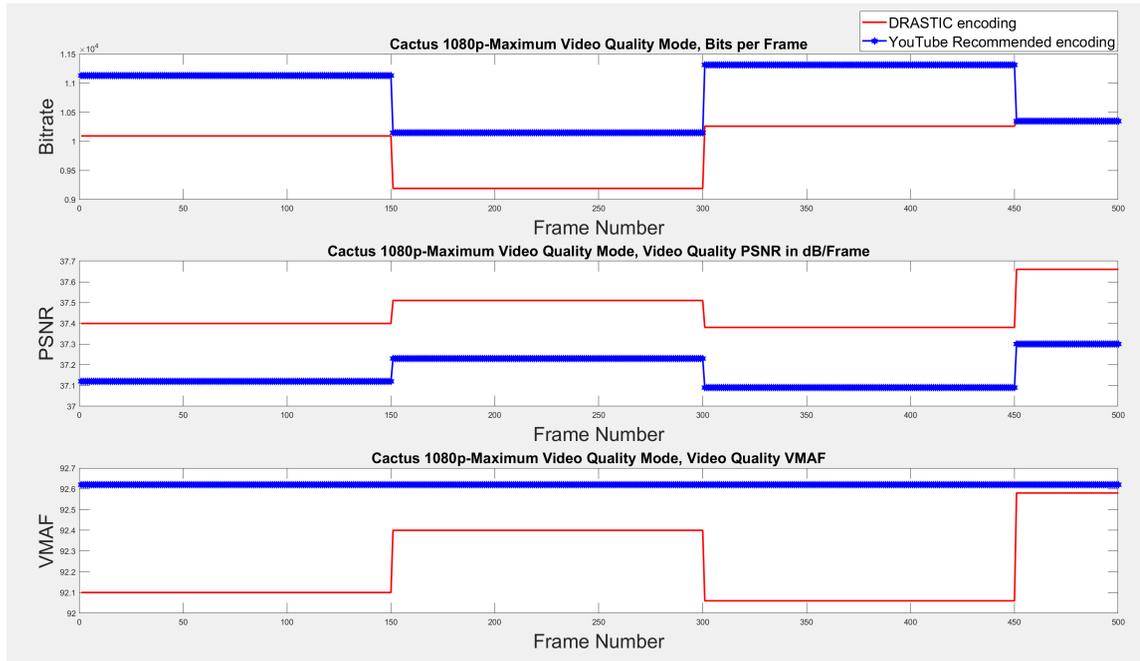

Figure 4.11: HEVC Test sequence, 1920x1080, Cactus Maximum Video Quality Mode.

### 4.3.4 Minimum Bitrate Mode

### Cactus Video HEVC Dataset

We use the default's average PSNR 37.11 dB as an acceptable video quality while maintaining a minimum FPS above 25 as constraints per segment. In Table 4.13, for the first segment DRASTIC achieves a PSNR of 37.074 dB with a bitrate of 9322.85 kbps and maintains a higher FPS of 49.21 and a VMAF of 90.58. The second segment achieves a higher PSNR of 37.54 with GOP B4 and a VMAF of 92.26 which is slightly higher than the first segment since the bitrate at this segment is 10178.66 kbps. The model equations for the minimum bitrate mode shows that all segments except the third uses B4 GOP and their model equations are given in Tables 4.14 and 4.15. As noticed, the model order here is quadratic as well similar to the maximum video quality mode.

| Seg ID | QP | Fil | GOP | Bitrate (kbps) | PSNR (dB) | FPS | VMAF |
|--------|-----|-----|-----|----------------|-----------|-----------|-------|
| Seg0 | 29 | On | B4 | *9322.85* | *37.074* | *49.21* | 90.58 |
| | | | | | >=37.1 | >=25 | |
| Seg1 | 28 | On | B4 | *10178.66* | *37.54* | *47.69* | 92.26 |
| | | | | | >=37.1 | >=25 | |
| Seg2 | 29 | On | B3 | *9467.78* | *37.12* | *37.09* | 90.81 |
| | | | | | >=37.1 | >=25 | |
| Seg3 | 28 | On | B4 | *10411.23* | *37.618* | *43.59* | 92.40 |
| | | | | | >=37.1 | >=25 | |
| *Avg* | | | | *9731.91* | *37.27* | *44.55* | *91.33* |

Table 4.13: DRASTIC Minimum Bitrate Mode for Cactus 1920x1080, 50 *fps*.

| Coefficients | $\beta_0$ | $\beta_1$ | $\beta_2$ | GOP.Str | Model Order | Adjusted $R^2$ |
|--------------|-----------|-----------|-----------|---------|-------------|----------------|
| log(PSNR) | 3.86 | -0.006686 | -7.314043e-05 | B4 | Quadratic | 0.99 |
| log(VMAF) | 3.822 | 0.0684533 | -0.0015321 | B4 | Quadratic | 0.99 |
| log(Bits) | 16.519 | -0.313125 | 0.00210119 | B4 | Quadratic | 0.99 |
| log(EncRate) | 0.4027 | 0.173966 | -0.0018987 | B4 | Quadratic | 0.93 |

Table 4.14: B4 GOP Model Equations for Minimum Bitrate Mode Mode.

At the third segment, there are more complex motions involved; hence, there is a GOP switch to B2 which attains a bitrate of 9467.78, PSNR 37.12 dB and VMAF of 90.81. The last segment follows with GOP B4 with the corresponding bitrate of 10411.23 kbps, PSNR 37.61 and a higher VMAF of 92.40. Overall, the average PSNR for all the segments achieved is 37.27 with a VMAF of 91.33 which is -1.29 than the default mode and very visually high quality video with a bitrate gain of 10 % and PSNR improvement of 0.11 as tabulated in Table 4.16.

| *Coefficients* | $\beta_0$ | $\beta_1$ | $\beta_2$ | GOP.Str | Model Order | Adjusted $R^2$ |
|---|---|---|---|---|---|---|
| log(PSNR) | 3.86 | -0.00661 | -7.489899e-05 | B3 | Quadratic | 0.99 |
| log(VMAF) | 3.80 | 0.069566 | -0.0015476 | B3 | Quadratic | 0.99 |
| log(Bits) | 16.65 | -0.319803 | 0.002179 | B3 | Quadratic | 0.99 |
| log(EncRate) | 0.706 | 0.153 | -0.001585 | B3 | Quadratic | 0.96 |

Table 4.15: B3 GOP Model Equations for Minimum Bitrate Mode.

| *Overall Bitrate Gain* | *Overall PSNR* | *Overall VMAF* |
|---|---|---|
| 10 % | 0.11 | -1.29 |

Table 4.16: Overall DRASTIC Gains from Minimum Bitrate Mode.

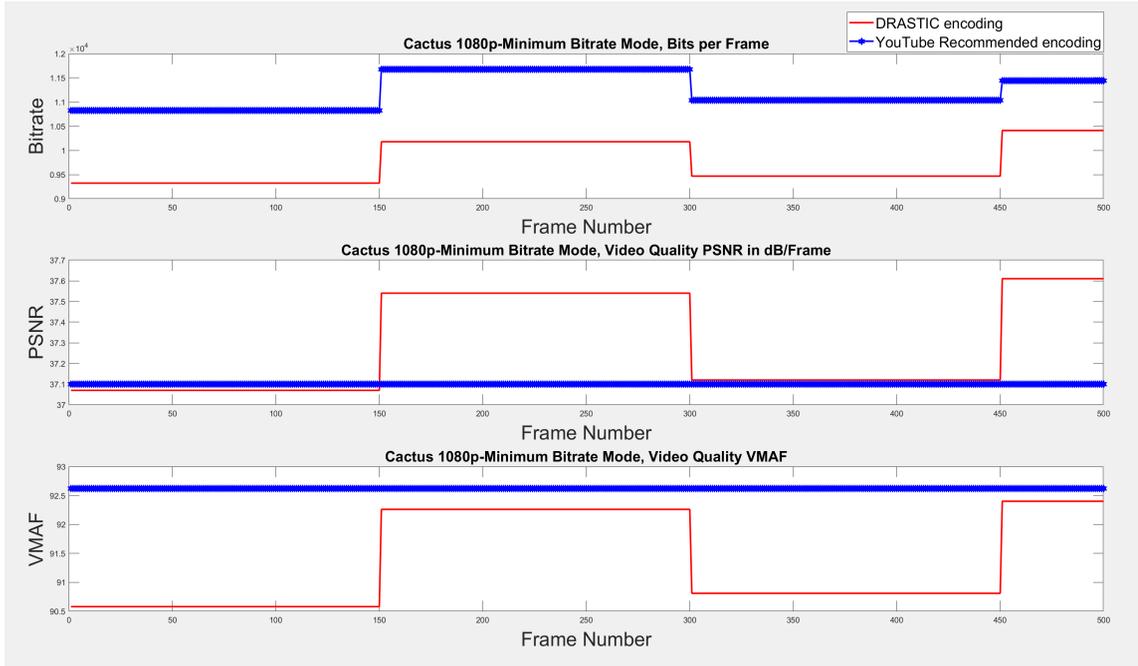

Figure 4.12: HEVC Test sequence, 1920x1080, Cactus Minimum Bitrate Mode.

The Cactus video in minimum bitrate mode achieved around 10 % in bitrate savings and loses 1.29 in VMAF visually this video is identical to the default recommended YouTube settings. So, we wanted to reduce the bitrate so that how far it has an impact perceptually affecting the video. We found in [65, 66], that a minimum of 6-point VMAF has to obtained to see any noticeable artifacts, meaning the VMAF reduction by six points away from the default recommended setting. With this setup, we gave a VMAF constraint by 6-points and gave a minimum VMAF of 87 as video quality constraint to the video. The resulting video saved around 41.5 % in bitrate savings with 6323.58 kbps and was visually identical to the typical setting. With DRASTIC already saving more bits, when PSNR was used as the objective video quality metric it was not substantially higher and is more of a mathematical observation. Whereas, the VMAF constraint reflects the subjective video quality metric and it is a direct reflection of how the video is perceived by the individual. We present the Cactus video low bitrate example here in Table 4.17 and the corresponding model

| *Overall Bitrate Gain* | *Overall PSNR* | *Overall VMAF* |
|---|---|---|
| 41.5 % | -0.86 | -6 |

Table 4.17: Overall DRASTIC Gains from Minimum Bitrate Mode.

equations and VMAF chart in Figure 4.13.

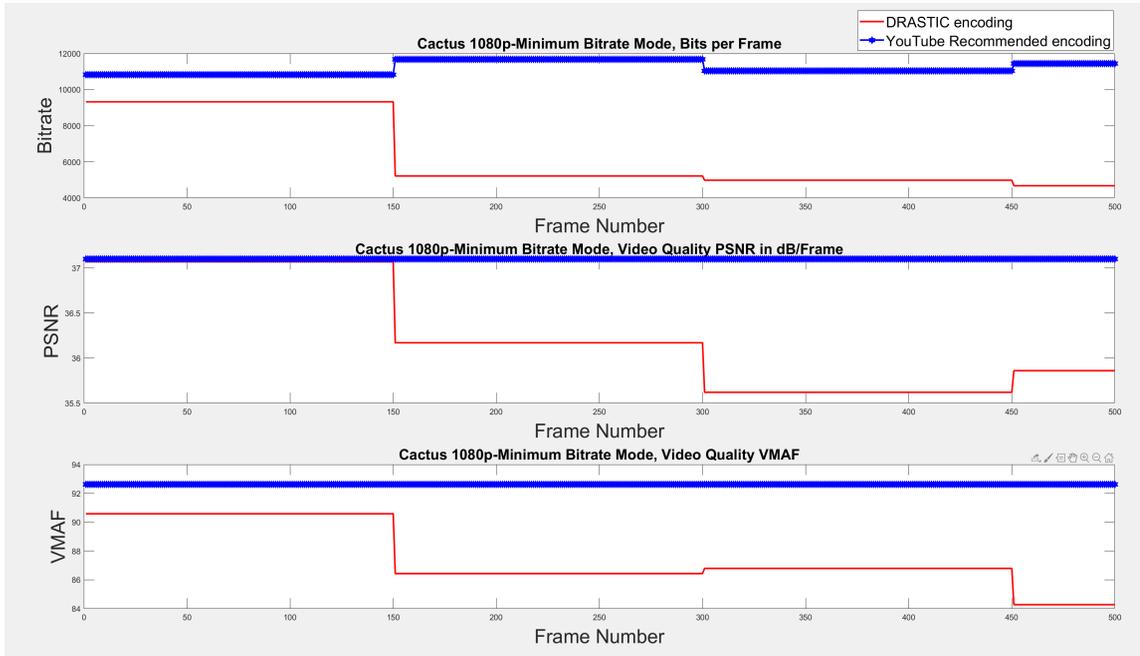

Figure 4.13: HEVC Test sequence, 1920x1080, Cactus Minimum Bitrate Mode - Low bandwidth with VMAF 86.

Even though, there is a reduction of 0.86 dB in PSNR, the video is perceptually similar to the default video. This is a practical illustration of how DRASTIC can handle an extremely low bandwidth scenario and still provide the video quality without any artifacts at those low bitrate conditions.

Notice that in Figure 4.14 DRASTIC provides an overall average score of 86. The cactus spines, tiger stripes, and faces on the poker card are perceptually similar compared to the default encoded video in Figure 4.15 with an overall average VMAF score of 92.

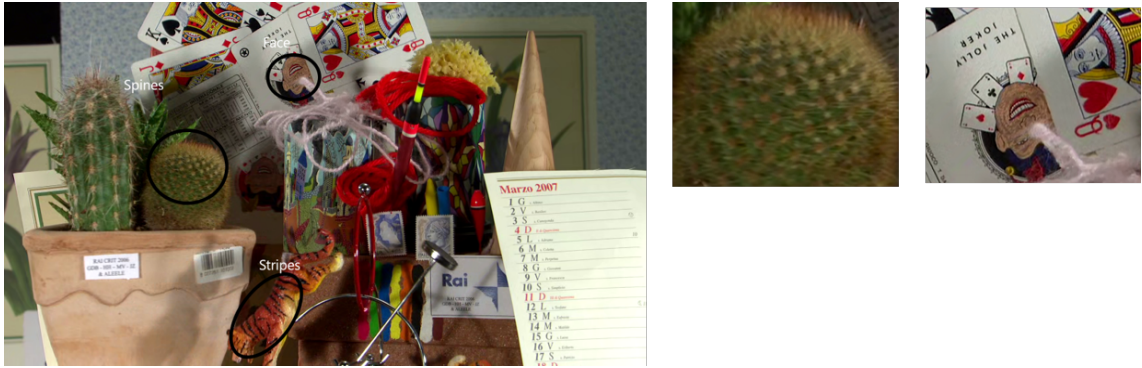

Figure 4.14: HEVC Test sequence, 1920x1080, Cactus Minimum Bitrate Mode - DRASTIC Low bandwidth with VMAF 86.

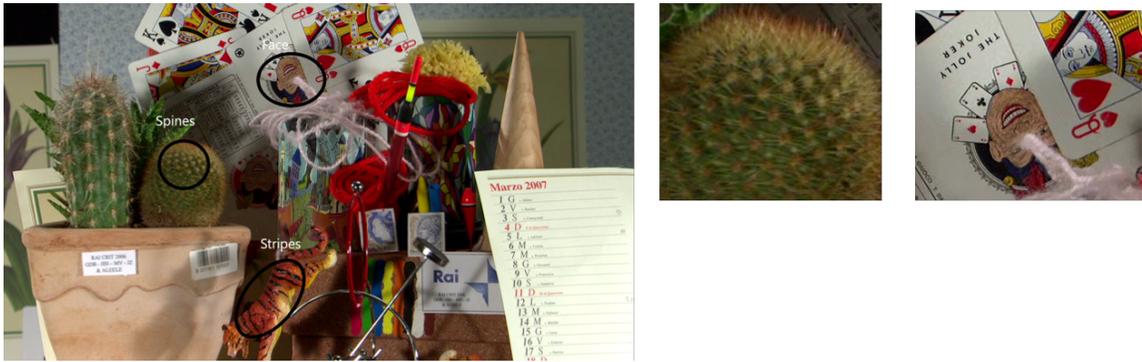

Figure 4.15: HEVC Test sequence, 1920x1080, Cactus Minimum Bitrate Mode - YouTube recommended VMAF 92.

| X265 Encoder Low Bandwidth example at 41.5% reduction in Bitrate | | | | | |
|---|---|---|---|---|---|
| QP=29.0 | fil=on | GOP=B4 | PSNR=37.07 | BitRate=9322.8 | VMAF=90.58 |
| QP=32.0 | fil=on | GOP=B3 | PSNR=36.17 | BitRate=5215.8 | VMAF=86.43 |
| QP=33.0 | fil=on | GOP=B3 | PSNR=35.62 | BitRate=4982.5 | VMAF=83.79 |
| QP=33.0 | fil=on | GOP=B4 | PSNR=35.86 | BitRate=4674.6 | VMAF=84.27 |
| *Avg* | | | *36.24 dB* | *6323.58 kbps* | *86* |

Table 4.18: Cactus Video Example of Extremely Low bandwidth Scenario with 6-point difference VMAF 86

## 4.4 Conclusion

We have proposed and demonstrated segment based adaptive video encoding systems using regression equations that provided significant bitrate savings with high video quality. The results also show DRASTIC's efficient adaptation at GOP level and provide much flexibility in terms of encoding rather than exhaustive computing or a sophisticated neural net. This segment-based encoding significantly has outperformed recommended bitrate approaches and provided better precision than recommendations by YouTube.

# Chapter 5

# Overview of Google VP9 Codec with Segment-based encoding at GOP level

## 5.1   Background of VP9 Video Coding Format

VP9 is the open source coding standard developed by Google [13], competitor to the H.265/HEVC [21] standard, and is considered to be the successor to VP8 [67, 68] codec which is the equivalent to H.264/AVC [48]. VP9 codec was mainly adopted by YouTube which is playable on internet browsers, video players and also to stream its videos [4, 43]. In contrast, HEVC was not adopted by none of the software and hardware vendors and the format could not be played on browsers or any other media player. VP9 [69] [61] was the only codec that supported media playing and widely supported in modern web browsers. VP9 supported HTML5 video tags which allowed the videos encoded in .webm/ivf container format allowed VP9 to be played with a .mkv (Matroksa Video format) video container. Originally VP9 challenged H.265/HEVC standard with a source codec used for the web, compared to VP8, some of the tools are unclear and adoption is affected by unsettled claims by multiple patent holders and patent pools. VP9 specification has been frozen in June 2013 but later was pushed by Google to optimize video distribution which made YouTube the only major adopter of the VP9 standard. Until 2016, Netflix [70] employed VP9 for the first time alongside with other encoders H.264/AVC and found potential bandwidth savings of 36% on average while the resulting video was quite similar to the video encoded with previous standards. In this chapter, we study VP9 codec and its internal tools and then apply them using the DRASTIC framework for segment-based encoding and analyze the results.

### 5.1.1 Block Partitioning

Let us start from the frame as VP9 divides each frame into 64x64 blocks called SuperBlocks (SBs). Compared to HEVC where the CTUs are partitioned as 64x64, 32x32 and all the way to 16x16, VP9 offers flexible partitioning sizes where a 64x64 SB can be split vertically or horizontally into either 64x32, 32x64, 32x64, 64x32 and a 32x32 can split similarly extending further into 8x8 which is the third level in this hierarchical split ranging from 8x4, 4x8, 4x4 etc. VP9 uses Tiles concept similar to HEVC where the frame is divided into group of SBs along their boundaries and it's always a power-of-2 so that a frame can be divided into a maximum of 4 tiles depending on resolution. For example, a 480p video can have only 2 tile-columns or 2 tile-rows (or 2 tile-columns) and it can be processed independently by 4 thread enabling multi-threading. For 720p and 1080p, the number of tile-columns is 4 and so the total thread would be 8 respectively. As of now, VP9 does not support *"slices"*.

### 5.1.2 Group of Pictures-GOPs

In VP9, there are 3 different types of frames:

- Golden Frame (I Frame) - A key frame or an Intra frame which is inserted between scene changes.

- AltRef frame (Non-displayable) - Alternate reference frames which is not displayed in the bitstream but used in compound prediction and functionally very similar to *B Frame.*

- Last frame (Previously Encoded frame) - is the last fully decoded frame and it is visible in the bitstream.

Let us look at an overview of the GOP structure in VP9 standard. The GOP structure shown in Figure 5.1 has displayed and non-displayed frame marker throughout the bitstream [71] which is how a VP9 encoded stream looks like internally, and uses .webm container which is a subset of MKV container format allowing it to be played on any browser. The bitstream starts with a Keyframe/Intra frame marked as *Displayed* 0/0-KeyFrame and indicated by *G*. The next frame is shown as *Not-Displayed* 1/1-Inter and indicated by *A* and finally the Last Frame shown here as *Displayed* 12/11-Inter and indicated by *L*. A typical GOP structure looks like Fig-ure 5.2a and 5.2b where there is a key-frame group comprising of two Golden frames (colored in red) inserted between different scenes and a Golden Frame group which comprises of a Golden frame (I - colored in red), Alternate reference frames (A - Colored in yellow) in which the Last Frame (L - colored in blue). The Last frame hereafter referred as *L frame* has been boosted (G* - colored in green) with high quality meaning lower QP on that particular frame and acts as a reference frame to other frames in the Golden frame group to enable better prediction. The Alternate reference hereafter referred as *AltRef* frames provide a round about to B-frame Prediction [21] or in VP9 called as Compound prediction where these frames are used in the references but are not displayed in the bitstream. The more alternate reference frames give the better prediction for a particular frame. Each *AltRef* is used as a reference point to a keyframe on a GOP interval which is defined as the minimum distance between two Intra frames or the Intra refresh interval.

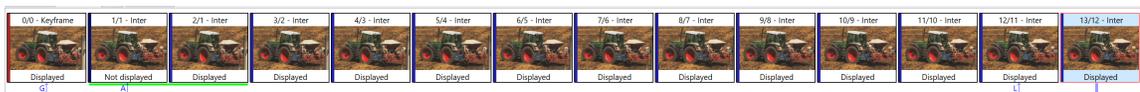

Figure 5.1: VP9 GOPs in WebM bitstream

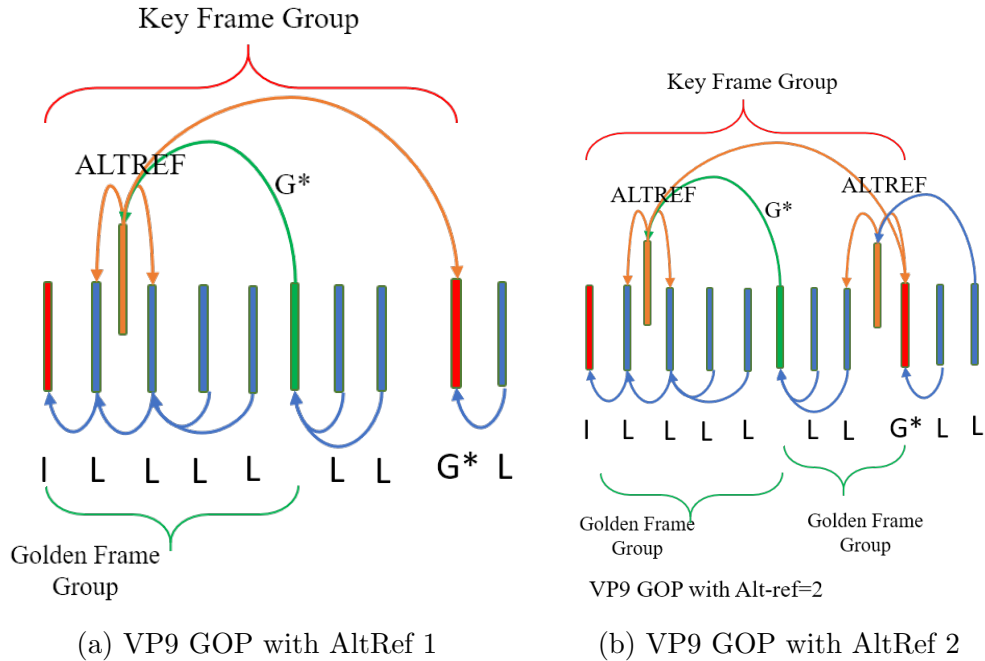

(a) VP9 GOP with AltRef 1    (b) VP9 GOP with AltRef 2

Figure 5.2: VP9 GOP Structures with Golden Frame Groups (a) Alternate Reference **AltRef1** frame. (b) Alternate Reference **AltRef2** frame.

As part of our study with VP9 codec, we introduced *AltRef* at different key intervals and came up with different GOP structures namely **ALT0, ALT1, ALT2, ALT4, ALT6** based on the number of alternate reference frames. VP9 reference specification [13] states that we can have a maximum number of 6 frames. When there is no *AltRef* frame represented by ALT0, the GOP is entirely made up of only Golden and Last frames which is very similar to Zero Latency(IP) mode in x265 [24]. By default, VP9 used ALT1 as its GOP structure which provides one of the fastest encodes as similar to low-delay webRTC applications and when all of the *AltRef* frames are used for referencing we call it ALT6 which provides efficient encodes in terms of bitrate and compression ratio. Additionally, these *AltRef* can be constructed from other past *AltRef* frames or future frames in the bit stream in order to reduce the total bitrate overhead.

### 5.1.3 Intra Prediction

Intra prediction is less complex in VP9 as it has 10 intra directions, 8 angular, one DC and True Motion (TM) compared to HEVC which provides 35 directions. DC mode in Figure 5.3 is where we take the average of all pixels of the current block. True motion refers to prediction mode where each pixel is subtracted from the top-left pixel array from its current block position.

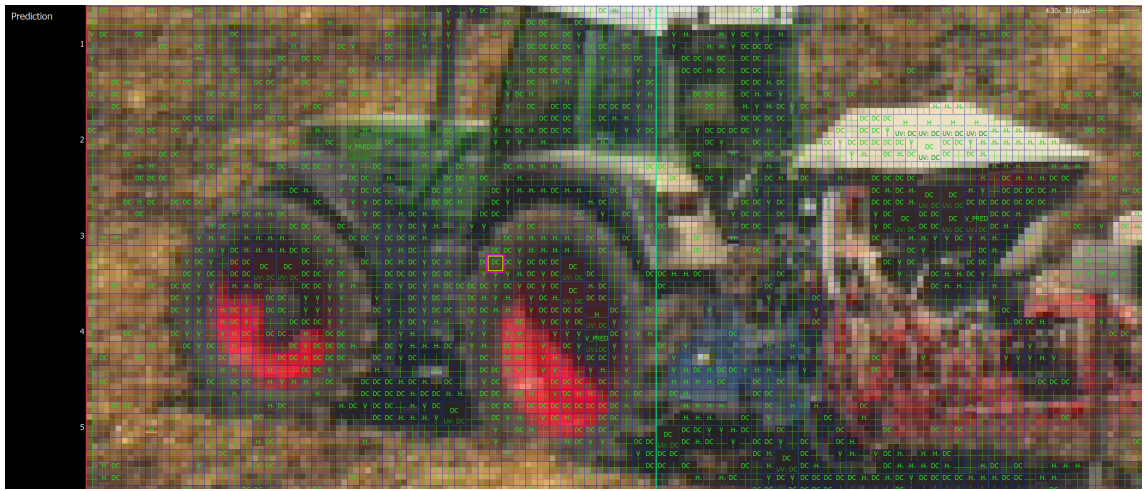

Figure 5.3: VP9 Intra modes.

Intra prediction in VP9 has block sizes upto 4x4, 8x8, 16x16, 32x32 as recursive splitting of intra blocks is allowed by reconstruction at the transform size specified. As in previous standards H.264, H.265 Intra prediction in VP9 also uses the top and left arrays both of which are reconstructed from the neighboring pixels and used for the prediction depending on the Intra angular mode. Always the left array [72] is same as the height of current block and the top array is twice the size of current block. For smaller luma block sizes like 4x4 in Figure 5.4, we use the last pixel value 67 to be extended further to complete the array. The same principle is applied to chroma intra prediction where we can see chroma Cb pixel block 122 chroma Cr pixel block 134 been extended to the right double the size of its current block.

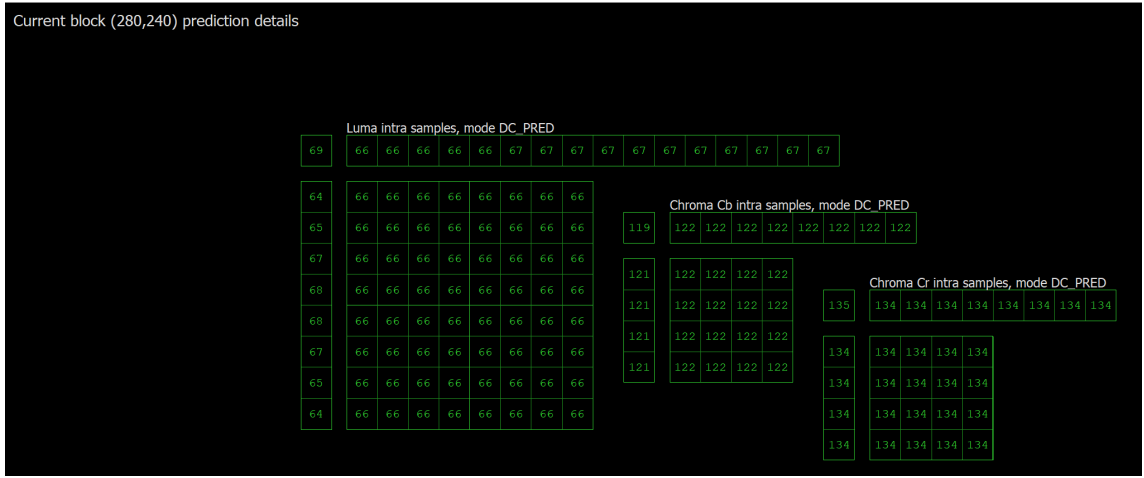

Figure 5.4: VP9 Intra prediction with Luma and Chroma in DC Mode.

## 5.1.4   Inter Prediction

Inter prediction [69] in VP9 is similar to other standards except it uses only 3 reference frames from a pool of 8 in the reference frame buffer shown in Figure 5.5. It supports block sizes from 4x4 up to 64x64 respectively with different prediction techniques.

*Compound prediction:* VP9 uses compound prediction which employs the AltRefs for prediction and can choose to have multiple motion vectors compared to other standards where there is only one motion vector transmitted per block. The types

| # | Frame | Ref | Refresh |
|---|-------|--------|---------|
| 0 | 9 | LAST | Yes |
| 1 | 0 | GOLDEN | Yes |
| 2 | 0 | ALTREF | No |
| 3 | 0 | Unused | No |
| 4 | 0 | Unused | No |
| 5 | 0 | Unused | No |
| 6 | 0 | Unused | No |
| 7 | 0 | Unused | No |

Figure 5.5: VP9 reference pool of frames.

of motion vectors are:

- Nearest MV - Candidate neighborhood Motion vectors from current frame.

- Near MV - Candidate Motion vectors which is co-located MVs in the previous frame.

- Zero MV - Where there is no motion.

- New MV - To be transmitted in the frame with a motion vector reference (Nearest MV and Near MV).

In Figure 5.6, inter prediction blocks show the inter mode in blue, with the motion vectors in orange and the different types of MVs and the reference frames (L for Last, G for Golden and A for AltRef). The AltRef is chosen because of its availability from future frames in the reference buffer which greatly enhances the flexibility of the prediction.

## 5.1.5 Transform Coding Tools

VP9 uses three transform types: DCT, ADST (Asymmetric DST) and WHT (Walsh Hadamard Transform), whereas HEVC uses DCT and DST for Intra 4x4 blocks.

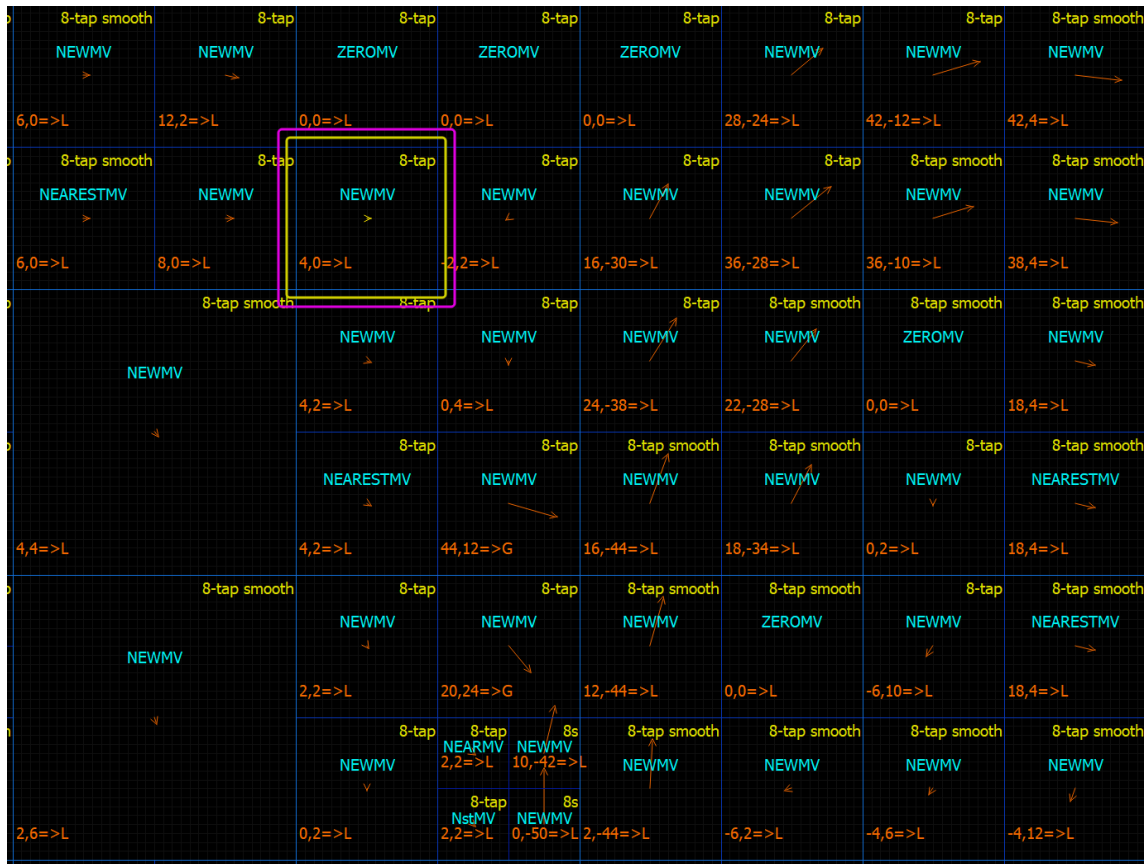

Figure 5.6: VP9 Motion Vectors.

VP9 uses a hybrid combination of both DCT and ADST depending on the video content. The transform block units sizes vary from 4x4, 8x8, 16x16, to 32x32. The quantization step size is quite large compared to HEVC where QP ranges from 0 to 51 while in VP9 it is from 0 to 63.

## 5.1.6 Loop Filters

Loop filters in VP9 are very similar to HEVC standard with different filters for sharpening, blurring the reconstructed image, and for noise reduction in the alternate reference frames. Filter strength can be adjusted to vary from 0 to 6.

## 5.1.7   Segmentation

A new addition to VP9 encoder was segmentation. The different encoding modules are segmented into eight types depending on different signals. For example, a particular block can carry the QP value, prediction mode, Motion vectors, Transform size, filter strength, etc. All of these features can be grouped into different segments and can be used to build a heat map or segment map identifying different portions of the video frame content.

## 5.2   Methodology

### 5.2.1   Segment-based Encoding with VP9 Configurations

In this section, we will employ the proposed method in Figure 4.1 using VP9 codec as the encoding framework. The pseudo-code in Figure 4.4 is similar to the x265 segment based encoding except here we build it upon VP9 codec configurations. In VP9 we start by splitting the video into 3 sec segments and encode them with different GOP, QP, filter combinations as a function of encoding configuration and for each objective video quality, bitrate, and encoding rate/time we obtain models to be used under their respective DRASTIC modes.

The number of VP9 encoding configurations is so large that deciding which configurations to use for encoding in VOD based applications was cumbersome. We used a familiar approach followed in x265 where we started off with different GOP structure and noticed whether they can be applied to VP9 encodings as well. We have summarized a whole list of VP9 encoding configurations which we used in the segment-based encoding in Table 5.1. Firstly, VP9 uses all different GOP structures and their CPU preset parameter which decides both the encode quality and the speed. The *best* setting for the CPU preset is similar to the placebo setting for X265. Thus, the *best* setting is extremely slow while providing excellent video quality. If the CPU preset is set to *good*, then the speed can be adjusted in the range of 0 to 5. At a set speed of 4 or 5, the encoder will turn off the Rate-Distortion Optimization(RDO) which will disregard the quality. We tried CPU preset set to *good*, with speed set to 3 and 2, but we found the encoder was still slow and was not utilizing all the cores. We then applied the CPU preset to *rt* which is quite faster and utilized the CPU cores and we changed the CPU preset to 4,8,12,16 and we found that the encode quality at CPU=8 and above settings gave fewer differences in the total bits and the storage size of each encoded video. So we decided to use the CPU preset to 8 with single pass encoding set as the quality as VP9 (build libVPx-Ver1.7) allows multiple rate-control methods and found this setting to be the best for VOD applications.

| Parameter | Value |
|---|---|
| Presets | realtime |
| Encoding Structure | ALT0,ALT1,ALT2, ALT4,ALT6 |
| CPU Used | 8 |
| DBF | On/Off |
| QP | 16 - 52 in steps of 4 |
| Tuning | PSNR |
| arnr-maxFrames | 7 |
| arnr-strength | 5 |
| arnr-type | 3 |
| Row-mt | 1 |
| Total encoding combinations per segment | 200 |

Table 5.1: VP9 Encoder Configurations for **rt** with our new GOPs.

Regarding the GOPs, we found the encoding structures from the bit stream as described in Section 5.1 we want to use: **ALT0, ALT1, ALT2, ALT4, ALT6** with different settings for the noise reduction for each alternate reference frame. The maximum number of references ARNR-maxframes for each *AltRef* was set to 7 with filter strength set to 5 and the ARNR-type to be 3. This ARNR setting was often used in VOD-specific frameworks [47, 49, 50, 73] as the literature recommended that these settings do have an impact on the video quality. For the loop filters, we set the deblocking filter similarly to x265, with a setting of On and Off and we changed the QP range from 0 to 52 in steps of 4 since VP9 offers a maximum QP up to 63. All of these configurations were suited to row based multi-threading since VP9 uses parallel tiles so that our encodes run quite faster to evaluate this exhaustive list of combinations. VMAF was also incorporated into the encoding pipeline which calculates the VMAF score using the perceptual model [63] VMAF 0.6.1 and the corresponding VMAF scores are stored to be used for the Pareto models.

With 200 encoding configurations, we fit the Pareto front using a small number of parameters. We measured Encoding Rate in the number of frames per second (FPS), Bitrate in kilobits per second and Video Quality using both PSNR & VMAF. The local model predicts the objectives based on the constraints and, depending upon the DRASTIC [26] mode, can provide estimates for the next 150 video frames. Different encoding combinations were considered before we finalized configurations that directly impact the encoding visually and the resulting compression ratio.

## 5.2.2 Forward Regression Models and Inverse Prediction in VP9

The model building process is quite similar to that in Section 4.1 except where we apply VP9 configurations to the model building process. Here are the model equations summarized,

$$\ln(\text{PSNR})_i = \alpha_0 + \beta_1 \cdot \text{QP}_i + \beta_2 \cdot \text{QP}^2{}_i + \beta_3 \cdot \text{QP}^3{}_i$$

$$\ln(\text{VMAF})_i = \alpha_1 + \beta_{11} \cdot \text{QP}_i + \beta_{12} \cdot \text{QP}^2{}_i + \beta_{13} \cdot \text{QP}^3{}_i$$

$$\ln(\text{Bits})_i = \alpha_2 + \beta_{21} \cdot \text{QP}_i + \beta_{22} \cdot \text{QP}^2{}_i + \beta_{23} \cdot \text{QP}^3{}_i$$

$$\ln(\text{FPS})_i = \alpha_3 + \beta_{31} \cdot \text{QP}_i + \beta_{32} \cdot \text{QP}^2{}_i + \beta_{33} \cdot \text{QP}^3{}_i$$

where $\beta_1, \beta_{i,1}, \beta_{i,2}, \beta_{i,3}$ represent QP coefficients and, $\alpha_0, \alpha_1, \alpha_2, \alpha_3$ denote the constants of the polynomial regression equation. The model building is a cumulative process since we have to exhaustively combine so many different encoding configurations and then obtain the resulting objectives along with its parameter setting and store them as tables. For each GOP structure encoded, we obtain the Pareto points which are used in the model building with various encoding combinations and the resulting optimal models are saved to be used for the next segment. For all of these model fittings, we want to build a polynomial model space and the order of the model equations were varied from linear, quadratic and cubic fit depending on the video.

With prior knowledge of x265 segment based modeling, we simplified the VP9 forward model building by choosing QP and deblocking filter as the two predictor variables for the model fitting. Similarly, the objectives had a very good adjusted R square value for each objective (PSNR, VMAF, Bitrate, FPS) quadratic model for Basketball Drill [2] video from HEVC Test sequences with a score of 0.99, 0.98. 0.99, 0.96, respectively. Another significant statistical test was the p-value which ranges from 0 to 1 and we found all models were following a similar trend as that of x265 with a small p-value of ($\leq 0.05$) [64] proving that the QP and Deblocking filter should be used for prediction.

Following the forward model for each GOP built, we satisfy the constraint optimization modes based on the selected DRASTIC operating mode to obtain the corresponding encoding configuration sets and constraints. For example, in minimum bitrate mode, we get constraints for quality and frame rates and then we apply them to the equations generated from the forward model with these as constraints. Using Newton's [74] inverse equation, we find the optimal QP and Deblocking filter output from these equations taking into account video quality and encoding frame rate violations. For the minimum bitrate mode, we finally encode the video with optimal QP and Filters for the given segment make sure that the PSNR predicted is above the acceptable video quality and we want for the video to look better without much artifacts. This mode of operation simulates low bandwidth scenarios, where the video quality drops and degrades.

In simple terms, we basically take the forward model equation from encoding rate and bitrate and solve for a QP that maximizes the quality of the video. The QP value generated by the prediction is a floating point value and we approximate using an integer as the encoders accept only integer based QP values. Also, we note that the coefficients of the model fittings especially that of QP is more significant and have negative values which clearly state that increase in QP will decrease the bitrate and the video quality. Additionally, the forward model prediction might have induced some error due to fitting and will affect the prediction process during Newton's inverse method. So to compensate for the error, we allow soft violations say 10% for bitrates and encoding frame rates and 3-5% for video quality respectively. By this, we generate multiple solutions for QP which in our case is the dominant predictive variable. So, we carefully determine the QP values generated by the Newton method by estimating whether they can obey the constraints and if in case of a failure, we perform a local search around the QP neighborhood which is in the case (QP+4, QP−4) and then repeat the prediction process again until the constraints are satisfied.

## 5.3 Results

### 5.3.1 Maximum Video Quality Mode Class C Basketball Drill 480p Video

We start with Class C 832x480p, *50fps* from HEVC dataset. Basketball Drill video is 10s clip where a group of players practice in loop with the ball and keep running throughout the video. In this optimization mode the objective is to maximize the video quality while conforming to bandwidth constraints in terms of typical upload data rates as recommended by YouTube [4]. For 480p with 50fps the recommendation is 4000 kbps. We present the default mode in Table 5.2 where the QP=34 approaches or almost above 4000 kbps.

In this mode, we gave the constraints for each segment from the default's individual segment objectives. For the first segment, we have 4051 kbps as in Table 5.6 from the default set as a constraint and DRASTIC chose ALT1 GOP achieves a bitrate 3823.33 kbps, PSNR of 38.84 dB, and VMAF of 96.13 respectively. Similarly, in the second and third segments there is GOP switch happening from ALT 1 to ALT2 and achieves bitrates of 4051.34 kbps, PSNR of 38.82 dB and VMAF of 96.93 respectively. Also, the FPS in this video segments which makes it a high FPS video example as VP9 at lower resolutions can encode at real-time at very high frame rate. But this functionality is not available in 1080 or even higher resolutions. For the third segment, DRASTIC has a slightly lower PSNR of 0.02 dB compared to the default which is 38.71 dB but with bitrates 4075.47 kbps and VMAF of 96.45 respectively. Overall, the bitrate gain is 4.2% and PSNR is 0.11 less than the default PSNR even though DRASTIC wins in two segments and the video is perceptually identical to the default encoded video. This is clearly reflected in the VMAF scores as the overall difference is 0.03 between default VMAF and DRASTIC VMAF as shown in Table 5.7. All of the GOP model equations are given in Tables 5.4, 5.3, 5.5.

| *Seg ID* | *CQP* | *Fil* | *GOP* | *Bitrate (kbps)* | *PSNR (dB)* | *VMAF* |
|----------|-------|-------|-------|------------------|-------------|--------|
| Seg0 | 34 | On | ALT0 | 4045.13 | 38.85 | 95.38 |
| Seg1 | 34 | On | ALT0 | 4172.05 | 38.76 | 96.36 |
| Seg2 | 34 | On | ALT0 | 4314.00 | 38.71 | 97.51 |
| Seg3 | 34 | On | ALT0 | 4329.03 | 38.35 | 97.25 |
| *Avg* | | | | **4192.25** | **38.83** | **96.50** |

Table 5.2: Typical Mode - YouTube Recommended Bitrate achieved by CQP.

| *Coefficients* | $\beta_0$ | $\beta_1$ | $\beta_2$ | GOP.Str | Model Order | Adjusted $R^2$ |
|----------------|-----------|-----------|-----------|---------|-------------|----------------|
| log(PSNR) | 3.853 | -0.00363 | -5.0348e-05 | ALT4 | Quadratic | 0.99 |
| log(VMAF) | 4.462 | 0.01256 | -0.0002654 | ALT4 | Quadratic | 0.98 |
| log(Bits) | 10.736 | -0.0566 | -0.000258 | ALT4 | Quadratic | 0.99 |
| log(EncRate) | 4.50 | 0.02354 | -0.000105 | ALT4 | Quadratic | 0.90 |

Table 5.3: ALT4 GOP Model Equations for Maximum Video Quality Mode.

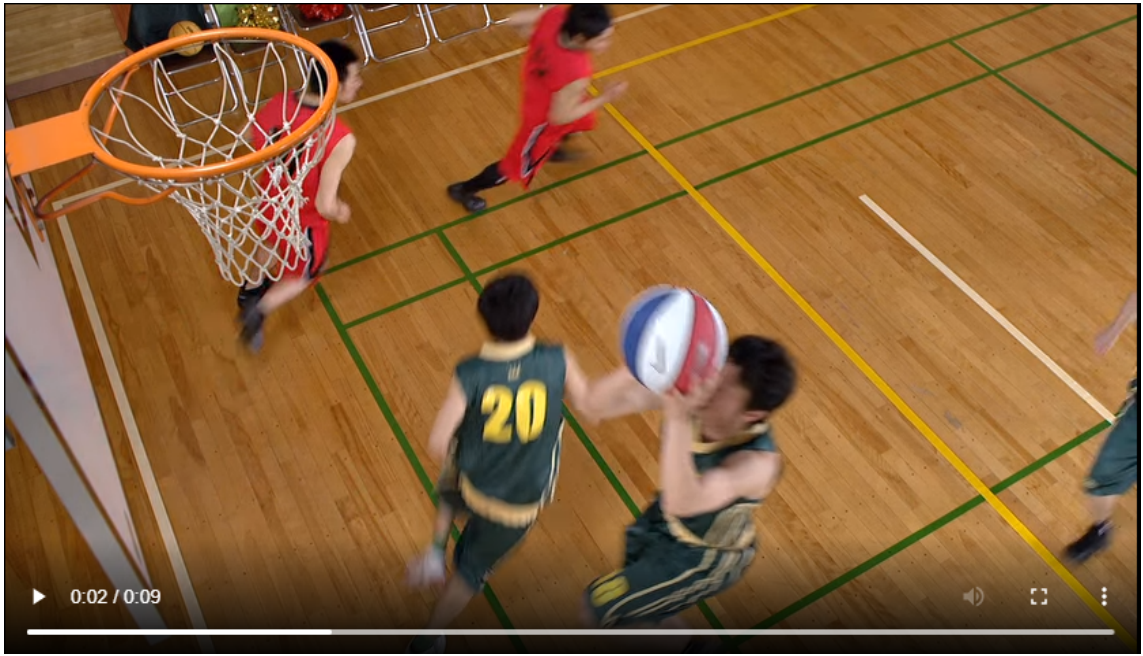

Figure 5.7: BasketballDrill from HEVC [2] Video Sequence, 832x480, 50fps.

| **Coefficients** | $\beta_0$ | $\beta_1$ | $\beta_2$ | GOP.Str | Model Order | Adjusted $R^2$ |
|---|---|---|---|---|---|---|
| log(PSNR) | 3.850 | -0.00356 | -5.211425e-05 | ALT1 | Quadratic | 0.99 |
| log(VMAF) | 4.455 | 0.01322 | -0.0002811 | ALT1 | Quadratic | 0.98 |
| log(Bits) | 10.657 | -0.056265 | -0.000255 | ALT1 | Quadratic | 0.99 |
| log(EncRate) | 4.332 | 0.034247 | -0.000245 | ALT1 | Quadratic | 0.93 |

Table 5.4: ALT1 GOP Model Equations for Maximum Video Quality Mode.

| **Coefficients** | $\beta_0$ | $\beta_1$ | $\beta_2$ | GOP.Str | Model Order | Adjusted $R^2$ |
|---|---|---|---|---|---|---|
| log(PSNR) | 3.851 | -0.00357 | -5.16617e-05 | ALT2 | Quadratic | 0.99 |
| log(VMAF) | 4.4597 | 0.012858 | -0.0002723 | ALT2 | Quadratic | 0.98 |
| log(Bits) | 10.684 | -0.05577 | -0.0002667 | ALT2 | Quadratic | 0.99 |
| log(EncRate) | 4.496 | 0.023265 | -9.7890e-05 | ALT2 | Quadratic | 0.89 |

Table 5.5: ALT2 GOP Model Equations for Maximum Video Quality Mode.

| *Seg ID* | *QP* | *Fil* | *GOP* | *Bitrate (kbps)* | *PSNR (dB)* | *FPS* | *VMAF* |
|---|---|---|---|---|---|---|---|
| Seg0 | 36 | Off | ALT1 | **3823.33** | 38.84 | **145.8** | 96.13 |
| | | | | <=4051 | | >=50 | |
| Seg1 | 36 | Off | ALT2 | **4051.34** | 38.82 | **143.25** | 96.93 |
| | | | | <=4172 | | >=50 | |
| Seg2 | 36 | Off | ALT1 | **4075.47** | 38.69 | **183.47** | 96.45 |
| | | | | <=4134 | | >=50 | |
| Seg3 | 36 | Off | ALT4 | **4298.60** | 38.40 | **114.11** | 96.26 |
| | | | | <=4329 | | >=50 | |
| *Avg* | | | | **4014.9** | **38.74** | **153.16** | **96.47** |

Table 5.6: DRASTIC Maximum Video Quality Mode for BasketballDrill 832x480, 50 *fps*.

| *Overall Bitrate Gain* | *Overall PSNR* | *Overall VMAF* |
|---|---|---|
| 4.2% | -0.11 dB | -0.03 |

Table 5.7: Overall DRASTIC Gains from Maximum Quality Mode

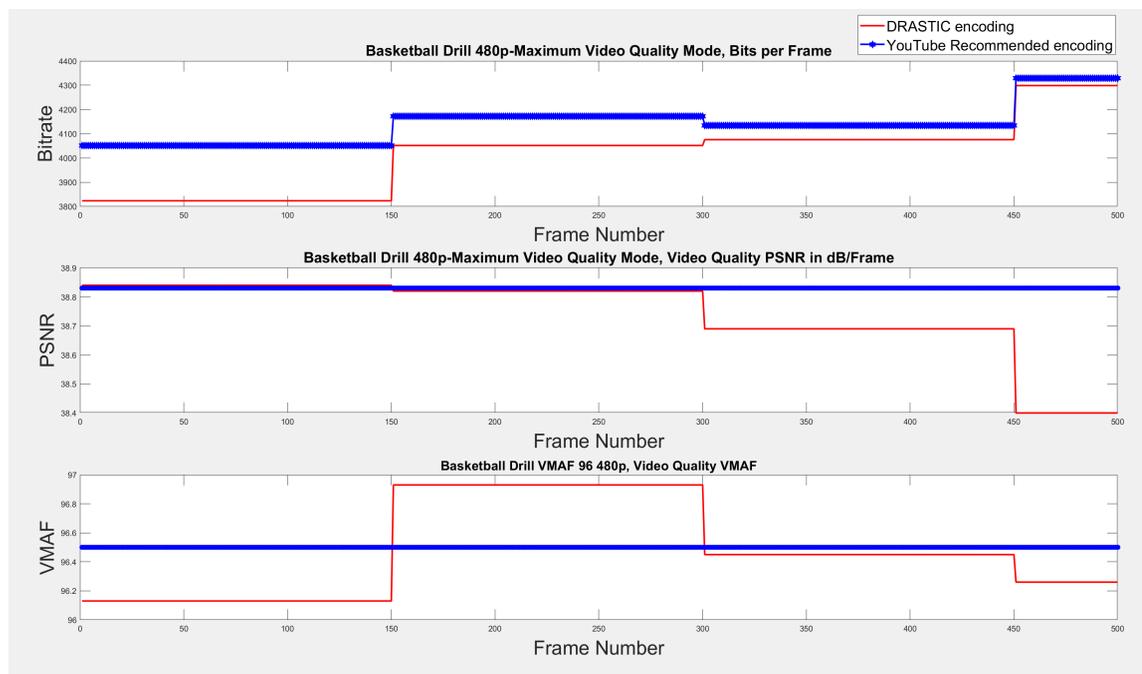

Figure 5.8: HEVC Test sequence, 832x480, Basketball Drill Maximum Video Quality Mode

### 5.3.2 Minimum Bitrate Mode
### Class C Basketball Drill 480p Video

We start with Class C 832x480p, *50fps* from HEVC dataset. Basketball Drill video is 10s clip where a group of players practice in loop with the ball and keep running throughout the video. In this optimization mode the objective is to minimize the bitrate without losing the visual quality. So we set the minimum acceptable PSNR from the default mode and set to 38.83 dB. For the first segments, the PSNR achieved by DRASTIC was 39.46 dB, 39.28 dB and 38.89 dB respectively which is higher than the acceptable threshold that we set as 38.83 dB. The corresponding bitrates and VMAF scores achieved by DRASTIC are 3926.41 kbps, 3870.24 kbps, 3671.57 kbps and 96.35, 96.55, 95.58 respectively. Throughout, these segments there is a GOP switch from ALT2 to ALT4 and the last segment has a switch again to ALT1 as seen in the Table 5.8. The fourth segment has a bitrate of 3276.9 kbps, PSNR of 38.26 dB and VMAF of 93.82. Overall, the bitrate savings are 10.11% with a PSNR gain of 0.28 dB and slight reduction in VMAF of -1.33 respectively. The model equations are tabulated in the Tables 5.9, 5.10 and 5.11 respectively.

| Seg ID | QP | Fil | GOP | Bitrate (kbps) | PSNR (dB) | FPS | VMAF |
|--------|----|-----|-----|----------------|-----------|-----|------|
| **Seg0** | 36 | Off | ALT2 | 3926.41 | 39.46 | 90.24 | 96.35 |
| | | | | | >=38.83 | >=50 | |
| **Seg1** | 37 | Off | ALT4 | 3870.24 | 39.28 | 88.05 | 96.55 |
| | | | | | >=38.83 | >=50 | |
| **Seg2** | 38 | Off | ALT4 | 3671.57 | 38.89 | 93.97 | 95.58 |
| | | | | | >=38.83 | >=50 | |
| **Seg3** | 40 | On | ALT1 | 3276.9 | 38.26 | 72.5 | 93.82 |
| | | | | | >=38.83 | >=50 | |
| **Avg** | | | | **3768.156** | **39.115** | **88.928** | **95.926** |

Table 5.8: DRASTIC Minimum Bitrate Mode for Basketball Drill 832x480, 50 *fps*.

| Coefficients | $\beta_0$ | $\beta_1$ | $\beta_2$ | GOP.Str | Model Order | Adjusted $R^2$ |
|--------------|-----------|-----------|-----------|---------|-------------|----------------|
| log(PSNR) | 3.869 | -0.00426 | -3.593591e-05 | ALT2 | Quadratic | 0.98 |
| log(VMAF) | 4.474 | 0.0112119 | -0.00023707 | ALT2 | Quadratic | 0.98 |
| log(Bits) | 10.331 | -0.0466433 | -0.0003759 | ALT2 | Quadratic | 0.99 |
| log(EncRate) | 2.498 | 0.1904638 | -0.002575699 | ALT2 | Quadratic | 0.92 |

Table 5.9: ALT2 GOP Model Equations for Minimum Bitrate Mode.

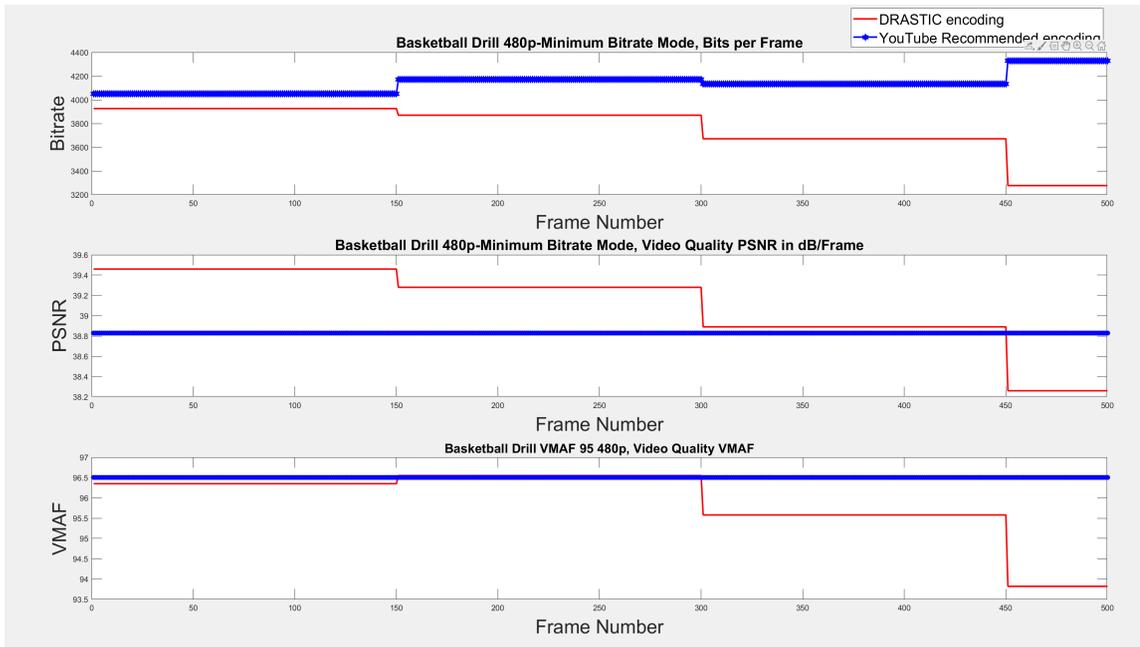

Figure 5.9: HEVC Test sequence, 832x480, Basketball Drill Minimum Bitrate Mode

| *Coefficients* | $\beta_0$ | $\beta_1$ | $\beta_2$ | GOP.Str | Model Order | Adjusted $R^2$ |
|---|---|---|---|---|---|---|
| log(PSNR) | 3.869 | -0.0042875 | -3.604318e-05 | ALT1 | Quadratic | 0.98 |
| log(VMAF) | 4.472 | 0.01140 | -0.000242 | ALT1 | Quadratic | 0.98 |
| log(Bits) | 10.30 | -0.046919 | -0.000365 | ALT1 | Quadratic | 0.99 |
| log(EncRate) | 2.587 | 0.186147 | -0.002526 | ALT1 | Quadratic | 0.91 |

Table 5.10: ALT1 GOP Model Equations for Minimum Bitrate Mode.

| *Coefficients* | $\beta_0$ | $\beta_1$ | $\beta_2$ | GOP.Str | Model Order | Adjusted $R^2$ |
|---|---|---|---|---|---|---|
| log(PSNR) | 3.871 | -0.00433 | -3.471101e-05 | ALT4 | Quadratic | 0.98 |
| log(VMAF) | 4.475 | 0.011089 | -0.000232 | ALT4 | Quadratic | 0.98 |
| log(Bits) | 10.390 | -0.04794 | -0.000366 | ALT4 | Quadratic | 0.99 |
| log(EncRate) | 2.451 | 0.19260 | -0.00260 | ALT4 | Quadratic | 0.93 |

Table 5.11: ALT4 GOP Model Equations for Minimum Bitrate Mode.

| *Overall Bitrate Gain* | *Overall PSNR* | *Overall VMAF* |
|---|---|---|
| 10.11 % | 0.28 dB | -1.33 |

Table 5.12: Overall DRASTIC Gains from Minimum Bitrate Mode

## 5.4 Conclusion

The contributions of this chapter under VP9 codec we have proposed and demonstrated segment based adaptive video encoding systems using regression equations that have significant savings on bitrate provided with a high video quality. The results also show it is an efficient adaptation at GOP level and provides much flexibility in terms of encoding rather than exhaustive computing or a sophisticated neural net. This Segment based encoding significantly has outperformed standard recommended bitrate approaches and precision comapred to approaches that rely on stored pre-encoded on a large system.

# Chapter 6

# SVT-AV1: A Scalable, Open Source AV1 Codec and Local Pareto Models at GOP level

## 6.1 Background of AOM/SVT-AV1 Video Coding Standard

Originally, AV1 development started as an extension to the libvpx-VP9, or VP10 and had features from Mozilla'a Daala [75] Codec and Cisco's Thor Codec with solid focus on a royalty free, open source codec that is completely optimized for the web and deployed for video streaming. So the Alliance for Open Media(AOM) was formed for both video and audio codecs that are openly available to the market and easily accessible for hardware developers to cater to the growing need of video applications like video conferencing, video on demand and live video gaming. Also, the goal was to provide better compression than the previous standards, H.264, VP9, HEVC and the new emerging MPEG based VVC/H.266 encoders.

This chapter will focus on one of the implementations of AV1 standard via SVT-AV1 which stands for Scalable Video Technology-AV1 [16] Codec and is the first video codec co-developed by the Alliance for Open Media (AOM) [14]. SVT-AV1 is a joint collaboration between Intel and Netflix, members of AOM primarily built for video on demand, video transcoding, live streaming applications. Additionally, the codec is performance optimized targeting towards real-time encodings and higher performance supporting 1080p and 4K videos. We will provide an overview of the SVT-AV1 codec tools and then provide a summary of our proposed approach of segment-based encoding using this new SVT-AV1 codec.

### 6.1.1 Block partitioning

Originally, AV1 had a recursive block partitioning [76] system similar to VP9 and HEVC (64x64) with block sizes of 128x128 and all the way down to 4x,4 allowing each block to be further subdivided using 10-way partitioning for high-resolution videos. For example, the 128x128 block sizes can be split using quad-tree partitions into 10-way splitting starting from horizontal, vertical splitting and T-Splitting down to 4x4 blocks. Compared to previous video coding standards in VP9, we only had 64x64 blocks with recursive splitting down to 4x4 but with limited sub-block level 8x8 divisions at the 4-way partitions. AV1 extensively improves the partitioning and has more flexibility and control over 8x8 sub-blocks.

### 6.1.2 Group of Pictures - GOPs

In SVT-AV1, the number of reference frames are extended from 3 to 7 with the addition or naming the individual frames in the candidate reference pool as in libaom. But the SVT-AV1 uses three-level Hierarchical B pictures in Figure 6.1 and four-level Hierarchical B pictures in Figure 6.2 in their implementation of the AV1 standard.

In addition to the candidate pool, the frames are named as they are referenced for prediction:

- Golden Frame (I Frame) - A key frame or an Intra frame which is inserted between scene changes.

- AltRef frame (Non-displayable) - Alternate reference frames which are not displayed in the bitstream but used in compound prediction and functionally very similar to *b Frame* in x265.

- AltRef2 frame (Non-displayable) - Alternate reference frames which are not displayed in the bitstream but used in compound prediction between the Golden and AltRef and functionally are very similar to *b Frame* in x265.

- BWD frame (Non-displayable) - Alternate reference frames used as an overlay between Altrefs and not displayed in the bitstream but used in compound prediction and functionally very similar to *B Frame*.

- Last, Last2, Last3 frames (Past Previously Encoded frame) - are the last fully decoded frames from the reference buffer and it is visible in the bitstream.

Frames BWD, ALT2 and ALT are temporally filtered from the future frames in the temporal buffer and are arranged hierachically B pictures similar to a Random Access B in HEVC or VVC standards. With the additional number of AltRefs, we can change the GOP structure. With the introduction of AltRefs, the hierarchical B pictures behave very well similar to the Random Access GOP which in 4 layers is the default GOP structure of SVT-AV1.

In the Hierarchical 3 layer short as *HL3*, we see both the display order and coding order of the frames arranged. By following the coding order, we encode the I frame at layer 1 first and then move to the BWDREF *(B)* which is a reference frame for another *B* frame at second layer. The second layer B positioned at 2 references both the I frame at coding order 0 and B frame at coding order 1. We then move to the third layer where there are two *b* frames which are actually called AltRefs where AltRef1 references I and B frames at coding orders 0 and 2. The next AltRef2 positioned at 4 references both the B's in layer 1 at coding order 1 and the other B at the second layer at coding order 2 respectively. This is an example of hierarchical B pictures arranged temporally. In our proposed segment based encoding, we utilized this hierarchical system and, with different number of AltRefs, we were able to produce different encodings with coding efficiency and performance which will be discussed later in the next sections.

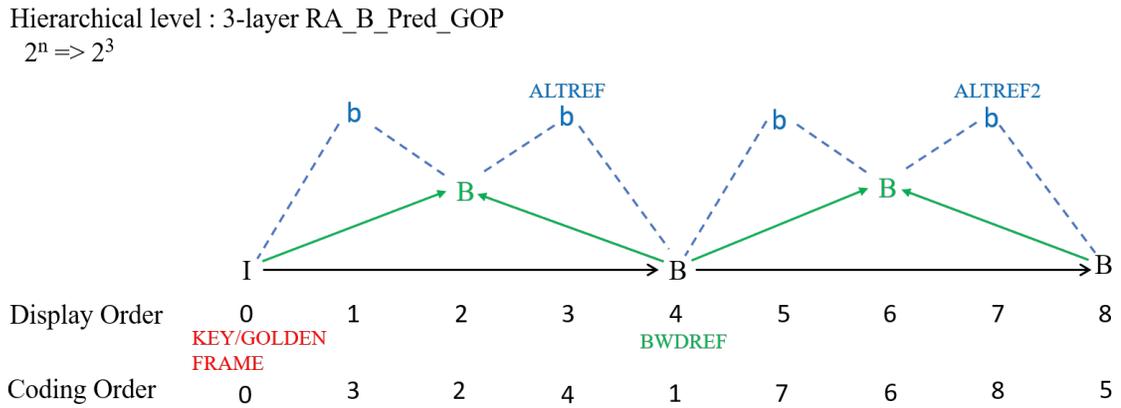

Figure 6.1: Hierarchical GOP Structure HL3 of SVT-AV1
[77]

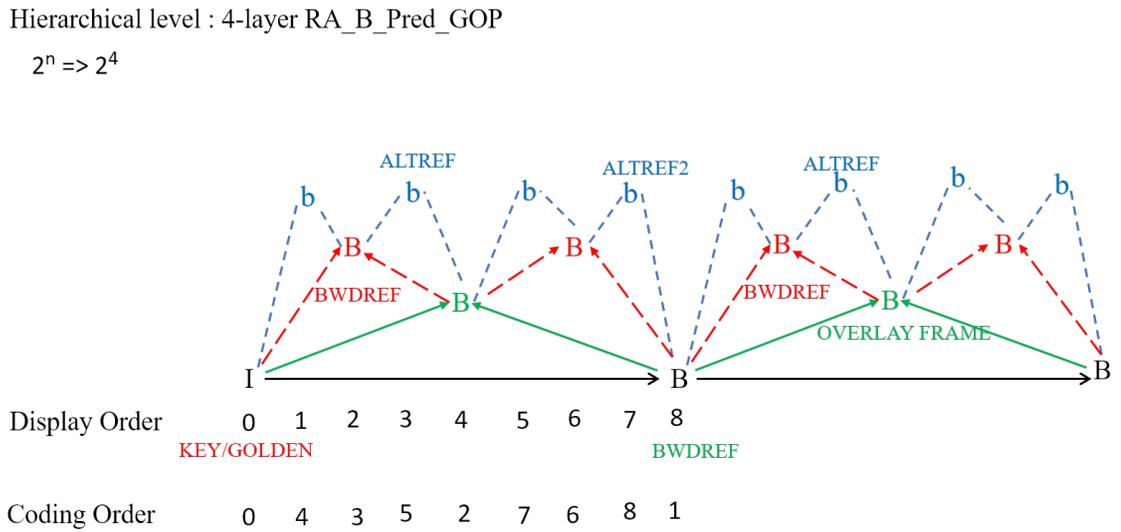

Figure 6.2: Hierarchical GOP Structure HL4 of SVT-AV1
[77]

### 6.1.3   Intra Prediction

AV1 has 56 directional prediction modes which are more than the 35 modes of HEVC, and the 10 Intra modes by VP9. Additionally, AV1 has 10 Intra smoothing modes, Chroma from luma prediction (CfL), Color Pallete coding and Intra block copying which is primarily applied in screen content coding. Also AV1 has extended its *Higher Directional angular* modes covering wider possibilities because of the increased block sizes 128x128 to provide accurate prediction along those directions. On top of VP9's 8 extrapolation directions, angle delta is enabled and also has extended modes realized using bi-linear interpolation of spatial references. Very similar to *True Motion(TM)* mode, there is a new tool that is added to the intra prediction known as Paeth predictor at the pixel level. *Chroma from Luma prediction- CfL* is a technique where the chroma AC components are predicted from the subsamples of corresponding luma AC coefficients. CfL uses linear prediction models that are conveyed in the bitstream, making the decoder implementation lighter and less sophisticated compared to the emerging VVC coding standard where a similar approach is employed for Intra prediction mode. *Intra Block Copying* can be applied to screen-content coding from patterns and textures from previously encoded frames. The Intra Block copy utilizes those previously reconstructed blocks in the same frame by signaling an intra frame motion vector and effectively captures the content of the screen-shots with a lot of text.

## 6.1.4   Inter Prediction

AV1 uses a block-based motion compensation for coding the motion vectors. AV1 supports overlapped block prediction and warped motion compensation supporting both translational and warped motion for the first time.

*Spatial MV prediction*: MVs of neighbors using the same reference frames are added to the pool. Compared to VP9, a deeper spatial neighborhood is searched here and separate pools for compound pairs are built from the reference frames.

*Temporal MV Prediction*: Temporal MV candidates are computed from motion trajectories through current block. Motion vectors throughout the current block are carried to the next frames by effectively indexing the motion trajectories in buffers and keeping a track of the projections. By this, when we decode the motion trajectories, the corresponding motion vector candidates for the current block are determined. This is capable of tracking motion at different frames especially tracking a particular object.

*Dynamic motion vector referencing*: VP9 only considers 2 MV candidates pulled from a fixed searching order In AV1, spatial and temporal MV candidates are indexed, prepared, scored, merged and ranked and AV1 supports 4 candidates. After they are indexed, they are sent to the bitstream.

*Overlapped Block Motion Compensation (OBMC)*: uses the assigned MVs per block. OBMC creates secondary predictions from neighboring MVs and blends them with block motion compensation to mitigate the effect of discontinued motion fields. AV1 OBMC is a 2-sided overlapped predictor in order to adjust for the flexible partitioning framework. Overlapping is operated in the top/left halves. AV1 uses predefined 1D smooth filters and the design keeps the memory bandwidth the same as the conventional compound prediction.

*Warped Motion Compensation*: In AV1, there are Warped motion models: 1.

Affine motion 2. Global warping (Frame level) 3. Local warping (Block level).

AV1 uses a 6-parameter affine motion model and allows for a limited degree of warping. It uses small warping that can be vectorized efficiently by one vertical shearing followed by a horizontal shearing for smaller motions.

Global warping Model is estimated from the encoder source by feature matching algorithms and the parameters are conveyed at frame level. The approach works very well for zoom, rotation and panning effects in videos where motion vectors can be extracted and analyzed for adaptive encoding.

The local warping model is estimated implicitly signal warping parameters for individual blocks. The motivation is to model real-time motions that cannot be simply represented by affine motion or cannot be estimated by homographic [78] motion models. Local warping models are estimated by using a linear curve fitting of neighborhood MVs and signal them into the bitstream which has an impact in the mode decisions as these parameters will be estimated after decoding. Combining both the local and global warp motion models is a great way of encoding but will be of higher computational complexity.

### 6.1.5 Transform Residual Coding

Transform partition sizes in AV1 range from 4x4 to 64x64 which is very similar to VP9, HEVC standards and allow for flexible partitions. The Transform kernels are extended in AV1 employing different versions of DCT, Asymmetric DST, flipped Asymmetric DST (flipADST), and Identity transform (IDTX). All these transforms are of bigger sizes because of the original blocks getting larger partitions and hence there are fewer kernels.

## 6.1.6   Loop Filtering

AV1 has three sets of filters in several stages: 1. Deblocking filter 2. Constrained Directional Enhancement Filter (CDEF) 3. Restoration loop Filter. The deblocking filter in AV1 uses the same filtering concepts as in VP9 and HEVC but is slightly better in terms of interpolation of both the luma and chroma samples. Followed by the deblocking filter is the CDEF or the De-ringing filter which removes ringing artifacts. Originally implemented from Cisco's Thor codec, it uses a low-pass directional filter in order to preserve edges. The final stage in the filtering process involves reconstructed pixels from the original video and then the filters are applied to improve the overall image quality. There are two kinds of restoration filters. One is a symmetric Wiener filter with effective weights applied with a 7-tap filtering mechanism which improves the image, followed by the self-guided projected filters which basically project onto the image itself. The final combination of all three filters makes one of the more complicated filtering mechanisms in the video coding standards.

## 6.2 Methodology

### 6.2.1 Segment-based Encoding with SVT-AV1 Configurations

In this section, we present the Segment-based encoding for SVT-AV1 encoder by following a similar approach as in Sections 4.2 and 5.2. We use the pseudo-code in Figure 4.4 to follow the Segment-based encoding by analyzing the encoding configurations and then we describe the modeling system. Originally, we started off with *libaom* encoder which has very similar encoding configurations to the VP9 codec but, we later moved onto SVT-AV1 since the latter is extremely fast, parallelized and multi-threaded. The *libaom* encoder, on the other hand, was slow and the encoding configurations do not have good documentation and the source code was quite complex to parse from the decoder side. We used a recent build of SVT-AV1 version 0.7 that had decent documentation but still there are many tools from the *libaom* that are yet to be implemented as the SVT-AV1 codec is still being finalized.

SVT-AV1 codec is complex and there are multiple configurations available to explore with different tuning options. SVT-AV1 codec is still in development and several tools from the AV1 standard are yet to be implemented. We decided to focus mainly on GOPs, filters and encoding presets for the use of segment-based coding as we did for x265. We have summarized a table of SVT-AV1 encoding configurations which we used in the segment-based encoding as shown in Table 6.1. For SVT-AV1, figuring out the GOPs was the first priority. SVT-AV1 combines the libaom reference frames with that of the Random access GOP configuration which is predominantly used for video transmission and streaming. The problem with libaom is its naming conventions to B frames and P frames in order to circumvent the patent issues and henceforth the big confusion before we started the encoding process. We confirmed the GOP structure from the SVT-AV1 documentation [16] that they are using random access but with Hierarchical level B pictures. At the time of this study, only 3 and 4 layer hierarchical levels are employed while the latest versions of the SVT-AV1 comes with hierarchical levels 2, 3, 4, 5 respectively.

We made use of the *AltRef* frames along with these hierarchical levels and we found out they have different encoding rate, bitrate and PSNR on different test encodes. We found that there is a similarity with x265 which has *B2 GOP*, VP9 which has *ALT2* and then in SVT-AV1 where we have hierarchical level 3 with 2 Altrefs hence named as *HL3ALT2*. Our naming convention was to use GOPs with both the hierarchical layers and vary the number of Altrefs together. Hence, we have 6 GOPs: HL3ALT0, HL3ALT2, HL3ALT8, HL4ALT0, HL4ALT2, HL4ALT8 of which HL4ALT8 is the default GOP from the SVT-AV1 documentation. In choosing the encode mode, we have a range of 0 to 7 with 0 being the slowest encode and 7 being the fastest encode. We chose 7 for encode mode that will be suitable for single pass encoding system applicable in VOD [47,49,50,73] systems. The QP ranges here from 0 to 63 in SVT-AV1 similar to VP9. We set the lowest QP to 16 and all the way to 52 in steps of 4. The arnr-maxframes is set to 7, arnr-strength to 5 and the arnr-filter type set to 3 as per the default settings in the SVT-AV1 bitstream specifications. Since, the SVT-AV1 codec employs a multiple filtering system, we enabled them all ON/OFF like deblocking and loop restoration filter both to be ON/OFF. And again VMAF is incorporated into the encoding pipeline using the VMAF 0.6.1 perceptual model [63] since VMAF is not built into any of the encoders as part of the system. Also, the VMAF score obtained by using this VMAF 0.6.1 model might be different if we use VMAF 0.6.2/0.6.3 and so for consistency, we used the VMAF 0.6.1 model for our perceptual video quality score.

| Parameter | Value |
|---|---|
| Encode Mode | 7 |
| Encoding Structure | HL3ALT0,HL3ALT2,HL3ALT8, |
| | HL4ALT0, HL4ALT2, HL4ALT8 |
| DBF | On/Off |
| Restoration Filter | On/Off |
| QP | 16 - 52 in steps of 4 |
| arnr-maxFrames | 7 |
| arnr-strength | 5 |
| arnr-type | 3 |
| Total encoding combinations | |
| per segment | 240 |

Table 6.1: SVT-AV1 Encoder Configurations for **rt** with our new GOP's

## 6.2.2   Forward Models and Inverse equation in SVT-AV1

We considered a total of 240 encodes per segment, and obtained the objectives PSNR in dB, VMAF, Bitrate in kbps, and Encoding rate in FPS. The model building process is quite similar to Figure 4.1 except here we apply these SVT-AV1 configurations. Here are the model equations summarized,

$$\ln(\text{PSNR})_i = \alpha_0 + \beta_1 \cdot \text{QP}_i + \beta_2 \cdot \text{QP}^2{}_i$$

$$\ln(\text{VMAF})_i = \alpha_1 + \beta_{11} \cdot \text{QP}_i + \beta_{12} \cdot \text{QP}^2{}_i$$

$$\ln(\text{Bits})_i = \alpha_2 + \beta_{21} \cdot \text{QP}_i + \beta_{22} \cdot \text{QP}^2{}_i$$

$$\ln(\text{FPS})_i = \alpha_3 + \beta_{31} \cdot \text{QP}_i + \beta_{32} \cdot \text{QP}^2{}_i$$

where $\beta_1, \beta_{i,1}, \beta_{i,2}$ represent QP coefficients and, $\alpha_0, \alpha_1, \alpha_2$ denote the constants of the polynomial regression equation. For the SVT-AV1 model building, we assumed it will follow a similar trend as QP and deblocking filter is taken as the predicted variable with respect to the objectives PSNR, Bitrate, VMAF and FPS correspondingly. We fit the model and the objectives had a very good adjusted R square value (for PSNR, VMAF, Bitrate, FPS) with a score of 0.99, 0.99, 0.99 and 0.99, respectively. The p-value for all these models are ($leq$0.05) [64], proving once again that the QP and Deblocking filter both had a significant effect on the fitted model objectives.

Forward models are built for each GOP and we solve the constraint optimization modes based on the selected DRASTIC operating mode to obtain the corresponding encoding configuration sets and constraints. We allow soft violations say 10% for bitrates and encoding frame rates and 3-5% for video quality, respectively. We initialize the QP search with QP=30 and then use Newton's method to derive the optimal models. For example, in maximum video quality mode, we have constraints set for encoding rate and bitrates. Since we have multiple QP values generated after solving the inverse equation, we approximate the generated QP value to the nearest integer and apply it to the encoder to obtain the optimal objectives. For example, if the QP value predicted is 27.5, then in maximum video quality mode it will be rounded to QP=27.0 in order to obtain higher video quality. If it's a minimum bitrate mode, then we will round it to QP=28.0 that will minimize the bitrate without sacrificing quality. If the generated QP value does not obey the constraints, then a local search performed with the nearest values of QPs in the range of (QP+4 , QP−4) and then we try to obtain a newer QP, and then repeat the process again until we satisfy the constraints.

## 6.3 Results and Discussions

### 6.3.1 Maximum Video Quality Mode
### Class E Kristen and Sara Video HEVC 720p Dataset

We take Kristin and Sara video as shown in Figure **??** from Class E 1280x720 from HEVC dataset and from YouTube [4] recommendation the bitrate is 5000 kbps. For the default mode we found that QP=20 achieves a total bitrate of 5070.03 kbps. Kristin and Sara video is a standard example of stationary video with a static back-ground and people interacting in a conversation without any complex motions. The default recommended settings and DRASTIC tables are summarized in Table 6.2 and 6.3 respectively. The GOP model equation tables are provided in Table 6.4 and 6.5 respectively.

| Seg ID | CQP | Fil | GOP | Bitrate (kbps) | PSNR (dB) | VMAF |
|--------|-----|-----|-----|----------------|-----------|------|
| **Seg0** | 20 | On | HL4ALT8 | 5031.22 | 45.61 | 97.2 |
| **Seg1** | 20 | On | HL4ALT8 | 5057.23 | 45.49 | 97.56 |
| **Seg2** | 20 | On | HL4ALT8 | 4714.28 | 45.51 | 94.12 |
| **Seg3** | 20 | On | HL4ALT8 | 6292.16 | 45.36 | 98.32 |
| **Avg** | | | | ***5070.03*** | ***45.51*** | ***96.49*** |

Table 6.2: Typical Mode - YouTube Recommended Bitrate achieved by CQP.

| Seg ID | QP | Fil | GOP | Bitrate (kbps) | PSNR (dB) | FPS | VMAF |
|--------|-----|-----|-----|----------------|-----------|-----|------|
| **Seg0** | 24 | Off | HL3ALT8 | *5024.1* | 45.54 | *25.40* | 97.05 |
| | | | | <=5070.03 | | >=25 | |
| **Seg1** | 24 | Off | HL3ALT8 | *4958.16* | 45.51 | *25.43* | 97.16 |
| | | | | <=5070.03 | | >=25 | |
| **Seg2** | 23 | Off | HL3ALT8 | *5044.39* | 45.61 | *25.77* | 97.13 |
| | | | | <=5070.03 | | >=25 | |
| **Seg3** | 24 | Off | HL4ALT8 | *4837.91* | 45.06 | *24.41* | 97.71 |
| | | | | <=5070.03 | | >=25 | |
| **Avg** | | | | *4991.78* | *45.50* | *25.41* | *97.17* |

Table 6.3: DRASTIC Maximum Video Quality Mode for Kristen and Sara 1280x720, 24 fps.

For the first and second segments with a average bitrate constraint of 5070 kbps, we obtain 5024.1 and 4958.16 kbps respectively. The corresponding VMAF scores are 97.05 and 97.16 almost perceptually similar to the default mode. Further, the second and third segments achieve bitrates of 5044.39 and 4837.91 kbps respectively. Overall, the average bitrate is 4991.78 kbps with a PSNR of 45.5 dB and VMAF of 97.17 respectively. With the threes segments using HL3ALT8 and last segment uses HL4ALT8 GOP, the corresponding model equations of them are shown in the Table 6.4 and 6.5. The overall gains for the maximum quality mode are provided in 6.6 and the corresponding chart in Figure 6.4

| *Coefficients* | $\beta_0$ | $\beta_1$ | $\beta_2$ | GOP.Str | Model Order | Adjusted $R^2$ |
|---|---|---|---|---|---|---|
| log(PSNR) | 3.812 | -0.000184 | -4.14199e-05 | HL3ALT8 | Quadratic | 0.99 |
| log(VMAF) | 4.566 | 0.002096 | -5.5965325e-05 | HL3ALT8 | Quadratic | 0.99 |
| log(Bits) | 10.7088 | -0.0975288 | 0.00041118 | HL3ALT8 | Quadratic | 0.99 |
| log(EncRate) | 2.362 | 0.03782 | -0.000386 | HL3ALT8 | Quadratic | 0.94 |

Table 6.4: HL3ALT8 GOP Model Equations for Maximum Video Quality Mode.

| *Coefficients* | $\beta_0$ | $\beta_1$ | $\beta_2$ | GOP.Str | Model Order | Adjusted $R^2$ |
|---|---|---|---|---|---|---|
| log(PSNR) | 3.809 | -0.000126 | -4.484814e-05 | HL4ALT8 | Quadratic | 0.99 |
| log(VMAF) | 4.563 | 0.002373 | -6.368074e-05 | HL4ALT8 | Quadratic | 0.99 |
| log(Bits) | 10.591 | -0.100708 | 0.000468 | HL4ALT8 | Quadratic | 0.99 |
| log(EncRate) | 2.424 | 0.031148 | -0.0002480 | HL4ALT8 | Quadratic | 0.97 |

Table 6.5: HL4ALT8 GOP Model Equations for Maximum Video Quality Mode.

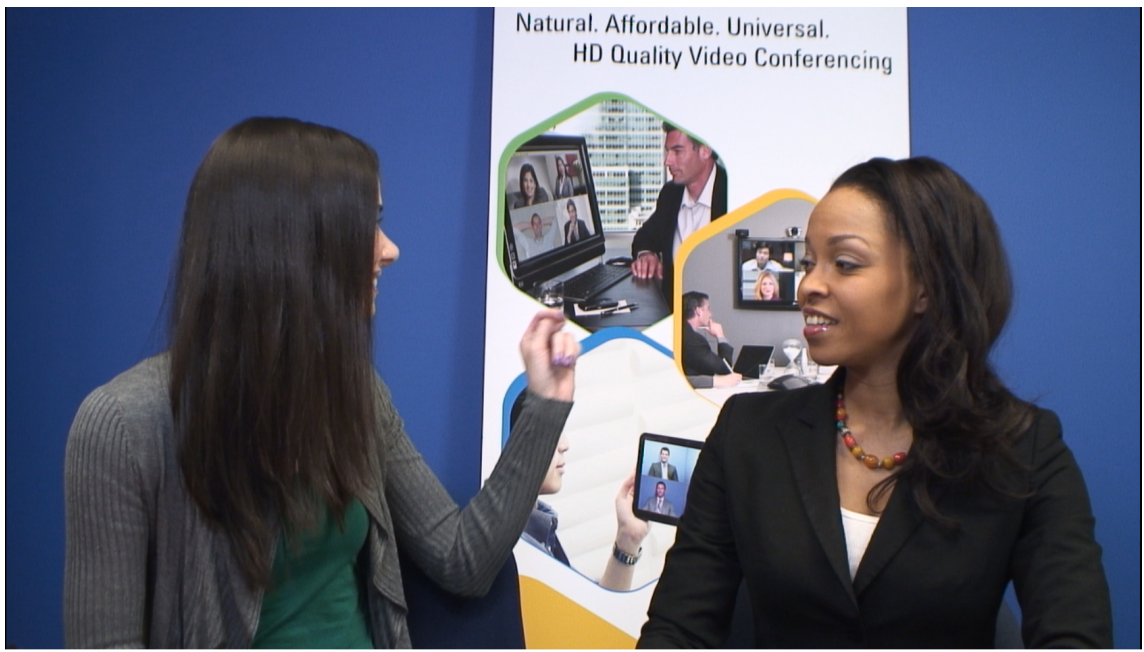

Figure 6.3: Kristen and Sara from HEVC [2] Video Sequence, 1280x720, 24fps

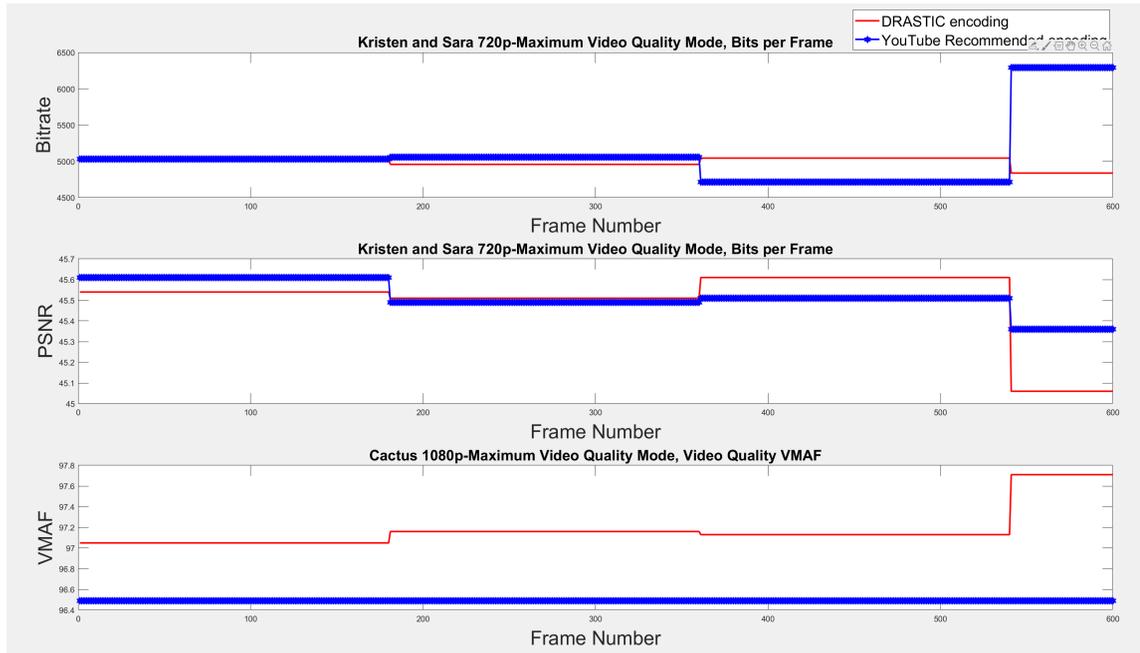

Figure 6.4: HEVC Test sequence, 1280x720, Kristen and Sara maximum Quality Mode

| *Overall Bitrate Gain* | *Overall PSNR* | *Overall VMAF* |
|---|---|---|
| 1.5 % | -0.01 dB | 0.68 |

Table 6.6: Overall DRASTIC Gains from Maximum Quality Mode.

## 6.3.2   Minimum Bitrate Mode -
### Class E Kristen and Sara Video HEVC 720p Dataset

In the minimum bitrate mode, we tried two approaches: 1. PSNR as the minimum acceptable quality metric 2. VMAF as the minimum acceptable quality metric. In the first approach, we saved 7% but with 6-point reduced VMAF from the default VMAF (92) as a constraint, we save around 61% bitrate savings respectively. We'll describe them briefly in the sections below.

| Seg ID | QP | Fil | GOP | Bitrate (kbps) | PSNR (dB) | FPS | VMAF |
|--------|----|-----|--------|----------------|-----------|--------|------|
| *Seg0* | 22 | off | HL4ALT8 | 4772.61 | *45.52* | *24.62* | 96.96 |
|  |  |  |  |  | >=45.51 | >=24 |  |
| *Seg1* | 22 | off | HL4ALT8 | 4623.82 | *45.58* | *25.67* | 97.11 |
|  |  |  |  |  | >=45.51 | >=24 |  |
| *Seg2* | 22 | off | HL4ALT8 | 4539.86 | *45.53* | *26.60* | 97.02 |
|  |  |  |  |  | >=45.51 | >=24 |  |
| *Seg3* | 22 | off | HL4ALT8 | 5338.25 | *45.18* | *29.47* | 97.81 |
|  |  |  |  |  | >=45.51 | >=24 |  |
| *Avg* |  |  |  | *4714.71* | *45.50* |  | *97.1* |

Table 6.7: DRASTIC Minimum Bitrate Mode for Kristen and Sara 1280x720, 24 *fps*.

| Coefficients | $\beta_0$ | $\beta_1$ | $\beta_2$ | GOP.Str | Model Order | Adjusted $R^2$ |
|--------------|-----------|-----------|-----------|---------|-------------|----------------|
| log(PSNR) | 3.814 | -0.000113 | -4.18589e-05 | HL4ALT8 | Quadratic | 0.99 |
| log(VMAF) | 4.5593 | 0.0019051 | -5.356909e-05 | HL4ALT8 | Quadratic | 0.99 |
| log(Bits) | 10.518 | -0.1047603 | 0.0005110 | HL4ALT8 | Quadratic | 0.99 |
| log(EncRate) | 2.454 | 0.0481 | -0.00052 | HL4ALT8 | Quadratic | 0.93 |

Table 6.8: HL4ALT8 GOP Model Equations for Minimum Bitrate Mode.

With PSNR 45.51 dB as a constraint, we have the first segment from DRASTIC obtaining a PSNR of 45.52 dB, bitrate of 4772.6 kbps and second segment up to 45.58 dB and bitrate of 4623.82 kbps, quite a marginal improvement. The third and the fourth segments have 45.53 dB, bitrate of 4539.86 kbps and 45.18 dB, bitrate of 5338.25 kbps respectively. On the VMAF scores, we have 96.96, 97.11, 97.02 and 97.81 for the first four segments in order. Overall, if we use PSNR as the video quality metric we save 7.0% in bitrate savings as seen in Table 6.9 and a marginal improvement of 0.01 dB in PSNR and 0.61 in VMAF respectively. The model equations and the overall graph is provided in Table 6.8 and Figure in 6.5 respectively.

| Overall Bitrate Gain | Overall PSNR | Overall VMAF |
|----------------------|--------------|--------------|
| 7.0 % | 0.01 dB | 0.61 |

Table 6.9: Overall DRASTIC Gains from Minimum Bitrate Mode

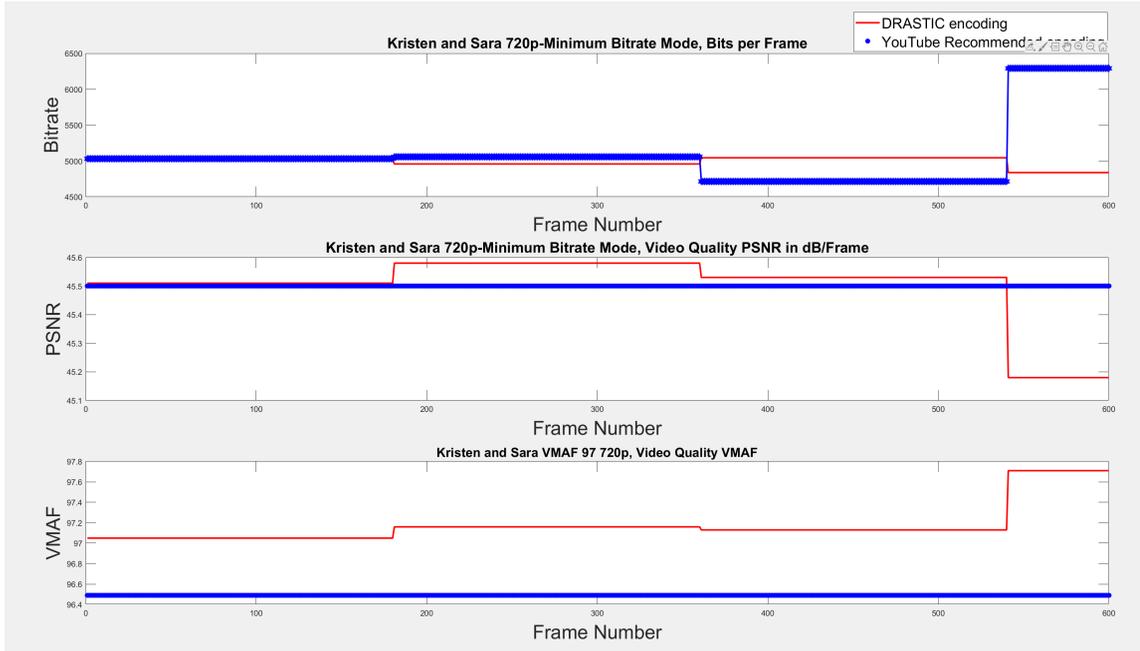

Figure 6.5: HEVC Test sequence, 1280x720, Kristen and Sara Minimum Bitrate Mode

Since, previously we do know that DRASTIC along with 6-point VMAF has a significant bitrate savings, we applied the same approach except we gave an acceptable video quality VMAF score of 96 instead of the default VMAF score of 97. We observed that there is 61% bitrate savings and the video is perceptually similar to the default YouTube recommended bitrates. The model equation and the overall gains is provided in the Tables **??**, 6.11, 6.12 and Figure in 6.6.

| SVT-AV1 Encoder Low Bandwidth example at 61% reduction in Bitrate | | | | | |
|---|---|---|---|---|---|
| QP=33.0 | fil=off | GOP=HL4ALT8 | PSNR=43.14 | BitRate=1981.59 | VMAF=95.86 |
| QP=33.0 | fil=off | GOP=HL4ALT8 | PSNR=43.15 | BitRate=1959.51 | VMAF=96.09 |
| QP=33.0 | fil=off | GOP=HL4ALT8 | PSNR=43.19 | BitRate=1782.8 | VMAF=96.02 |
| QP=33.0 | fil=off | GOP=HL4ALT8 | PSNR=42.75 | BitRate=2349.3 | VMAF=96.67 |
| *Avg* | | | *43.11 dB* | *1952.1 kbps* | *96.05* |

Table 6.10: Kristen and Sara Video Example of Low bandwidth Scenario with 1-point difference VMAF 96

| *Coefficients* | $\beta_0$ | $\beta_1$ | $\beta_2$ | GOP.Str | Model Order | Adjusted $R^2$ |
|---|---|---|---|---|---|---|
| log(PSNR) | 3.814 | -0.000145 | -4.13e-05 | HL4ALT8 | Quadratic | 0.99 |
| log(VMAF) | 4.56 | 0.001 | -5.2509e-05 | HL4ALT8 | Quadratic | 0.99 |
| log(Bits) | 10.534 | -0.10598 | 0.000529 | HL4ALT8 | Quadratic | 0.99 |
| log(EncRate) | 2.41 | 0.05 | -0.00057 | HL4ALT8 | Quadratic | 0.96 |

Table 6.11: HL4ALT8 GOP Model Equations for Minimum Bitrate Mode - Low Bandwidth

| *Overall Bitrate Gain* | *Overall PSNR* | *Overall VMAF* |
|---|---|---|
| 61.0 % | -2.4 dB | -0.95 |

Table 6.12: Overall DRASTIC Gains from Minimum Bitrate Mode

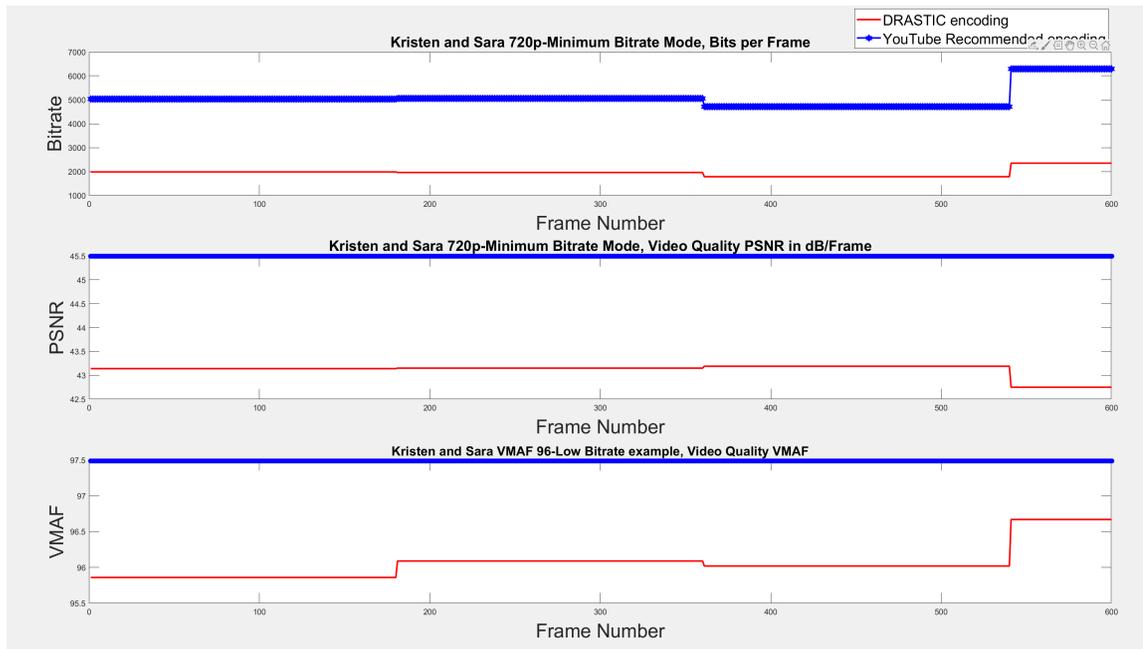

Figure 6.6: HEVC Test sequence, 1280x720, Kristen and Sara Minimum Bitrate Mode - Low Bandwidth example

## 6.4 Conclusion

The paper presents an adaptive encoding method that uses video content to determine constraints on video quality for real-time encoding. The basic approach is demonstrated on identifying camera motions but could be extended to cover other types of video content. Overall, the approach shows that substantial bitrate savings can be attained depending on the length of the activity of interest.

# Chapter 7

# Emerging VVC Encoding Standard with VMAF Metric Evaluation

## 7.1  VVC Video Coding Tools

Versatile Video Coding (VVC) is the next iteration of the H.265/HEVC video compression standard following the termination of JEM [79, 80] which was the original successor to HEVC standard. VVC or H.266 has numerous innovative tools added to mainly address the growing needs of video streaming, 360 videos with HDR content, omni-directional and support for 8K resolution. Like in previous standards, new tools have been added to the encoder at each stage right from block partitioning, Intra and Inter prediction modes, Transforms and Quantization, Entropy coding and Deblocking loop filters to provide higher coding gains and compression efficiency.

### 7.1.1  Block Partitions

A video frame is partitioned into Coding Tree units (CTUs) of block sizes 128x128 organized as tiles quite similar to the HEVC standard and can be grouped together as Tile-groups. A tile is a sequence of CTUs that covers a rectangular region of a picture [81] and is partitioned according to the raster-scan order. Note that a single Coding Tree Blocks CTB can have flexible partitions into Coding Units (CUs), Prediction Units (PUs) and Transform Units (TUs) in HEVC [21]. VVC uses a Quadtree [82] split into following nested partition types 1) Quad Split which is a recursive splitting of squares of 4 sub-blocks, 2a) Binary split where the block division occurs by 2 parts either horizontal or vertical, 2b) Ternary [83] split where the blocks are split as rectangular block divisions recursively as shown in Figure 7.1 using these different levels of splitting, fine details, textures, and spatio-temporal motions can be captured with greater flexibility.

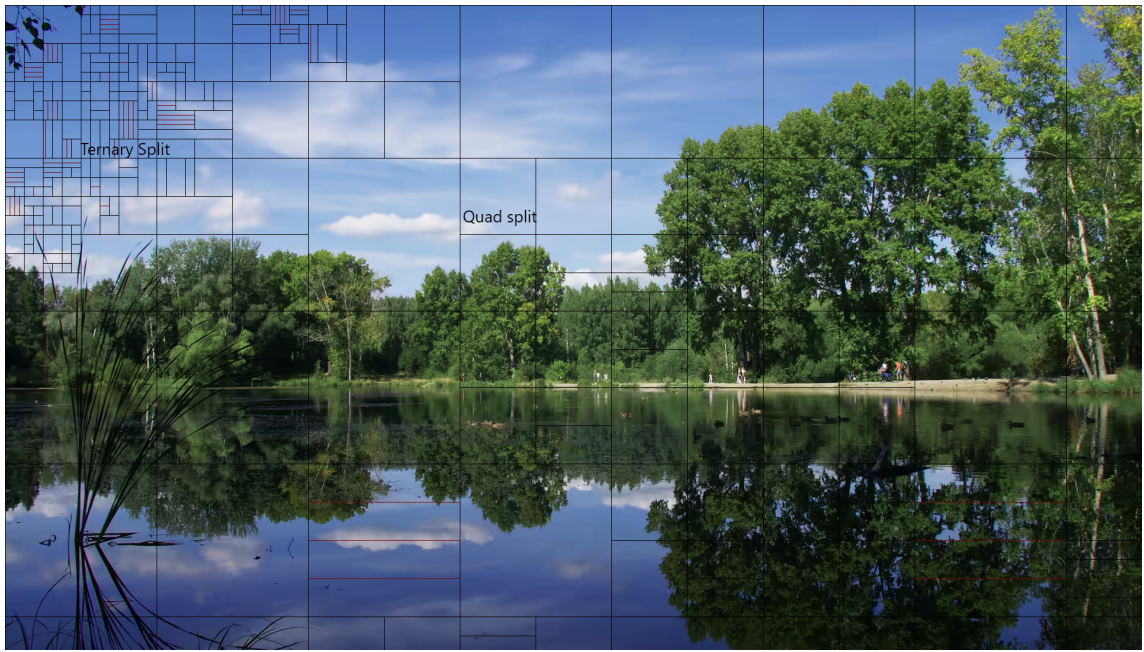

Figure 7.1: CTU with Multiple Partitions-Quad and Ternary Splits in VVC.

## 7.1.2  Intra Prediction

Intra Coding in HEVC has 33 directions in addition to DC and planar modes for a total of 35 directions. VVC extends the angular directions even further to 65 with planar and DC mode making a total of 67 directions. Extended directions mean more precise predictions for rectangular and non-square blocks. Intra coding in HEVC provided square blocks for prediction and have to be powers of 2 since the maximum size of CTU went up to 32x32. In VVC, rectangular blocks are provided and in DC mode, only the longer side of the block is taken to calculate the average across the blocks. Even though VVC is still standardized with some coding tools not officially ratified, there are promising new additions that are included to Intra coding.

*Cross-component linear model Intra Prediction*, for instance, is a new addition to VVC which is very similar to Chroma from Luma (CfL) prediction mode in AV1 standard where both the luma and chroma components carry block information in case of edge of a block to the bitstream. In cross-component prediction, this is exploited by direct prediction of the chroma components from the reconstructed luma block using a simple linear model with two parameters calculated from the intra reference pixels. Using these approaches, there is a great reduction in cross-component redundancy wherein luma and chroma samples are calculated for each angular direction as in HEVC. This is avoided in VVC because of the efficient use linear model prediction [84] taken from the reconstructed luma samples. In HEVC, for angular prediction we have one top and left neighboring samples for intra prediction at any time, while, in VVC, it has been extended to *Multi Reference Line Prediction* where two or more samples from both top and left are available to predict the current block.

With multiple lines of samples available, it is necessary to provide efficient filtering especially at the neighboring block edges and this is done by *Mode Dependent Intra Smoothing* which enhances the angular directional prediction accuracy by employing 4-tap interpolation filters. Note that HEVC used only 2-tap interpolation for intra smoothing but since the intra angular directions are more wider, they need additional smoothing to improve the prediction accuracy. Another add-on to Intra coding is the *Intra sub partitions* wherein the intra block itself is predicted using one of the intra mode and the prediction error signal is transformed and quantized and reconstructed after inverse-transform and finally stored in the intra picture buffer. This reconstructed sub-partition is then used as a reference for the other sub-partitions for that block. It is quite a divisional way of prediction from sub-partitions within a partition. For this to work, all the sub-partitions have to be predicted by the same angular mode.

### 7.1.3   Inter Prediction - OBMC & Affine Motion

Inter coding in VVC is similar to HEVC except the motion compensation can now account for non-translational motion models which was never considered due to higher computational complexity. VVC features several new additions to the inter prediction tools mainly focusing on multiple motion merge modes. A prominent addition to VVC is the introduction of OBMC and affine motion.

*Overlapped Block Motion Compensation* [85] is not a new technique but rather dates back to MPEG-4 standard but was not officially ratified because of the complex calculations involved with motion vectors (MVs). OBMC is where the predicted block is associated with a single vector MV0 corresponding to the blocks center, while corner MVs are taken from causal (already decoded) neighbor blocks. MV is most reliable in the center of the block (where prediction errors tend to be smaller than those at the corners). For a block, it's better to assign several MVs (its own and nearby blocks) and to blend reference samples for better prediction. Blending is executed in two separable stages: firstly, according to vertical direction and then according to horizontal direction. It is highly effective when there are artifacts produced at the boundary edges of a block due to low bandwidth situations and for these cases, the prediction due to OBMC is better than traditional motion compensation methods.

The motion compensation in HEVC standard accounts only for translation motions but in practice there are videos which have a lot of circular, zoom and rotations which cannot be accommodated by traditional motion compensation. VVC offers a new kind of motion modeling known as *Affine motion* where a particular block for tackling rotation or zooming using 4-parameter or 6-parameter equations for better prediction. This affine modeling works because at any block where rotation occurs, rotation persists throughout the video frames and hence the motion vector modeling can be propagated throughout the video. Note that AV1 also uses affine modeling but uses a global and local warping mechanism combined with affine motion modeling, resulting in a quite complex but more accurate prediction.

### 7.1.4   Transforms and Quantization

VVC supports large transform block sizes up to 64x64 where HEVC covered up to 32x32 block sizes for TUs. The primary advantage of having larger block sizes in Transform Units is that they provide better prediction in high resolutions like 1080p and 4K. HEVC by default used DCT-II for residual coding in both intra and inter-coded blocks and DST for 4x4 Intra coding blocks specifically and VVC supports DCT-VIII and DST-VII. In HEVC, the QP value had a range from 0 to 51 and here in VVC, QP has been increased to 63 similar to AV1 standard (which also covers a range from 0 to 63).

### 7.1.5   Loop Filters & Entropy Coding

HEVC standard provides two different kinds of filtering processes during the frame reconstruction: 1. Deblocking filter and 2. Sample Adaptive Offset (SAO). VVC has one more filter called adaptive loop filter (ALF) which was originally considered for HEVC standardization but left out in the final version. Later, it was picked as one of the primary filters in the JEM reference encoder and VVC adopted it making it three-loop filtering system. A similar chain of three filters is found in AV1 standard: 1. Deblocking filter 2. Constrained Directional Enhancement Filter (CDEF) and 3. Restoration filter. Both of these filtering mechanisms reduce artifacts, ringing effects (Ringing effect is atype of noise artifact in image processing) and enhance the image quality significantly.

## 7.1.6 Group of Pictures & Coding Performance

Video frames are coded as Group of Pictures (GOPs) based on configurations similar to HEVC and there is no change in VVC. These configurations are 1. All Intra (AI) 2. Low Delay (P/B) and 3. Random Access (P/B) and can be used to code any video content depending on our focus of application areas. Since our focus is in video streaming, we will focus on the Random-Access configuration which is the most relevant when it comes to video transmission and broadcasting.

Here, we summarize the VVC coding performance of Class-B BasketballDrive 1920x1080 video from HEVC test sequence and the results for chosen QPs (22, 27, 32, 37, 42) encoded with Low Delay P and Random Access B GOP structure as tabulated in Table 7.1. We also calculated both subjective and objective video quality metrics PSNR, SSIM and VMAF and the corresponding bitrates. Note that *YUV_PSNR*, *Y_PSNR*, *U_PSNR*, *V_PSNR* are obtained from encoding logs and then we calculate PSNR611 which is the weighted average of the luma, chroma blue and chroma red, respectively. Also known as Global PSNR, we obtain the corresponding VMAF scores using the VDK [63] VMAF tool. Currently, VVC is extremely slow and the encoding time is quite high and needs more optimization to cut down total encoding time complexity.

| VVC_LOWDELAY_P | | | | | | | | | | |
|---|---|---|---|---|---|---|---|---|---|---|
| TOTAL FRAMES | QP | Y-PSNR | U-PSNR | V-PSNR | YUV-PSNR | SSIM | VMAF | BITRATE in kbps | ENC.TIME in sec | PSNR611 |
| | 22 | 39.3615 | 43.8341 | 45.1895 | 40.4481 | 0.995347 | 99.973599 | 15946.1664 | 306453.796 | 40.649075 |
| | 27 | 37.4569 | 42.4572 | 43.0289 | 38.5489 | 0.989731 | 97.277038 | 4917.3664 | 160578.477 | 38.7784375 |
| | 32 | 35.5552 | 41.1415 | 41.1733 | 36.6538 | 0.978608 | 87.189286 | 2235.6088 | 103082.973 | 36.95575 |
| | 37 | 33.5596 | 39.8365 | 39.3749 | 34.6896 | 0.959044 | 73.570743 | 1124.8256 | 66745.178 | 35.071125 |
| | 42 | 31.3267 | 38.6067 | 37.5532 | 32.5137 | 0.925036 | 57.533403 | 561.5584 | 44171.891 | 33.0150125 |
| | | | | | | | | | | |
| VVC_RANDOM_ACCESS_B | | | | | | | | | | |
| | | | | | | | | | | |
| TOTAL FRAMES | QP | Y-PSNR | U-PSNR | V-PSNR | YUV-PSNR | SSIM | VMAF | BITRATE in kbps | ENC.TIME in sec | PSNR611 |
| | 22 | 39.3933 | 44.2801 | 45.36 | 40.4635 | 0.99588 | 99.95928 | 14849.0832 | 406862.829 | 40.7499875 |
| | 27 | 37.8607 | 43.2769 | 44.1146 | 39.022 | 0.99244 | 98.15996 | 4916.7528 | 229959.143 | 39.3194625 |
| | 32 | 36.2269 | 42.1619 | 42.3841 | 37.4023 | 0.985021 | 90.588555 | 2239.5712 | 153585.472 | 37.738425 |
| | 37 | 34.2932 | 40.8477 | 40.5033 | 35.4875 | 0.970336 | 78.513701 | 1106.396 | 90391.86 | 35.888775 |
| | 42 | 32.2309 | 39.7232 | 38.7187 | 33.4714 | 0.944693 | 64.073401 | 572.7824 | 48703.368 | 33.9784125 |

Table 7.1: VVC results for BasketballDrive 1920x1080, Class B HEVC Video Sequence.

With this brief overview of VVC encoding described, we will now jump to the BD-PSNR, BD-VMAF rate comparison of VVC, x265, libVPx, SVT-AV1 Codecs and subjective video quality assessments in the forthcoming sections 7.2 and 7.3 .

## 7.2   BD-PSNR & BD-VMAF Comparison Results for x265, VP9, SVT-AV1, VVC Video Coding Standards

### 7.2.1   Introduction

Several comparison studies [28], [86–90] of the encoders from different encoding standards have been done over the past years. These comparisons take into account the coding gains in terms of compression efficiency, performance over different bitrates. In this subsection, we provide a background of various codecs compared and our approaches to objectively measure the performance of the emerging VVC encoder, AOM/SVT-AV1, HEVC/x265 and VP9 codecs for Video-on-Demand (VoD) streaming applications.

Our motivation for this section is derived from the IEEE Spectrum article in [91] where different codecs from encoding standards are compared [12] for medical video applications. Experimental evaluation based on different medical video datasets showed that VVC outperforms all other challenging competitor codecs and delivers a better compression efficiency than HEVC.

From the literature, we studied different codec comparison methods for different video coding standards and note that the conclusions of their results are completely different. For example, in [92] which included JEM encoder (which was an extension and successor to HEVC standard), the authors found that HEVC gave better compression than the AV1 standard. In [93] the authors provide a comprehensive comparison of HM, JEM, AV1, x264, x265, VVC, from different standards and claimed to have taken a balanced methodology for maximum coding efficiency and using Intra Coding tools implementations from all the codecs, and claim that the results found VVC does a better job than all other encoders.

Consistently, there have been different inconclusive results that have been reported throughout the literature and it is not clear which encoder has a dominant coding/quality trade-off and efficiency. In [94], it was reported the coding efficiency for AV1/VP9 was lower than H.264 and H.265 encoders achieving bit-rate gains up to 10.5% and 65.7% for the same video quality. In terms of encoding complexity, HEVC encoder was 10 times faster than x265 at the same time providing a better coding efficiency of 12.7%.

In [61], the authors reported the average bitrate gains of AOM/AV1 outperforming AVC by 48%, HEVC by 17% and VP9 by 13%. In [95], the authors reported that AV1 achieved average bit-rate savings up to 17% compared to VP9, JEM (Predecessor to VVC) savings of 30% relative to HM. Facebook [96], on the other hand, reported that AOM/AV1 surpassed its predecessor VP9 by 30%, delivering bitrate savings. Overall, they reported gains up to of 50.3% for x264 main profile, 46.2% for x264 high profile and finally 34.0% for VP9.

In [97], the authors used SSIM to compare the codecs and reported that AV1 delivered a 10% bitrate reduction compared to HEVC for the same PSNR and SSIM quality scores while JEM outperformed HM and AV1 by 25.4%. Netflix had a large scale comparison [98] of x264, x265, VP9 codecs which aimed for VOD applications and claimed that x265 and libvpx-VP9 had significant BD-rate reductions with gains increasing for 720p and 1080p. At the low resolution of 360p, bit-rate reductions were up to 30.8% and 22.6% which increased significantly up to 43.4% and 43.5% at 1080p, respectively.

Another study from Netflix [99], which compared encoders with video-on-demand adaptive streaming as its application scenario, showed X.265/HEVC, VP9, AV1 codecs perform consistently with higher performance gains over H.264/AVC with BD-rate savings from 32.03% to 41.46% for VMAF, and from 32.13% to 44.82% for PSNR. AOM/AV1 outperformed all other codecs in terms of compression, while VP9 was marginally worse than X.265 in both PSNR and VMAF.

The following sections describe the video datasets, video codec setup and BD-metrics for both PSNR and VMAF for VVC and SVT-AV1 video codec from AOM, VP9 and x265 with VMAF as the perceptual video quality assessment metric for use in the video streaming domain. We wanted to compare these codecs for applications in adaptive streaming and we evaluated the codecs using three video datasets: UT-LIVE [1] , HEVC [2], Tampere [3]. In this chapter, we will first explain the various codec configurations of each encoder and the encoding tests, and then we move to subjective video quality assessment.

## 7.2.2 Video Content Description

Video datasets that are used in the codec evaluation and their content description are provided in the following Tables I, II and III. Each dataset is further broken down into its resolution, dimensions, frame rates and video content describing var-ious activities and spatio-temporal motions. The first dataset is from the HEVC Test sequences from Classes A-E with resolutions 2500x1600, 1920x1080, 832x480, 416x240, 1280x720 and with frame rates 24, 25, 30, 50, 60 *fps*, respectively. The second dataset is from Ultra Video group from the Tampere University with resolution 1920x1080 and with frame rates 30, 50, 60 *fps*, respectively. The third dataset is from UT LIVE VQA database with a resolution of 768x432 and with frame rates 50, 25 fps respectively.

| Video | Class | Dimension (WxH) | FPS | Video Content description |
|---|---|---|---|---|
| Blowing Bubbles | D | 416x240 | 50 | Medium motion, Zoom out, Textured background |
| BQ Square | D | 416x240 | 60 | Camere tilted movement, Non-uniform motion |
| Basketball Pass | D | 416x240 | 50 | High motion , Panning movement, Textured background |
| Race Horses | D | 416x240 | 30 | Medium motion, Camera tilted |
| Basketball Drill | C | 832x480 | 30 | High motion |
| Party Scene | C | 832x480 | 50 | Camera Zoom in, Medium motion |
| RaceHorses | C | 832x480 | 50 | Medium motion, Camera tilted moving across the background |
| BQMall | C | 832x480 | 30 | Medium motion, Camera panning, People walking across |
| Vidyo1 | E | 1280x720 | 60 | Stationary, Three People, face expressions |
| Vidyo2 | E | 1280x720 | 60 | Stationary, Single person |
| Vidyo3 | E | 1280x720 | 60 | Stationary, Single person |
| Four People | E | 1280x720 | 60 | Stationary, Four people conversing |
| Kristen and Sara | E | 1280x720 | 60 | Two People conversing static background |
| Johnny | E | 1280x720 | 60 | Single person with static background |
| Basketball Drive | B | 1920x1080 | 50 | High Motion, Camera following the player |
| Cactus | B | 1920x1080 | 50 | Complex circular, rotational motions with a static background |
| Kimono | B | 1920x1080 | 24 | Medium motion, Camera panning across the frame with a scene change |
| Park Scene | B | 1920x1080 | 24 | Medium motion, Camera pans across following the bicyclists |
| BQ Terrace | B | 1920x1080 | 60 | Medium motion, Camera tilts at an angle and then focuses on the road |
| Traffic | A | 2560x1600 | 30 | Stationary, several cars moving on a busy road |
| PeopleonStreet | A | 2560x1600 | 30 | Stationary, several people Crossing the road |

Table 7.2: HEVC Video Dataset
[2]

| Video | Class | Dimension (WxH) | FPS | Video Content description |
|---|---|---|---|---|
| Beauty | B | 1920x1080 | 60 | Stationary camera focusing on Smooth face and textured hair |
| Bo | B | 1920x1080 | 60 | Camera following a boat moving across the sea |
| HoneyBee | B | 1920x1080 | 60 | Honey bee moving across purple flowers |
| Jockey | B | 1920x1080 | 30 | High motion, Camera tracking a single jockey |
| ReadysetGo | B | 1920x1080 | 30 | High motion, Camera tracking multiple jockeys on the racecourt |
| ShakeandDry | B | 1920x1080 | 30 | Stationary camera focusing on dog shaking out the water |
| Yachtride | B | 1920x1080 | 60 | High motion, Camera following the boat and large swirls of surrounding water |

Table 7.3: Tampere Video Dataset
[3]

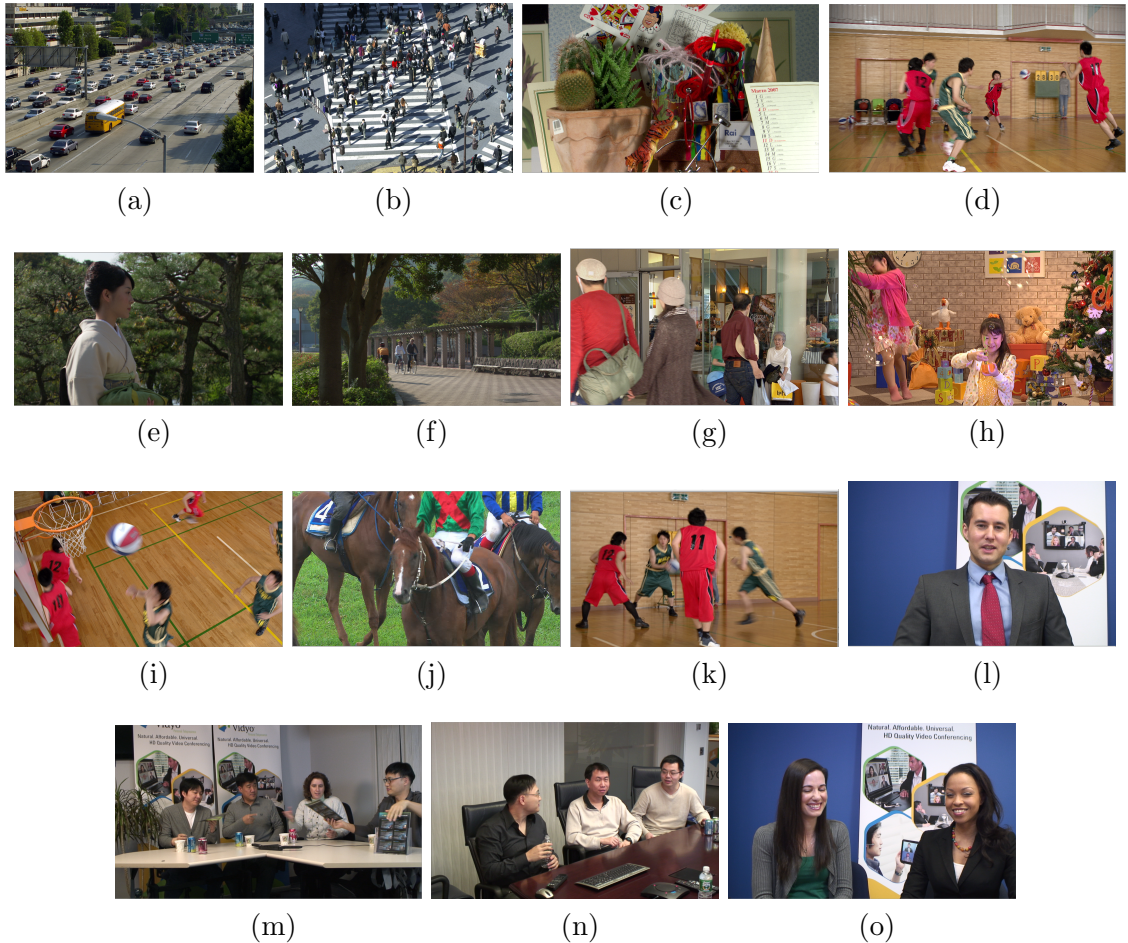

Figure 7.2:  HEVC Video Dataset with different Video resolutions and activities for BD-Rate [1] (a), (b) Traffic, People video with resolution 2500x1600, from Class A with 25 fps, (c), (d), (e), (f) Cactus, Basketball Drive, Kimono, Parkrun of 1920x1080 from Class B with 50 fps, (g), (h), (i), (j) BQMall, PartyScene, BasketballDrill, Racehorse of 832x480 from Class C, (k), (l), (m), (n), (o) Johnny, Four people, Vidyo1 and KristenandSara of 1280x720 from Class E with 60 fps, respectively.

## 7.2.3   Video Codec Configuration Setup

Each video codec chosen from a video coding standard and its encoding parameters are enlisted in Table 7.5. For x265, encoder we selected its default settings and *Ultra-fast* preset enabled for Video On Demand (VOD) adaptive streaming applications. For VVC standard, we used Random Access B GOP with an Intra-period of 32 for

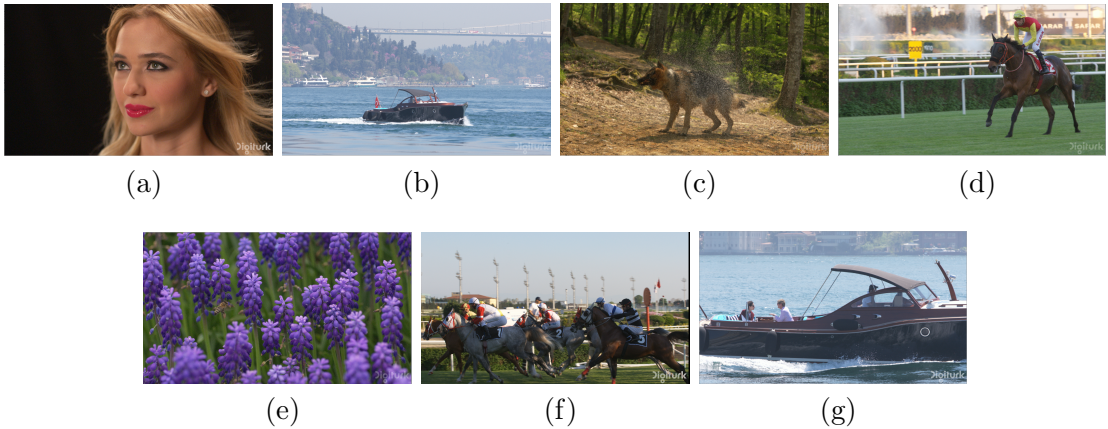

Figure 7.3: Tampere Video, 1920x1080 Dataset (a) Beauty,(b) Bo, (c) Shake and Dry, (d) Jockey, (e) Honeybee, (f) Ready set go, (g) Yacht videos with resolution of 1920x1080, respectively [3].

| *Video* | *Class* | *Dimension (WxH)* | *FPS* | *Video Content description* |
|---|---|---|---|---|
| Bluesky | Custom | 768x432 | 25 | Circular motion across the blue sky |
| MobileCalendar | Custom | 768x432 | 50 | Zooms out, non-uniform motion happens |
| Tractor | Custom | 768x432 | 25 | Camera follows the tractor and then zooms on the big wheels |
| Pedestrian | Custom | 768x432 | 25 | Stationary, People walking on a street |
| Parkrun | Custom | 768x432 | 50 | Camera tracks a man and then becomes stationary |
| Riverbed | Custom | 768x432 | 25 | Swirling water flow, Camera stationary |
| Rushhour | Custom | 768x432 | 25 | Static camera on a busy road |
| Sunflower | Custom | 768x432 | 25 | Highly textured sunflower and moving camera following the bee |
| Shields | Custom | 768x432 | 50 | Camera tracking the shield, Stops and then zooms in |
| Station | Custom | 768x432 | 25 | Camera Zooms out |

Table 7.4: UT LIVE VQA, Dataset at a resolution of 768x432
[1]

24 fps, 30 fps and 25 fps and 48 for 50 fps videos. In both x265 and VVC, Deblocking filter and SAO were enabled for highest video quality. For both x265 and VVC, the QP values are set in a range of 22, 27, 32, 37.

In VP9, we selected the default configurations with a *–lag-in-frames=25* set with *–end-usage=3* which is for fixed QP setting along with the alternate reference frames enabled. Each alternate reference frame has a maximum filter strength set to 5 and maximum number of reference frames is set to 7 with the alternate noise reduction set to 3. For SVT-AV1, the default encoding mode/preset is set to the highest quality and faster encoding speed with number of hierarchical levels 4. Both the restoration filter and loop filter are enabled in this case along with fixed QP setting. Also, the range of QP values for both VP9 and SVT-AV1 codecs are set to 27, 35, 46, 55. These values are chosen specifically in order to match the rate-quality values at any given bitrate for a fair comparison.

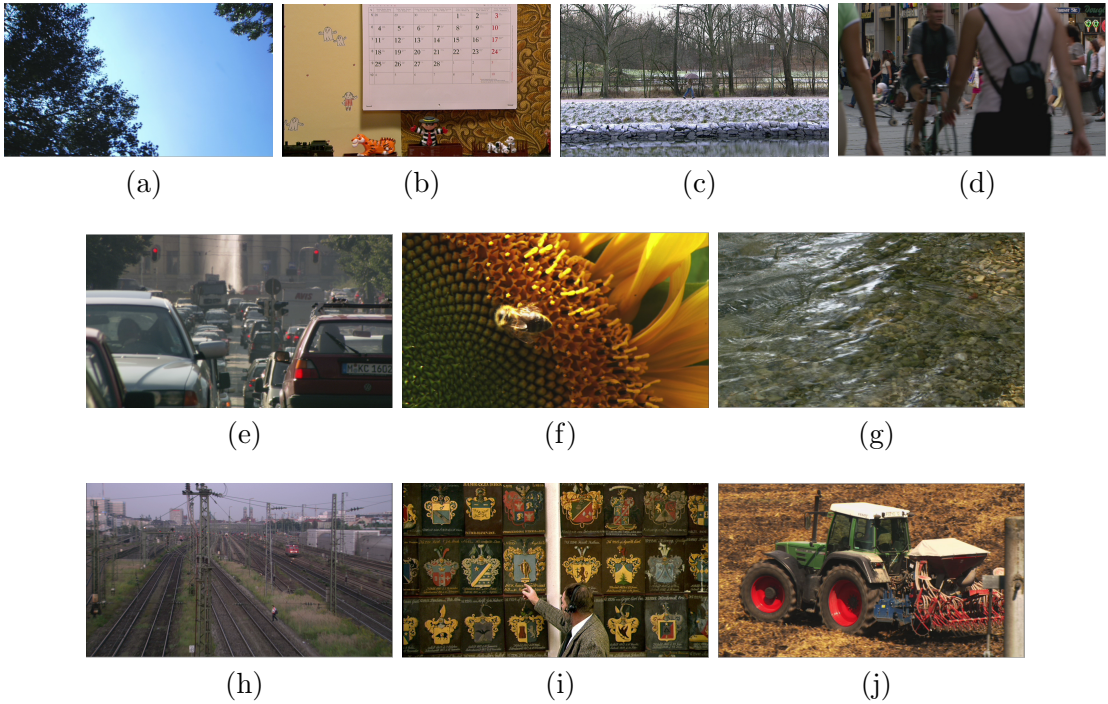

Figure 7.4: UT LIVE 768x432 Video Dataset (a) Beauty,(b) Bo, (c) Shake and Dry, (d) Jockey, (e) Honeybee,, (f) Ready set go, (g) Yacht videos with resolution 1920x1080, respectively.

| *Codec* | *Version* | *QP range* | *Encoding Parameters* |
|---------|-----------|------------|------------------------|
| X265 | 2.1 | 22, 27, 32, 37 | –psnr, –ssim, –sao, –deblock |
| VVC | 7.1 | 22, 27, 32, 37 | –psnr, –ssim, –sao, |
| VP9 | 1.8 | 27, 35, 46, 55 | –psnr, –i420 , –arnr-maxframes=7, –arnr-strength=5, –arnr-type=3, –end-usage=3, –bit-depth=8, –enable-altref=1, –lag-in-frames=25 |
| SVT-AV1 | 0.7 | 27, 35, 46, 55 | –psnr, –i420 , –arnr-maxframes=7, –arnr-strength=5, –arnr-type=3, –end-usage=3, –hierarchical-level 4, –enable-altref=1, –bit-depth=8, –restoration-filtering=0, –dlf=0 |

Table 7.5: Video Codec Configurations.

## 7.2.4 BD-Rate Bjontegaard Metrics

The BD-metric or the Bjontegaard Delta measure is an effective method to compare the performance of two codecs over different rate-quality points. In order to measure the BD-rates, we compute a RD points at four different QP levels and then we fit a cubic polynomial to the points for each of the codec. All the bitrate values are converted into log values. Then, the overlap or the area between the two fitted curves is computed, which infers the rate of average bitrate change to an equivalent PSNR or vice-versa.

### *BD-PSNR*

We take PSNR as our standard objective video quality measure since it is mostly used in video encodings as an objective benchmark metric even though it does not correspond well enough visually. Most encoders calculate the average PSNR using three different components from the original raw file which is in YUV format. *PSNR_Y* refers to luma component or the brightness intensity as a human eye is more susceptible to notice change in brightness than colors. *PSNR_U & PSNR_V* refers to chrominance components. We do a weighted average of all of them to measure the overall objective PSNR also known as PSNR611 or Global PSNR using:

$$PSNR_{611} = (6 * PSNR_Y + PSNR_U + PSNR_V)/8$$

### *BD-VMAF*

We take VMAF for our subjective video quality metric as it is currently the most popular metric that corresponds how the subject perceives the quality of the video. Several studies [100], [101], [1], [34], [102], [103], [104] have been conducted earlier that used SSIM, VIF, MSSSIM, VMAF for the perceptual quality. In our study, we use VMAF as the primary video quality measurement.

## 7.3   Results & Discussion

### 7.3.1   BD-PSNR & BD-VMAF for 240p HEVC Dataset

We begin with chosen configurations to encode for each encoder VVC, SVT-AV1, libVPx, x265 representing VVC, AV1, VP9, HEVC encoding standards respectively. Let's start with the HEVC Dataset which has 5 different classes spanning multiple resolutions and begin with 240p as shown in Tables 7.6 and 7.7 and Figures 7.5 and 7.7. Here, at the lowest resolution, VVC outperforms SVT-AV1 by 53.61%, x265 by 71.11% and 71.79% for VP9, respectively for PSNR. Notice that x265 requires 18.049% more bits than VP9 for the same quality level. SVT-AV1 requires 44.23% less bitrate than x265 and 34.811% less than VP9, which is significantly better for a future codec promising to cater to the low-bitrate streaming scenarios. The corresponding PSNR vs Bitrate in log scale is also provided in Figure 7.6 and 7.8 which shows VVC dominance in 240p resolution. Based on BD-VMAF, VVC achieves 59.97% bitrate reduction against SVT-AV1, 77.82% reduction against x265 and 75.06% reduction against VP9.

| *Bitrate savings Relative to* | | | | |
|---|---|---|---|---|
| *Encoding* | *VVC* | *SVT-AV1* | *x265* | *VP9* |
| *VVC* | - | 53.61% | 71.11% | 71.79% |
| *SVT-AV1* | | - | 38.79% | 40.93% |
| *x265* | | | - | 3.65% |

Table 7.6: BD-PSNR HEVC VIDEO DATASET 416x240p

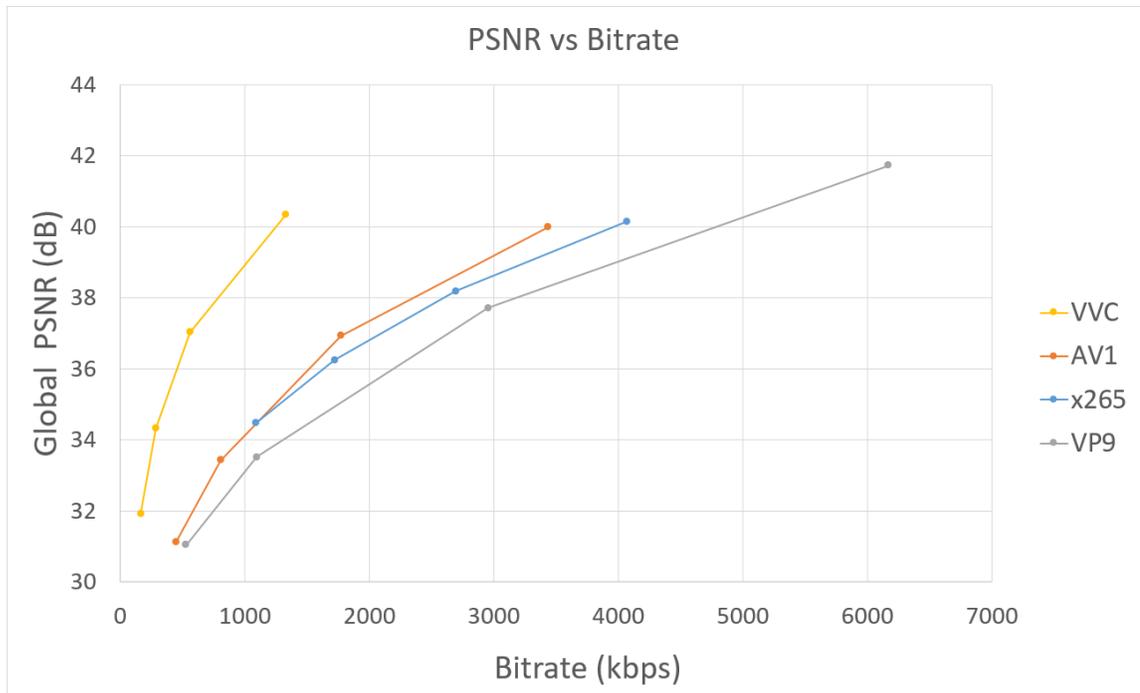

Figure 7.5: HEVC Dataset 240p RD Curves (PSNR vs Bitrate) of Median Values

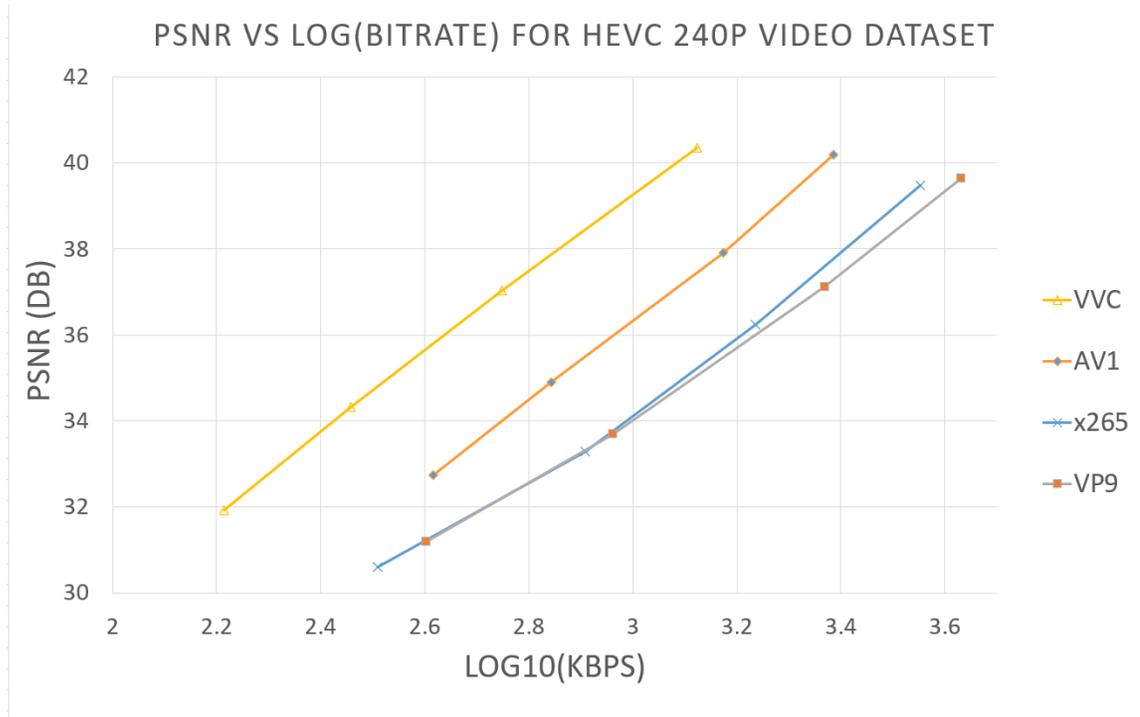

Figure 7.6: HEVC Dataset 240p RD Curves PSNR vs Log(Bitrate) of Median Values

| *Bitrate savings Relative to* | | | | |
|---|---|---|---|---|
| *Encoding* | *VVC* | *SVT-AV1* | *x265* | *VP9* |
| *VVC* | - | 59.97% | 77.28% | 75.06% |
| *SVT-AV1* | | - | 44.23% | 34.81% |
| *x265* | | | - | 18.05% |

Table 7.7: BD-VMAF HEVC VIDEO DATASET 416x240p

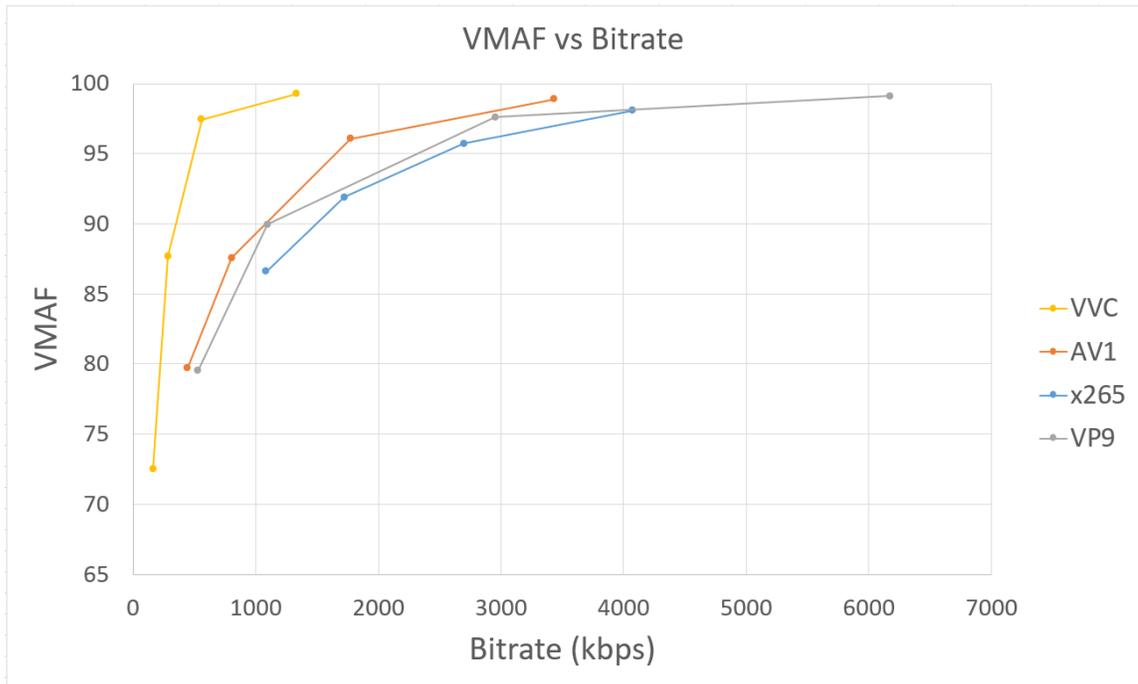

Figure 7.7: HEVC Dataset 240p RD Curves (VMAF vs Bitrate) of Median Values

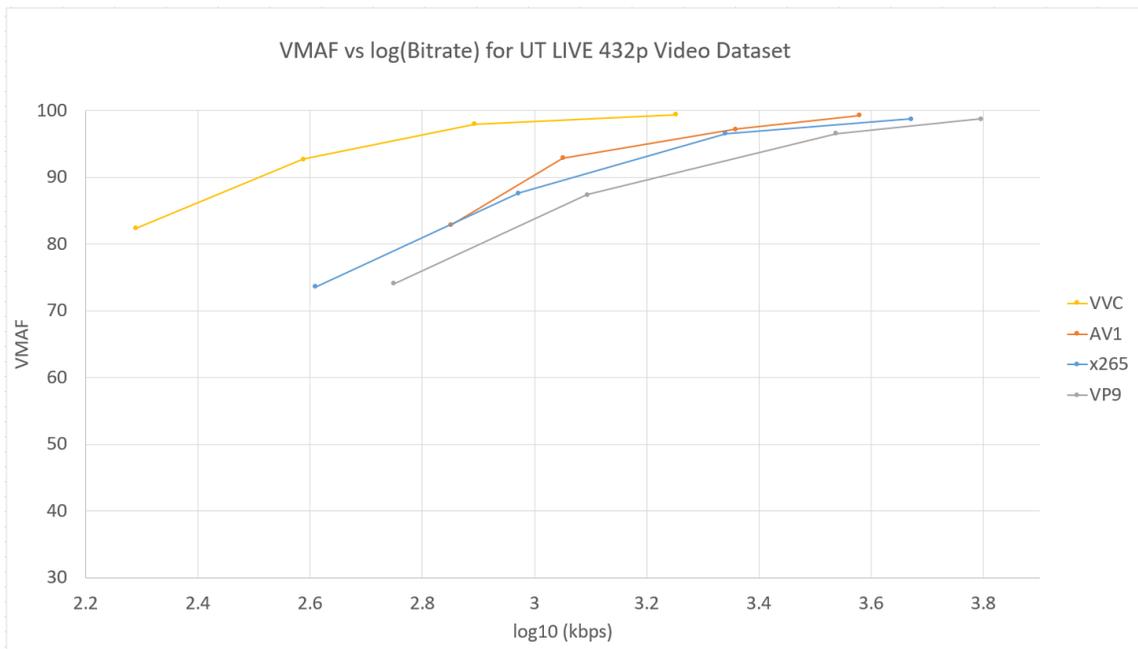

Figure 7.8: HEVC Dataset 240p RD Curves VMAF vs Log(Bitrate) of Median Values

## 7.3.2    BD-PSNR & BD-VMAF for UT LIVE Dataset

The dataset from UT LIVE VQA has a custom video resolution of 768x432 and VVC provides significant bitrate reductions against AV1 by 64.8%, x265 by 68.94% and 74.90% based on BD-VMAF, respectively. AV1 has lower BD-VMAF as shown in Table 7.9 with savings of around 7.30% against x265 and 12.80% against VP9. The corresponding rate curves for VMAF is described in Figure 7.11. For BD-PSNR rates in Table 7.8, VVC has better bitrate gains compared to the BD-VMAF of about 56.17%, 67.50% and 75% for AV1, x265 and VP9, respectively. AV1 saves around 23.33% against x265 and 36.52% against VP9. In Figure 7.9, we can see x265 performs fair with 25.30% bitrate gains against VP9 which is clearly evident from the log scale of the Bitrate curve in Figure 7.10 and 7.12.

| *Bitrate savings Relative to* | | | | |
|---|---|---|---|---|
| *Encoding* | *VVC* | *SVT-AV1* | *x265* | *VP9* |
| *VVC* | - | 56.17% | 67.50% | 75.00% |
| *SVT-AV1* | | - | 23.33% | 36.52% |
| *x265* | | | - | 25.30% |

Table 7.8: BD-PSNR UT LIVE VIDEO DATASET 768x432p

| *Bitrate savings Relative to* | | | | |
|---|---|---|---|---|
| *Encoding* | *VVC* | *SVT-AV1* | *x265* | *VP9* |
| *VVC* | - | 64.80% | 68.94% | 74.90% |
| *SVT-AV1* | | - | 7.30% | 12.80% |
| *x265* | | | - | 17.90% |

Table 7.9: BD-VMAF UT LIVE VIDEO DATASET 768x432p

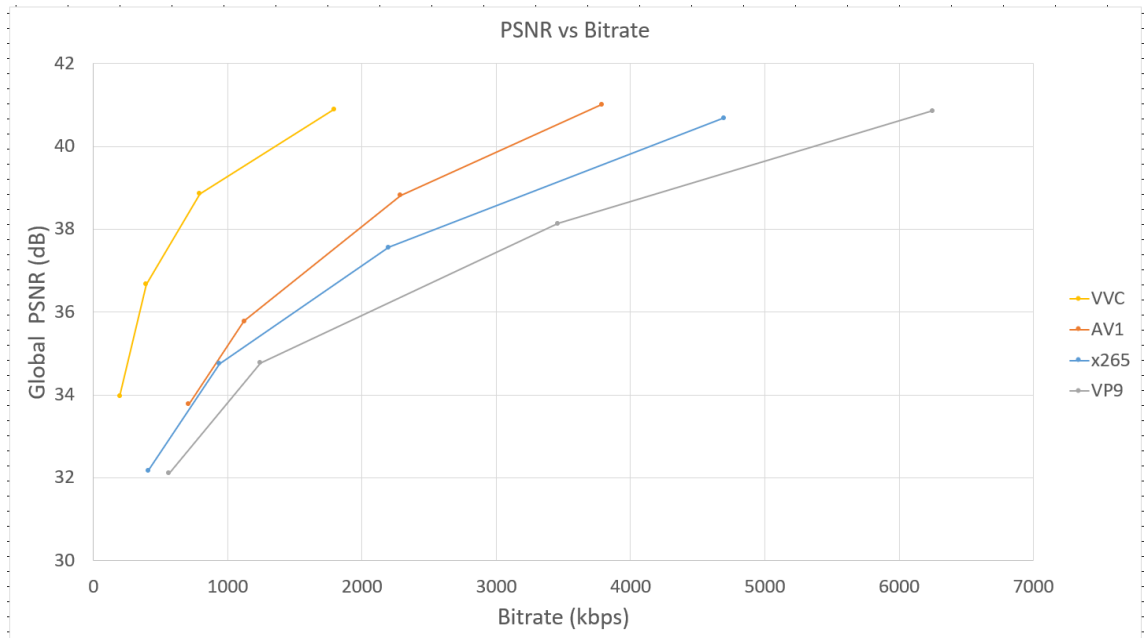

Figure 7.9: UT LIVE RD Curves (PSNR vs Bitrate) of Median Values

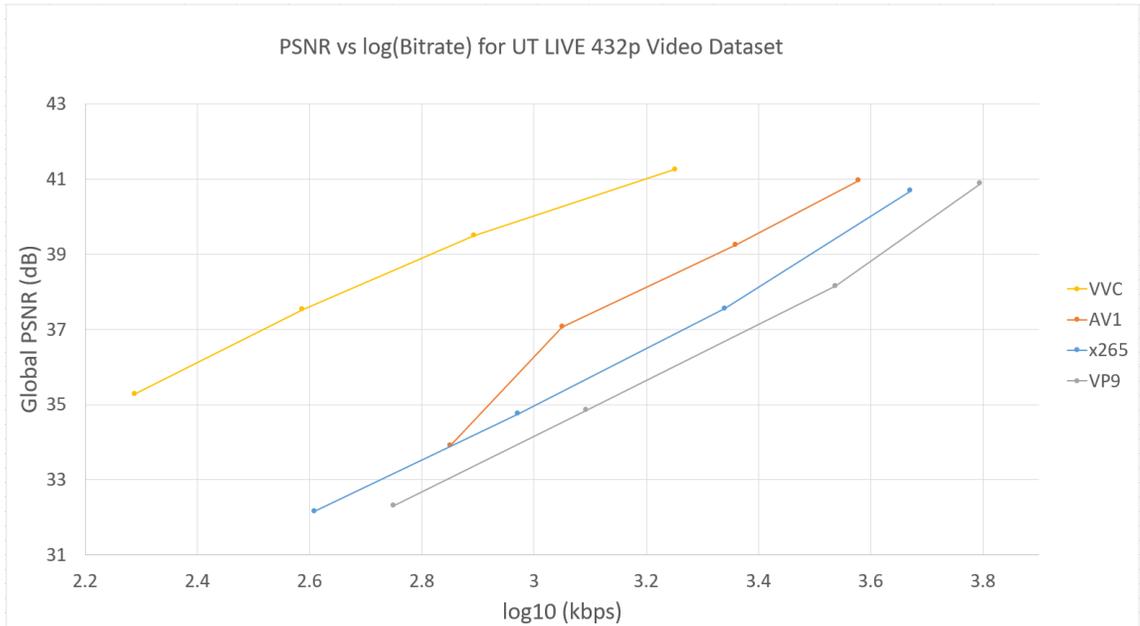

Figure 7.10: UT LIVE Dataset 432p RD Curves PSNR vs Log(Bitrate) of Median Values

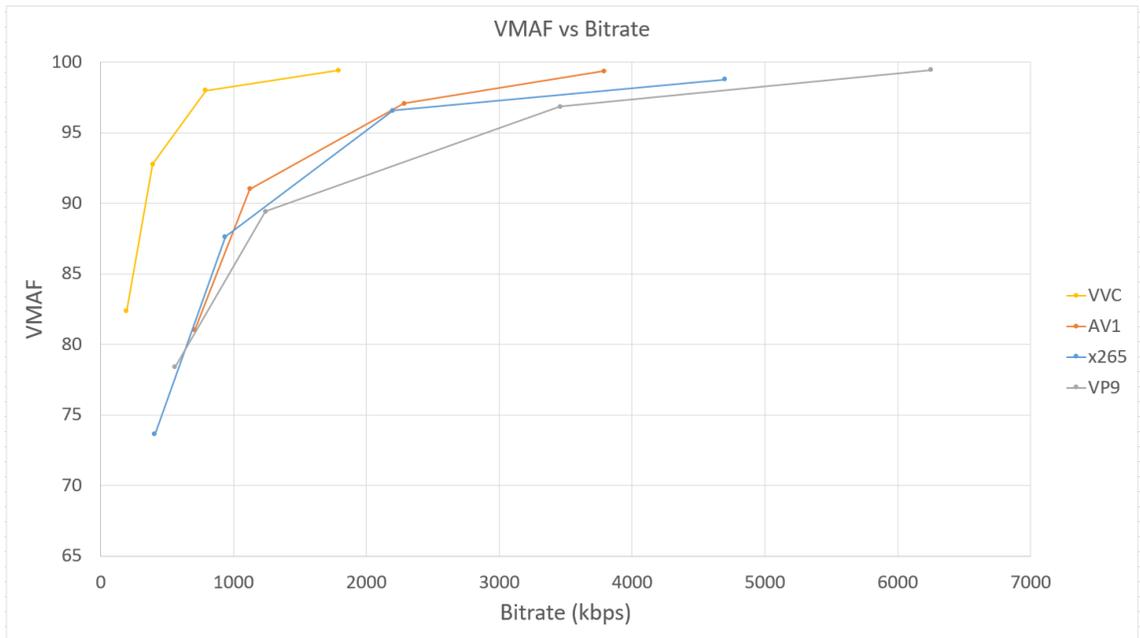

Figure 7.11: UT LIVE RD Curves (VMAF vs Bitrate) of Median Values

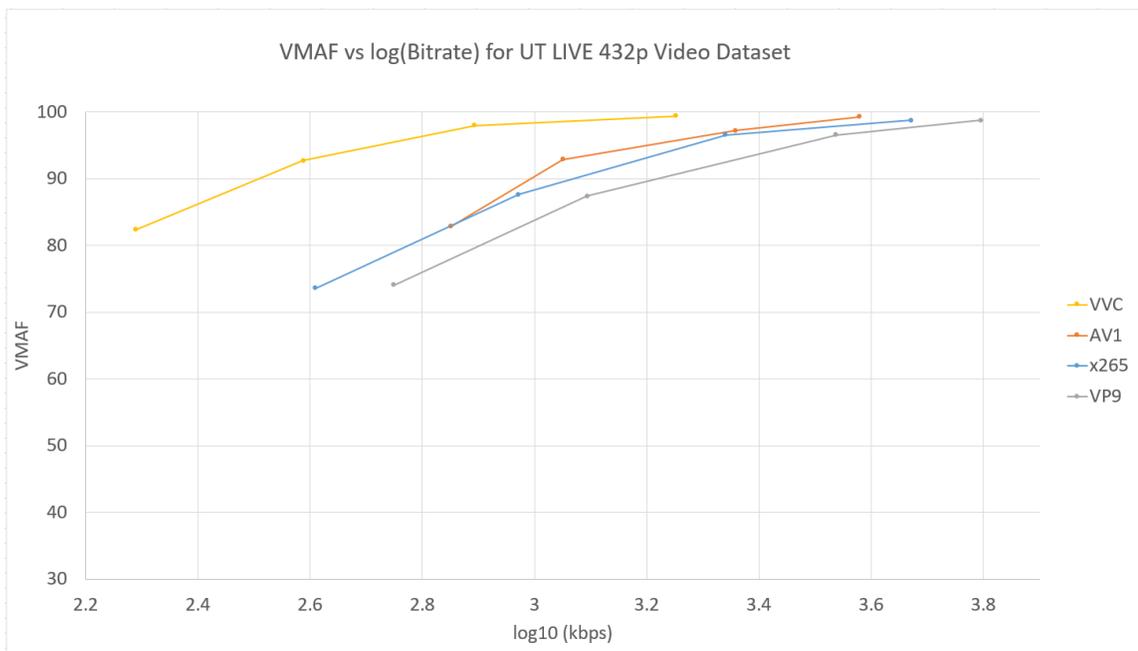

Figure 7.12: UT LIVE RD Curves VMAF vs Log(Bitrate) of Median Values

### 7.3.3    BD-PSNR & BD-VMAF for 480p HEVC Dataset

At 480p, where we have higher camera motions and textures, we observe that in Tables 7.10 and 7.11. VVC achieves BD-PSNR gains up to 56%, 70.3% and 73% against SVT-AV1, x265 and VP9, respectively. In terms of BD-VMAF, VVC beats SVT-AV1 by 59.63%, x265 by 71.54% and VP9 by 79.77%, respectively. VVC consistently beats VP9 with significant higher reduction in bitrates at approximately 80%, which proves that VVC savings have been quite higher than VP9. SVT-AV1 saves around 50.6% over VP9. On the other hand, x265 saves around 32.5% less reduction than SVT-AV1. It is important to note that the VMAF RD curves in Figures 7.13, 7.15, 7.14 and 7.16 respectively. As we see at higher bitrates, from all the codecs SVT-AV1, x265, VVC and VP9 the start of RD curve, SVT-AV1 at the highest bitrate provides a visual quality indistinguishable as VVC which is repeated by x265 and finally by VP9.

| *Bitrate savings Relative to* | | | | |
|---|---|---|---|---|
| *Encoding* | *VVC* | *SVT-AV1* | *x265* | *VP9* |
| *VVC* | - | 56% | 70.3% | 73% |
| *SVT-AV1* | | - | 31.06% | 38.44% |
| *x265* | | | - | 11.10% |

Table 7.10: BD-PSNR HEVC VIDEO DATASET 832x480p

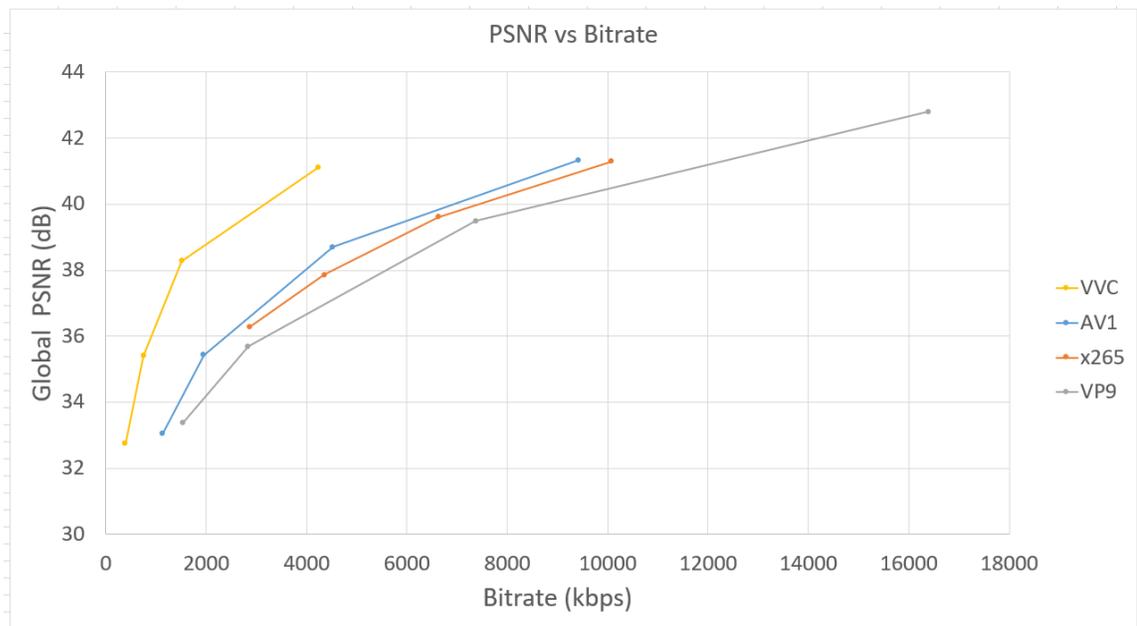

Figure 7.13: HEVC Dataset 480p RD Curves (PSNR vs Bitrate) of Median Values

| *Bitrate savings Relative to* | | | | |
|---|---|---|---|---|
| *Encoding* | *VVC* | *SVT-AV1* | *x265* | *VP9* |
| *VVC* | - | 59.63% | 71.54% | 79.77% |
| *SVT-AV1* | | - | 27.06% | 50.6% |
| *x265* | | | - | 32.50% |

Table 7.11: BD-VMAF HEVC VIDEO DATASET 832x480p

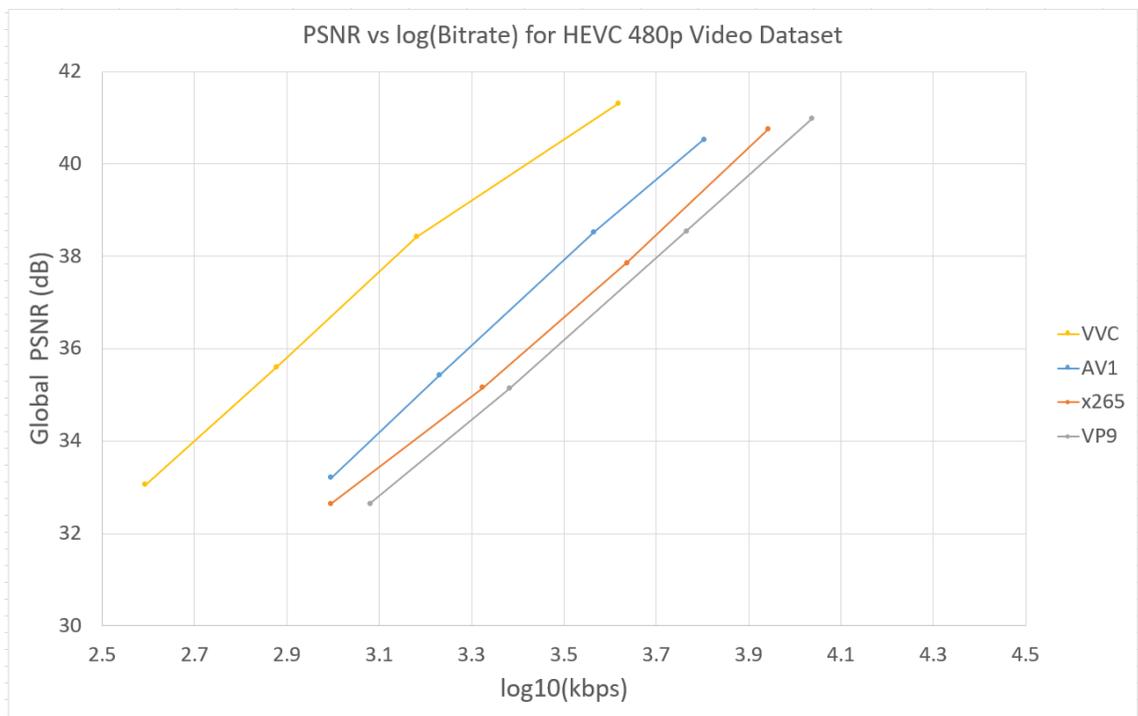

Figure 7.14: HEVC Dataset 480p RD Curves PSNR vs Log(Bitrate) of Median Values

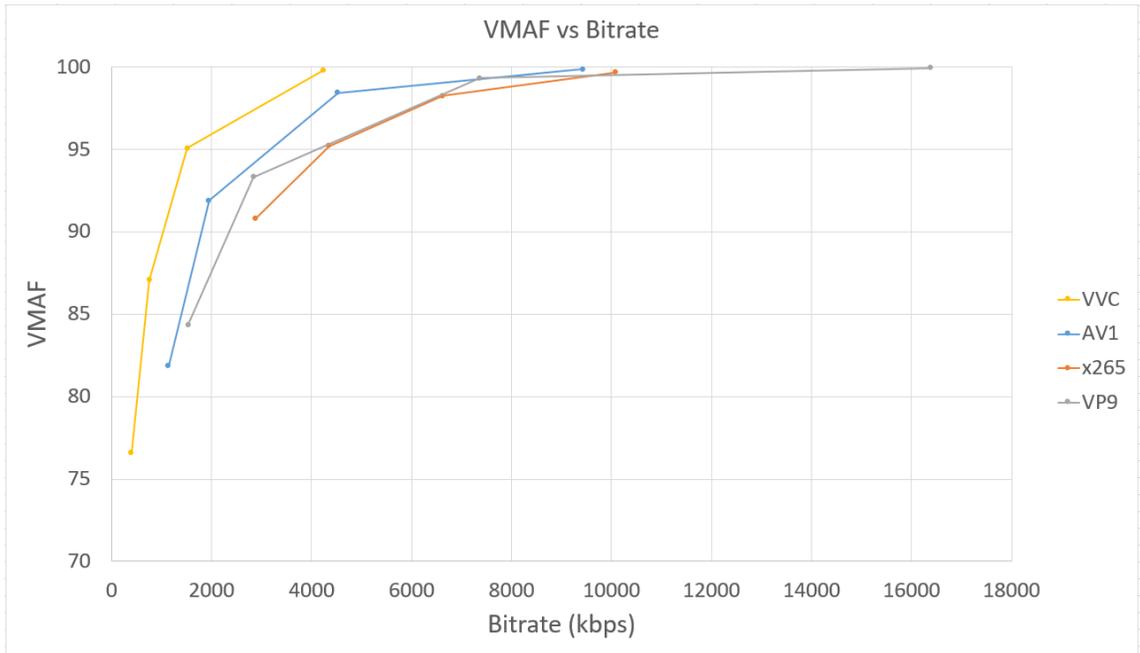

Figure 7.15: HEVC Dataset 480p RD Curves (VMAF vs Bitrate) of Median Values

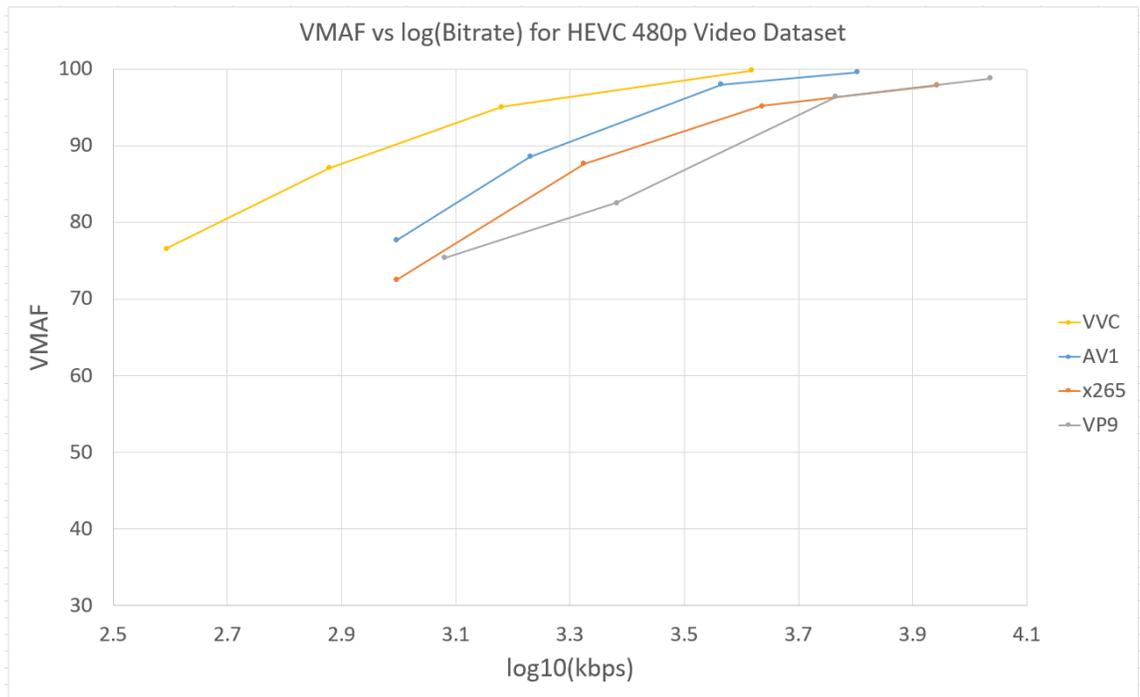

Figure 7.16: HEVC Dataset 480p RD Curves VMAF vs Log(Bitrate) of Median Values

## 7.3.4  BD-PSNR & BD-VMAF for 720p HEVC Dataset

Class-E videos where most of the video content is low motion and static background corresponds to video teleconferencing applications. BD-PSNR for Table 7.12, VVC saves up to 44.52%, 57.8% and 74.72% against SVT-AV1, x265 and VP9, respectively. Whereas, RD curves for VVC have huge bitrate gains as seen in Figure 7.17 than 240p or 480p as evident from the log scale of the RD curve in Figure 7.18 and 7.20. Results for BD-VMAF in Table 7.13 show that VVC saves around 62.98% versus AV1, 73.42% versus x265 and 76.37% versus VP9. Also at 720p, AV1 in Figure 7.19 saves approximately 30% & 38.64% against both x265 and VP9, which is quite promising as most video streaming content is recommended to stream in SD instead of HD in emergency crisis situations.

| *Bitrate savings Relative to* | | | | |
|:---:|:---:|:---:|:---:|:---:|
| *Encoding* | *VVC* | *SVT-AV1* | *x265* | *VP9* |
| *VVC* | - | 44.52% | 57.80% | 74.72% |
| *SVT-AV1* | | - | 35.83% | 62.27% |
| *x265* | | | - | 39.27% |

Table 7.12: BD-PSNR HEVC VIDEO DATASET 1280x720p

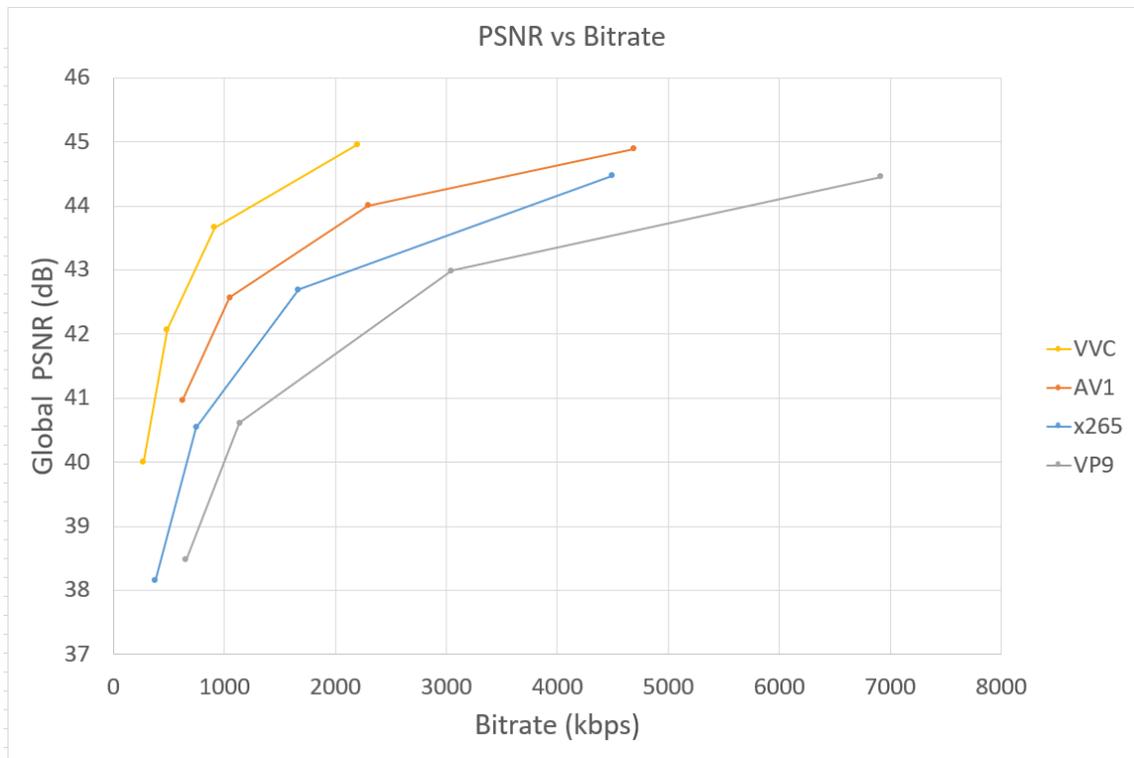

Figure 7.17: HEVC Dataset 720p RD Curves PSNR vs Bitrate) of Median Values

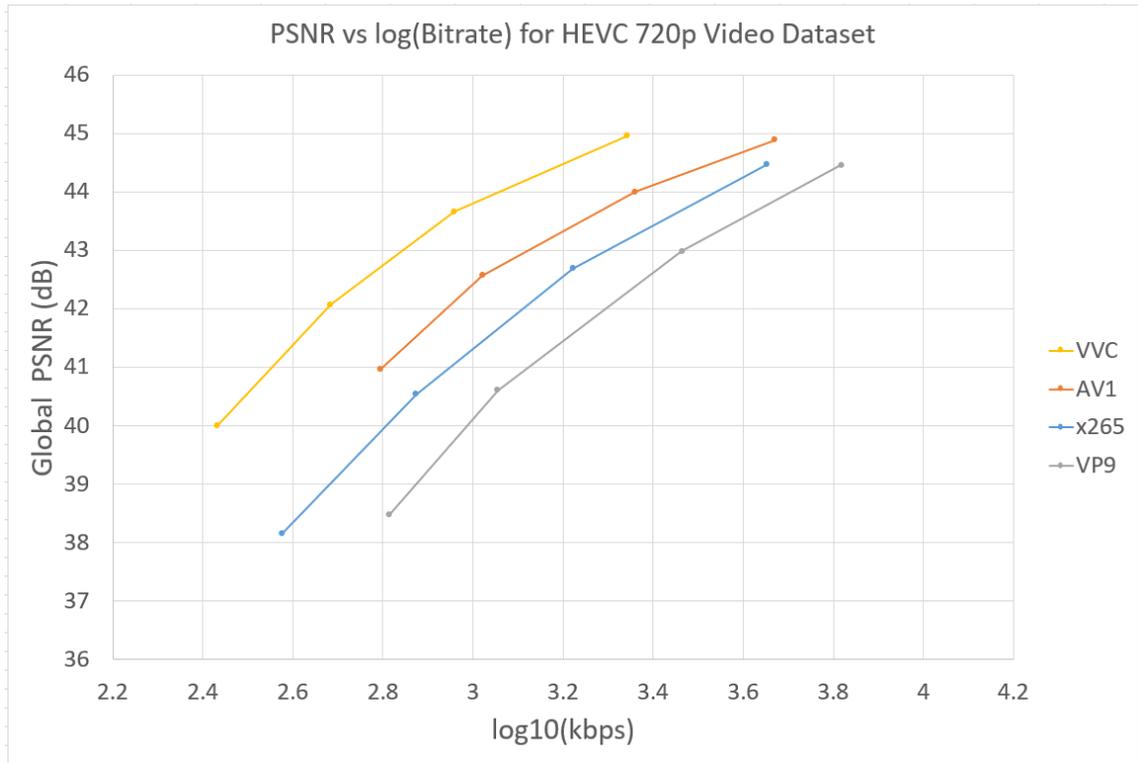

Figure 7.18: HEVC Dataset 720p RD Curves PSNR vs Log(Bitrate) of Median Values

| *Bitrate savings Relative to* | | | | |
|---|---|---|---|---|
| *Encoding* | *VVC* | *SVT-AV1* | *x265* | *VP9* |
| *VVC* | - | 62.98% | 73.42% | 76.37% |
| *SVT-AV1* | | - | 29.85% | 38.64% |
| *x265* | | | - | 16.18% |

Table 7.13: BD-VMAF HEVC VIDEO DATASET 1280x720p

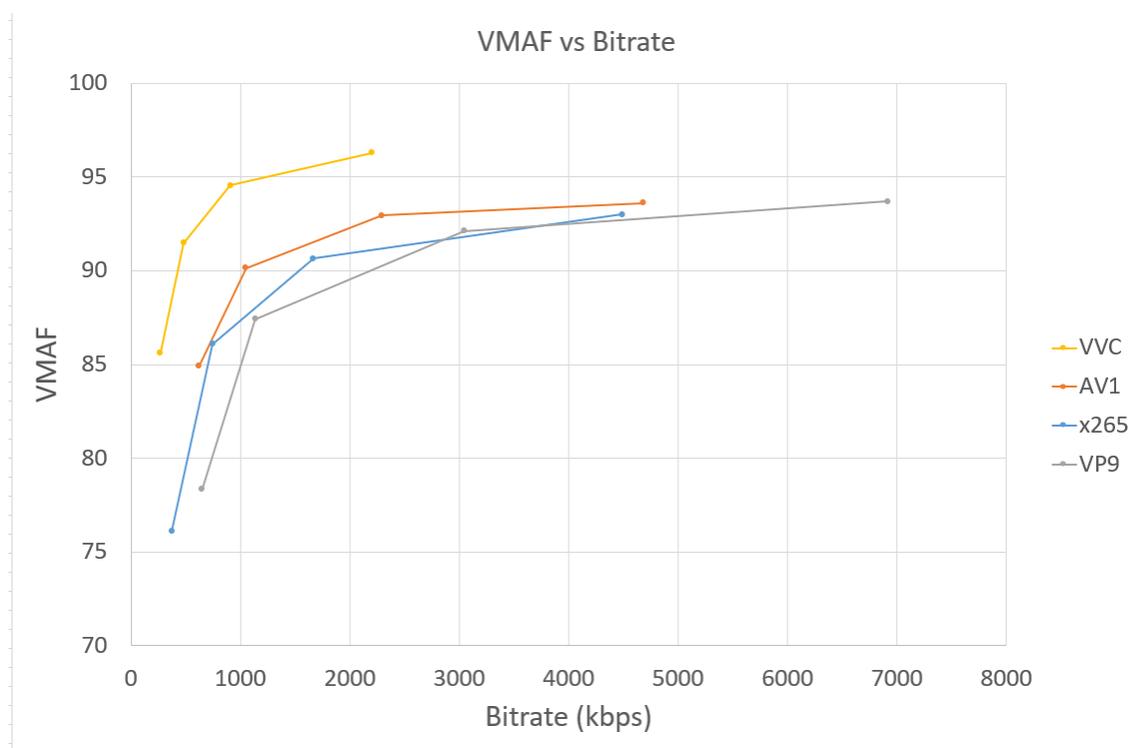

Figure 7.19: HEVC Dataset 720p RD Curves (VMAF vs Bitrate) of Median Values

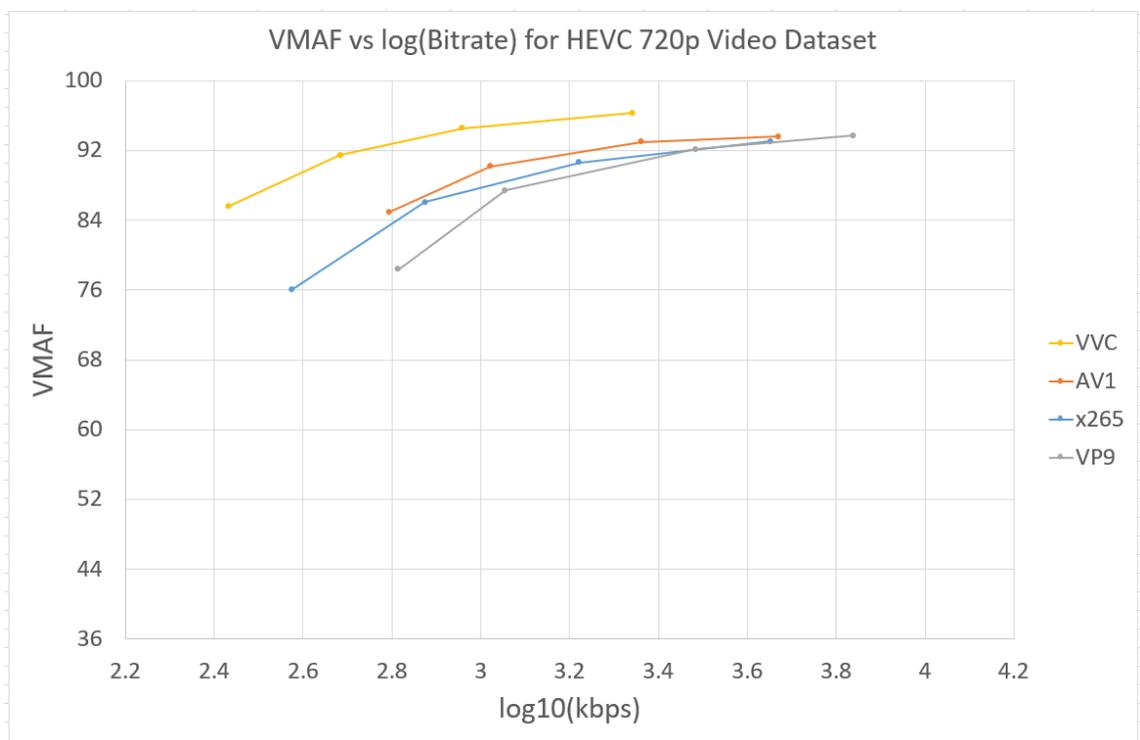

Figure 7.20: HEVC Dataset 720p RD Curves VMAF vs Log(Bitrate) of Median Values

## 7.3.5    BD-PSNR & BD-VMAF for 1080p HEVC & Tampere Dataset

For 1080p videos, we split the BD-tables based on datasets as we have both HEVC and Tampere in 1920x1080 resolutions. So let's have a look at them carefully. Based on BD-PSNR comparisons for both HEVC Table 7.14 and Tampere Table 7.16, VVC provided more savings in HEVC than Tampere as the latter dataset has complex camera motions, highly textured objects. VVC saves approximately 50% over AV1, 67% over x265 and 75.80% over VP9. In Tampere dataset, the bitrate gains drop significantly because of the complex motions involved in all of the videos. As we observe, VVC saves only 8.28% against AV1, 26% against x265 and a higher saving of 48.71% against VP9. From Figure 7.21 and 7.22, we see x265 gave better savings in Tampere than HEVC against VP9. On close observation, x265 saves around 27.60% against VP9 and in Tampere it gave 31.43% bitrate reductions.

| Bitrate savings Relative to | | | | |
|---|---|---|---|---|
| Encoding | VVC | SVT-AV1 | x265 | VP9 |
| VVC | - | 49.8% | 67.00% | 75.80% |
| SVT-AV1 | | - | 32.60% | 51.00% |
| x265 | | | - | 27.60% |

Table 7.14: BD-PSNR HEVC VIDEO DATASET 1920x1080p

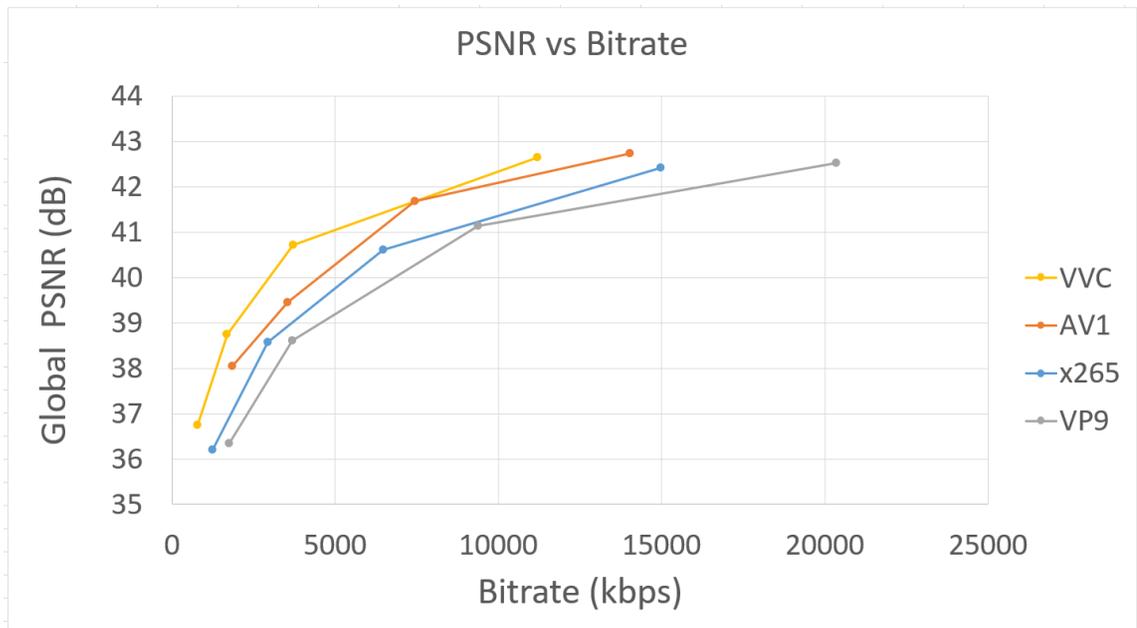

Figure 7.21: HEVC Dataset 1080p RD Curves PSNR vs Bitrate) of Median Values

In terms of BD-VMAF for 1080p, we observe the same trend as in BD-PSNR with less savings for the Tampere dataset as in Table 7.17 than HEVC. For the Tampere dataset, we have VVC savings around 15% against AV1, 18.18% against x265 and 33.23% against VP9, respectively. Also, AV1 gave fewer savings at 5.47% against x265, which is the lowest among all other resolutions and 23.9% against VP9. Even in HEVC dataset in Table 7.15, AV1 had 13.73% and 26.77% bitrate reductions against x265 and VP9. The only consistent performer in the HEVC dataset is VVC as in figure 7.23 and 7.24 which saves 54.2%, 59.8%, 67.8% against AV1, x265 and VP9, respectively.

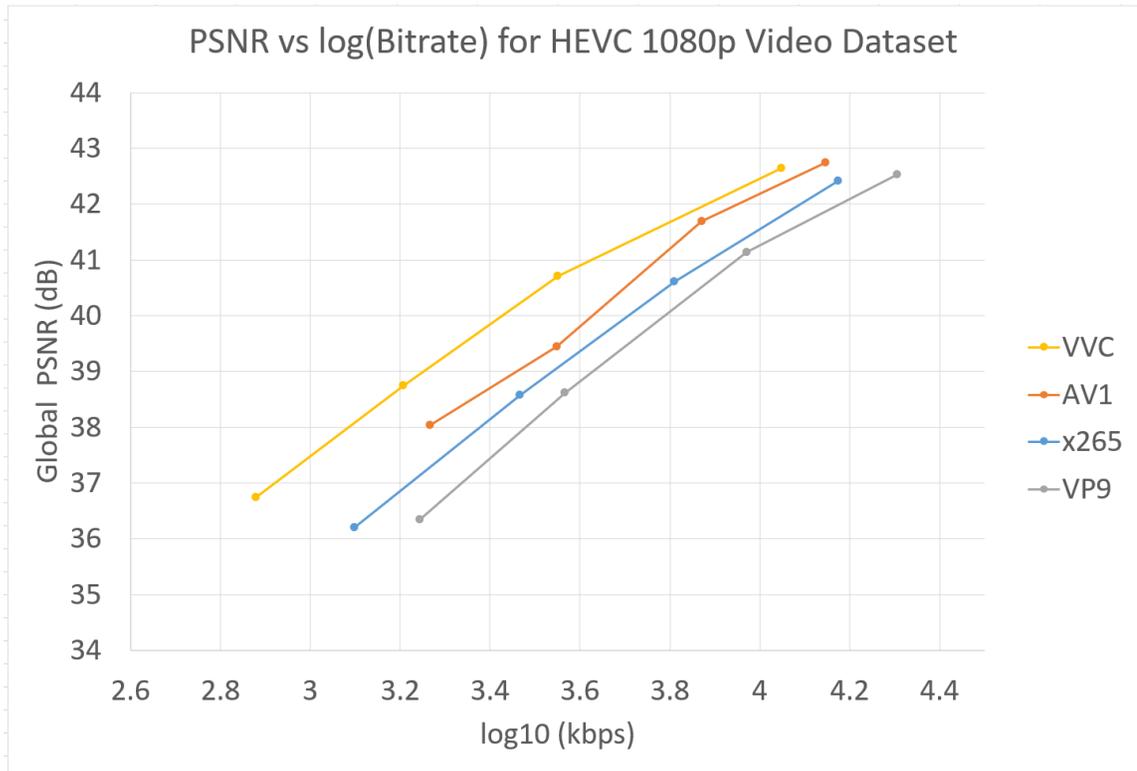

Figure 7.22: HEVC Dataset 1080p RD Curves PSNR vs Log(Bitrate) of Median Values

| Bitrate savings Relative to | | | | |
|---|---|---|---|---|
| *Encoding* | *VVC* | *SVT-AV1* | *x265* | *VP9* |
| *VVC* | - | 54.2% | 59.8% | 67.80% |
| *SVT-AV1* | | - | 13.73% | 26.77% |
| *x265* | | | - | 17.84% |

Table 7.15: BD-VMAF HEVC VIDEO DATASET 1920x1080p

| Bitrate savings Relative to | | | | |
|---|---|---|---|---|
| *Encoding* | *VVC* | *SVT-AV1* | *x265* | *VP9* |
| *VVC* | - | 8.28% | 26% | 48.71% |
| *SVT-AV1* | | - | 35.14% | 53.00% |
| *x265* | | | - | 31.43% |

Table 7.16: BD-PSNR TAMPERE VIDEO DATASET 1920x1080p

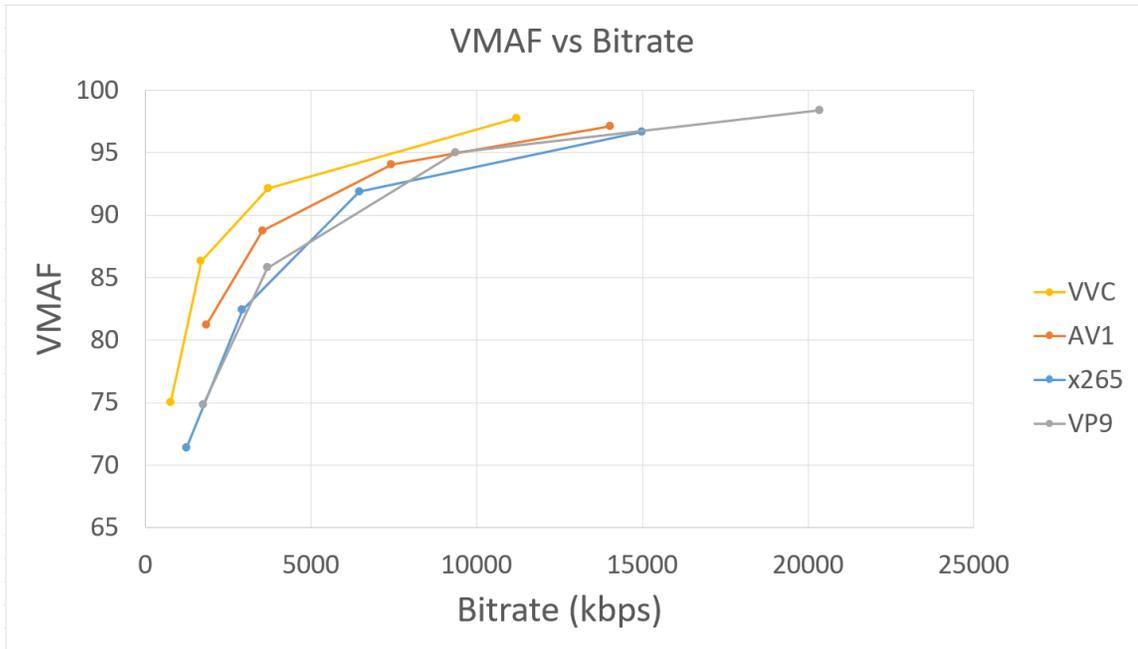

Figure 7.23: HEVC Dataset 1080p RD Curves (VMAF vs Bitrate) of Median Values

| *Bitrate savings Relative to* | | | | |
|---|---|---|---|---|
| *Encoding* | *VVC* | *SVT-AV1* | *x265* | *VP9* |
| *VVC* | - | 15.20% | 18.18% | 33.23% |
| *SVT-AV1* | | - | 5.47% | 23.9% |
| *x265* | | | - | 19.21% |

Table 7.17: BD-VMAF TAMPERE VIDEO DATASET 1920x1080p

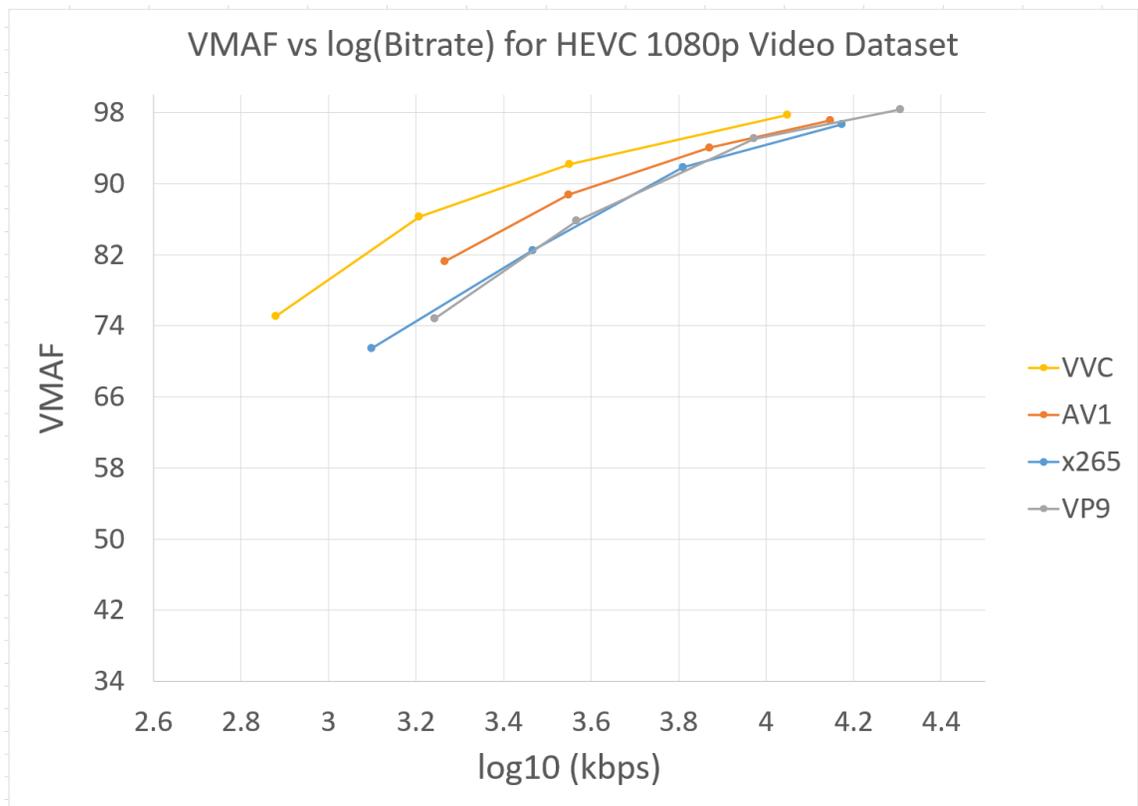

Figure 7.24: HEVC Dataset 1080p RD Curves VMAF vs Log(Bitrate) of Median Values

## 7.3.6    BD-PSNR & BD-VMAF for 1600p HEVC Dataset

For Class - A, 2560x1600, mostly used in traffic surveillance and stationary cameras, we have two video People on Street and Traffic. We will describe them separately to see how far they save in both BD-PSNR and BD-VMAF gains. For People video, we can see that VVC saves 50% against AV1, 62.98% against x265 and 62.26% against VP9 based on BD-PSNR Table 7.18 and the corresponding rate-curves can be seen for BD-PSNR in Fig 7.25 and 7.26.

In the BD-VMAF Table 7.20, we see that VVC has considerably higher gains up tp 77.86% against AV1, 69.44% against x265 and 65.27% against VP9, respectively. AV1 has savings up to 38.79% and 47.35% for both x265 and VP9 respectively. The corresponding RD-curves for BD-VMAF is shown in Figure 7.31 and 7.30.

| *Bitrate savings Relative to* | | | | |
|---|---|---|---|---|
| *Encoding* | *VVC* | *SVT-AV1* | *x265* | *VP9* |
| *VVC* | - | 49.57% | 62.98% | 62.26% |
| *SVT-AV1* | | - | 26.62% | 25.50% |
| *x265* | | | - | 1.57% |

Table 7.18: BD-PSNR HEVC PEOPLE VIDEO 2500x1600p

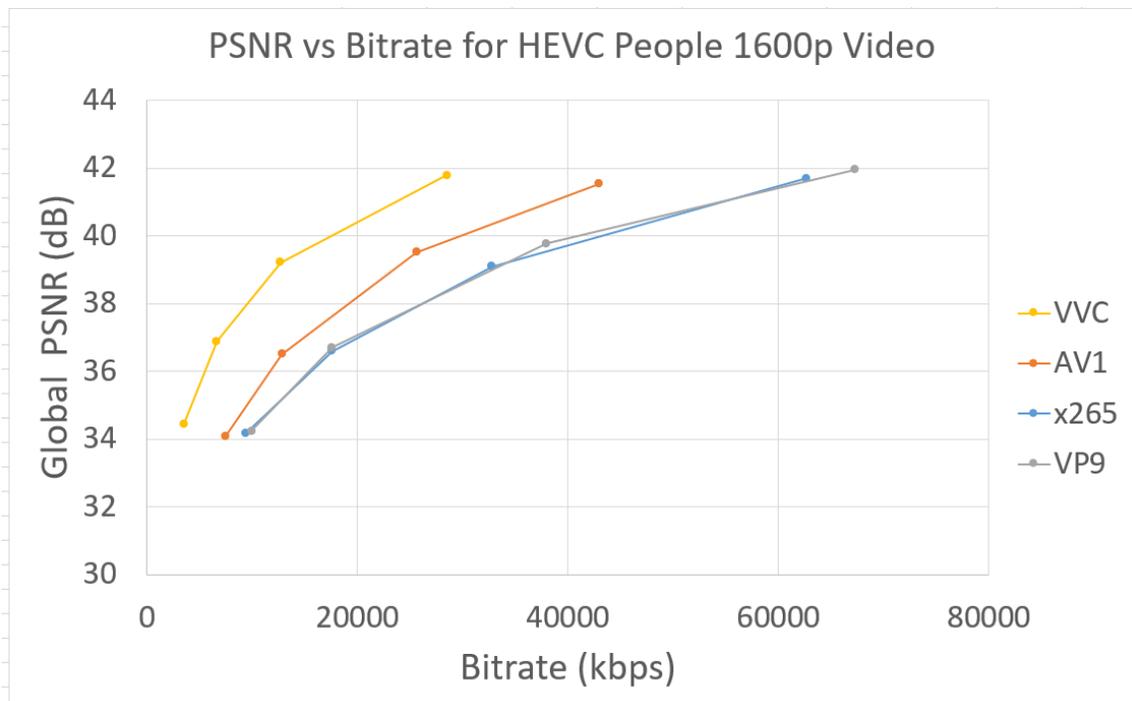

Figure 7.25: People 1600p RD Curve for (PSNR vs Bitrate) of Median Values

The Traffic video is characterized by very slow moving vehicles mostly found in surveillance imagery and transportation videos. From the BD Table 7.19, VVC saves up to 50.78% against AV1, 67.59% against x265 and 76.18% against VP9, respectively. AV1 has 33.88% reductions against x265 and 51.82% bitrate reductions against VP9. The corresponding log scale RD curves is shown in Figures 7.28 and 7.27.

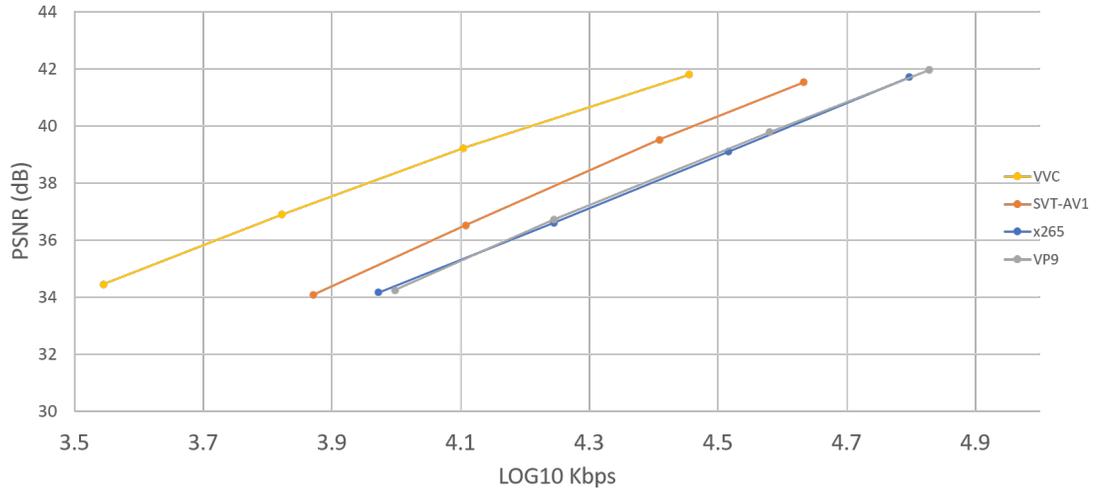

Figure 7.26: People 1600p RD Curve for PSNR vs Log(Bitrate) of Median Values

| Bitrate savings Relative to | | | | |
|---|---|---|---|---|
| Encoding | VVC | SVT-AV1 | x265 | VP9 |
| VVC | - | 50.78% | 67.59% | 76.18% |
| SVT-AV1 | | - | 33.88% | 51.82% |
| x265 | | | - | 26.55% |

Table 7.19: BD-PSNR HEVC TRAFFIC VIDEO 2500x1600p

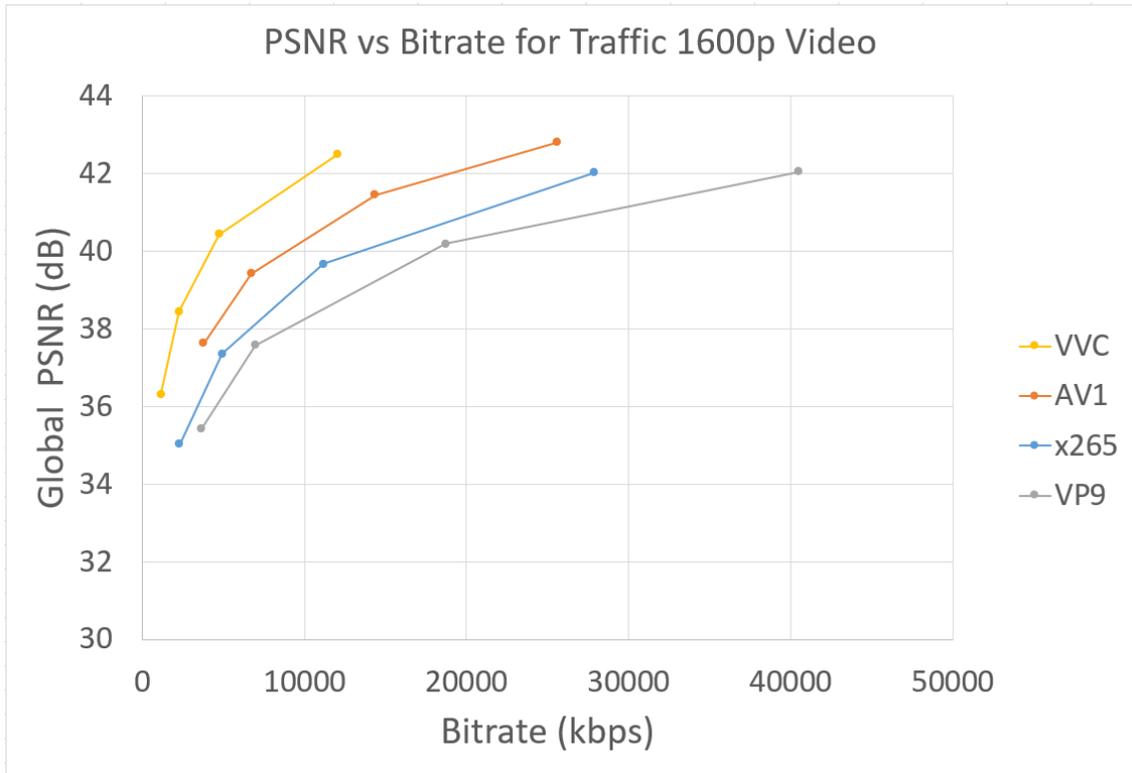

Figure 7.27: Traffic 1600p RD Curves (PSNR vs Bitrate) of Median Values

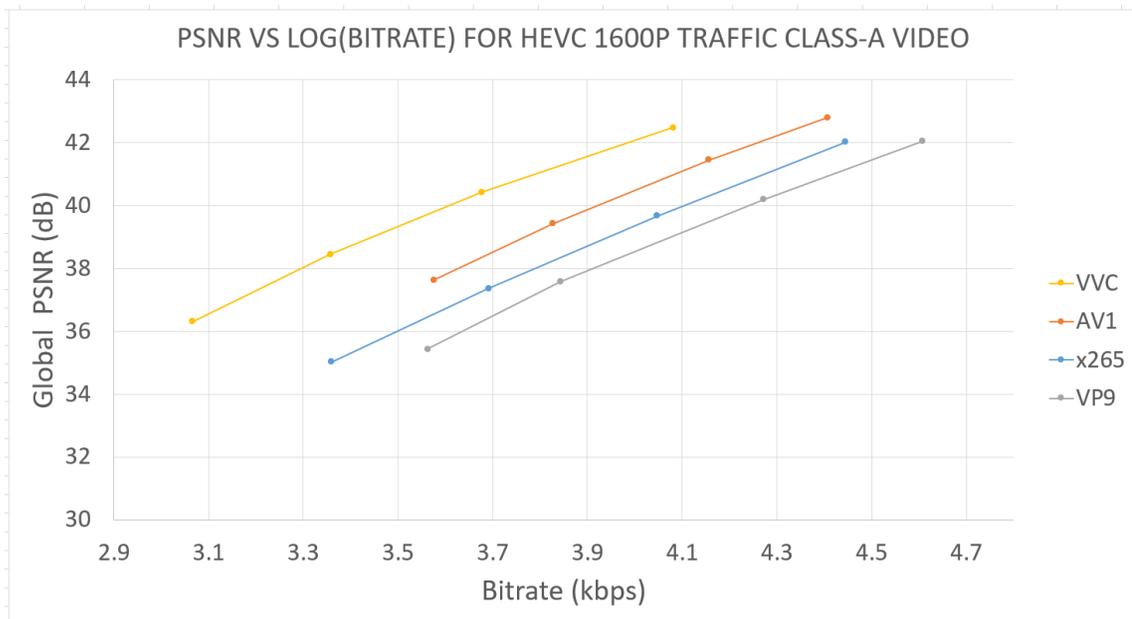

Figure 7.28: Traffic 1600p RD Curves (PSNR vs log Bitrate) of Median Values

| Bitrate savings Relative to | | | | |
|---|---|---|---|---|
| Encoding | VVC | SVT-AV1 | x265 | VP9 |
| VVC | - | 77.86% | 69.44% | 65.27% |
| SVT-AV1 | | - | 38.79% | 47.35% |
| x265 | | | - | 16.60% |

Table 7.20: BD-VMAF PEOPLE VIDEO 2500x1600p

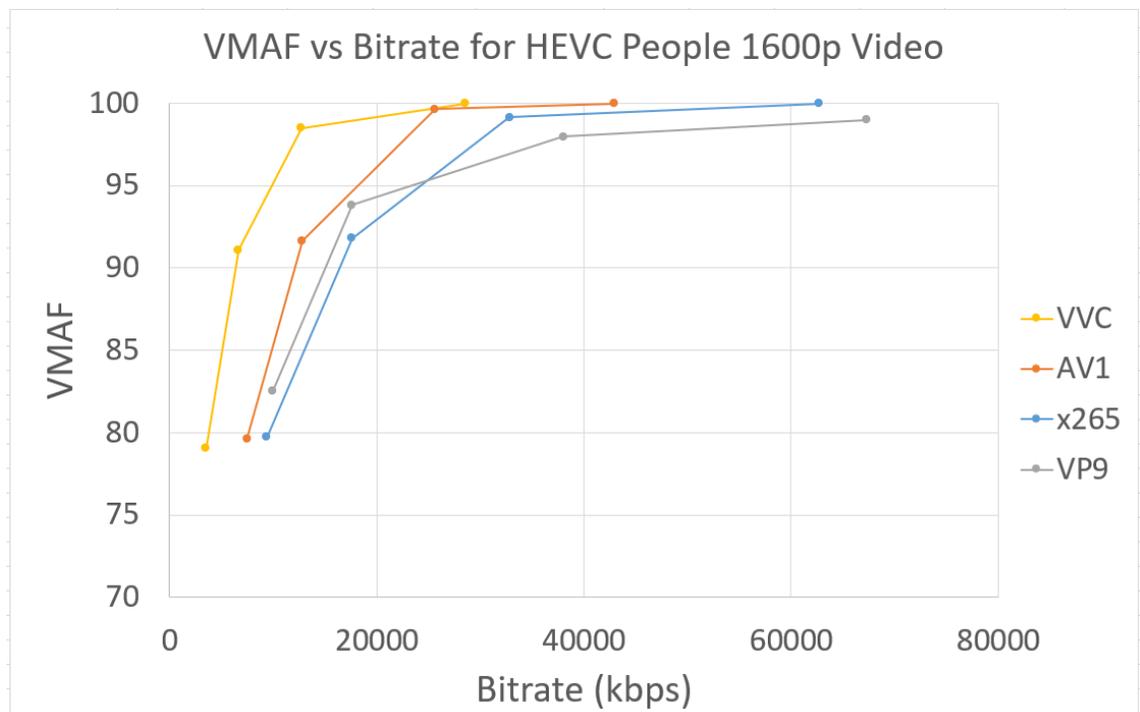

Figure 7.29: People 1600p RD Curves (VMAF vs Bitrate) of Median Values

The BD-VMAF gains for Traffic video are tabulated in Table 7.21 with VVC saving around 77.86% against AV1 and 69.44% against x265 and 65.27% against VP9, respectively. Here, AV1 saves lesser bitrate reduction of 18.32% against x265 than People video because of very slow motion occurring in this video and saves 29.17% against VP9. The RD curves for the VMAF is given in figure 7.31 and 7.32.

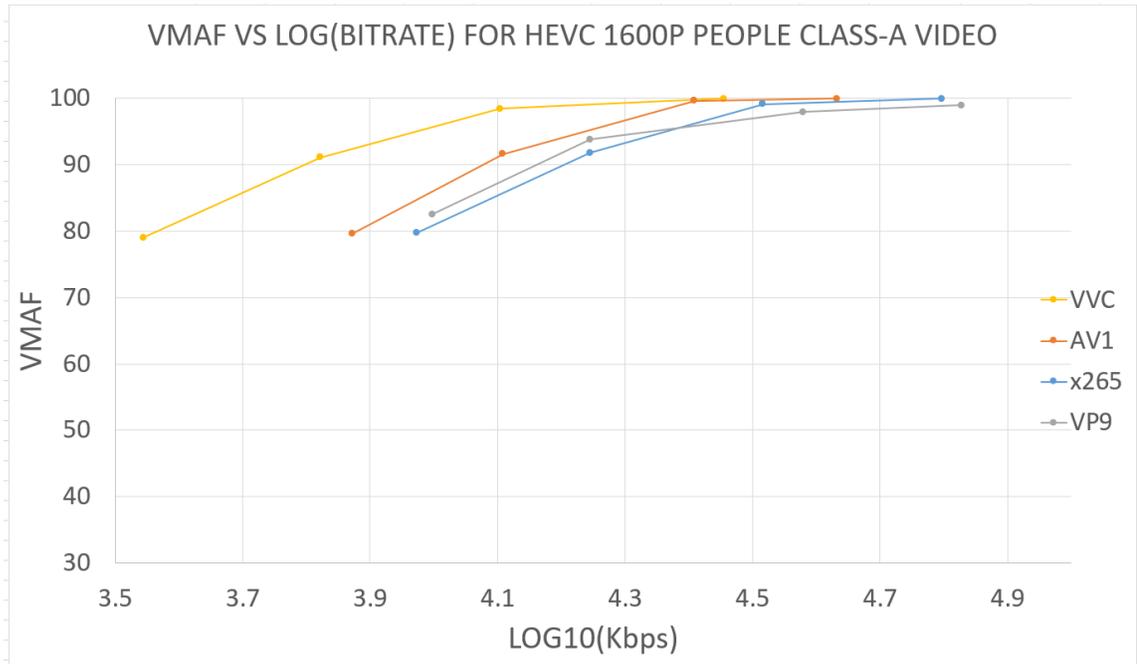

Figure 7.30: People 1600p RD Curves VMAF vs Log(Bitrate) of Median Values

| *Bitrate savings Relative to* | | | | |
|---|---|---|---|---|
| *Encoding* | *VVC* | *SVT-AV1* | *x265* | *VP9* |
| *VVC* | - | 58.47% | 66.10% | 71.14% |
| *SVT-AV1* | | - | 18.32% | 29.17% |
| *x265* | | | - | 15.32% |

Table 7.21: BD-VMAF TRAFFIC VIDEO 2500x1600p

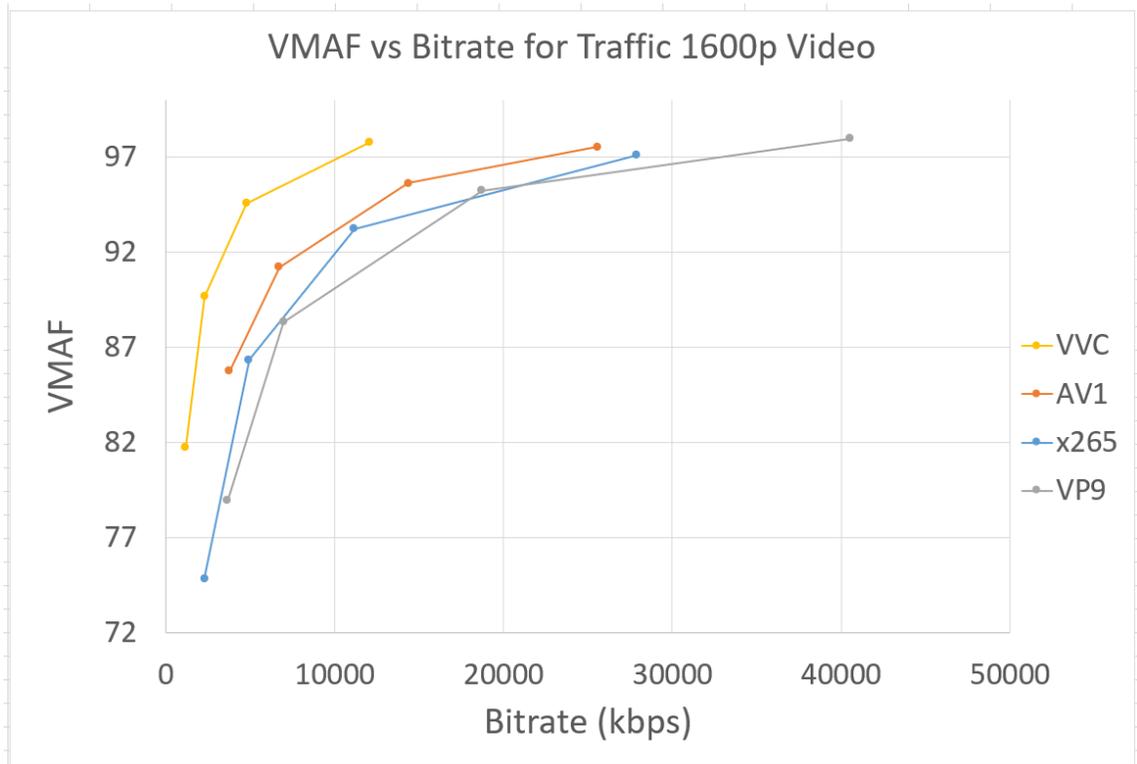

Figure 7.31: Traffic 1600p RD Curves (VMAF vs Bitrate) of Median Values

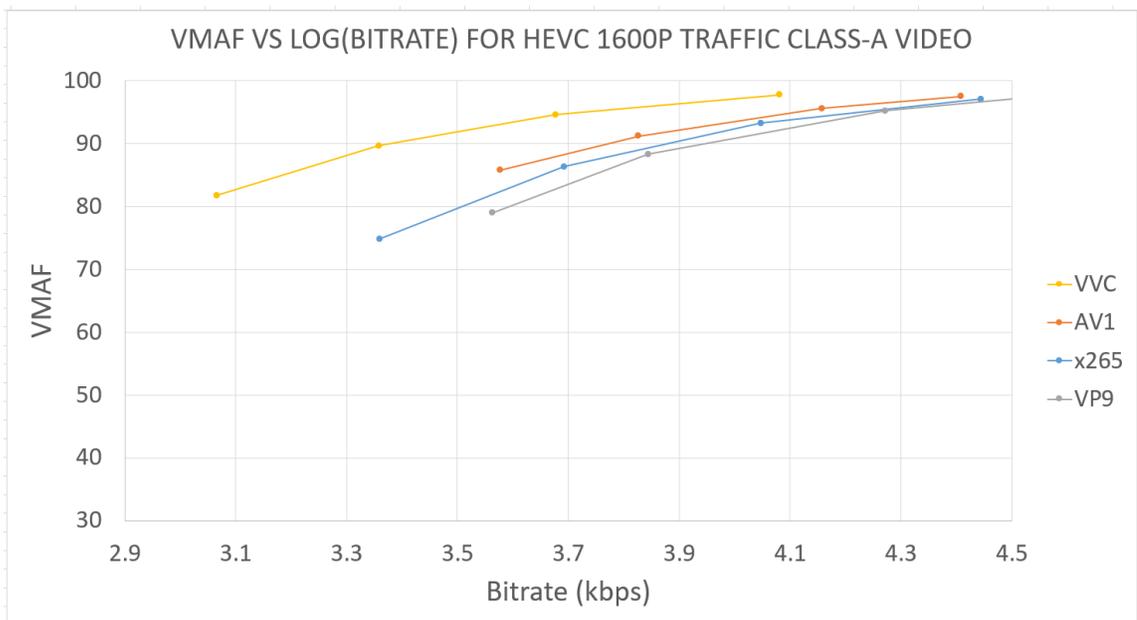

Figure 7.32: Traffic 1600p RD Curves VMAF vs Log(Bitrate) of Median Values

### 7.3.7 Overall Performance of VVC vs All Codecs

This section compared different video encoding standards from both objective and subjective video quality and we have summarized the results for three different video datasets.

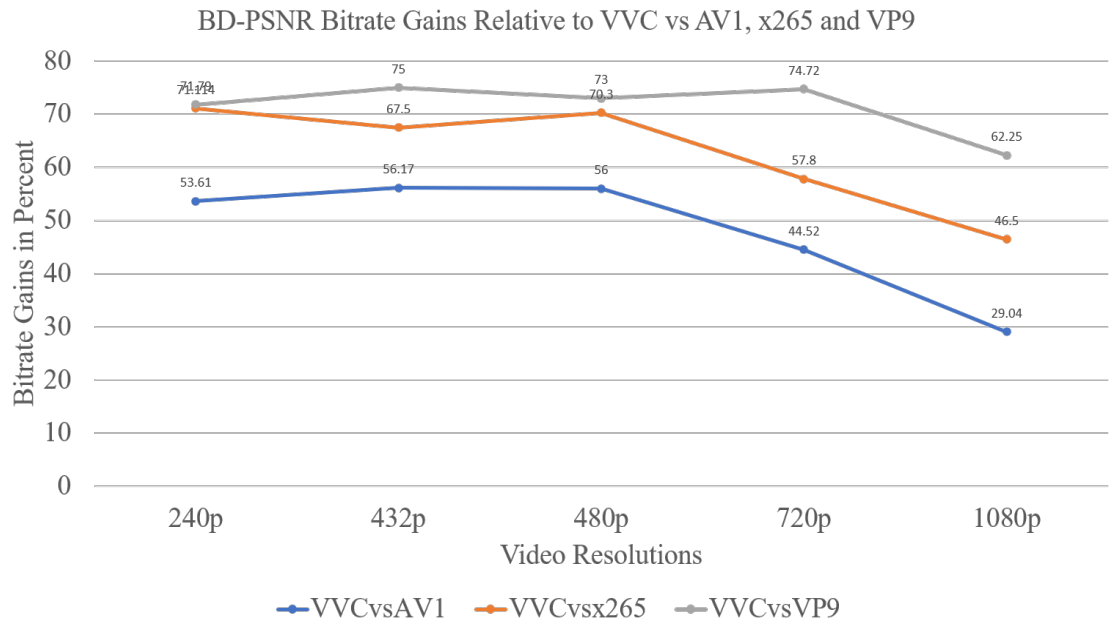

Figure 7.33: BD-PSNR Bitrate Gains of VVC vs All Codecs per Resolution

From Figures 7.33 and 7.34, it is clear VVC has the best results consistently winning and having huge bitrate gains from the lowest 240p to the highest 1600p. On the other hand, VVC encoding is extremely slow and has huge encoding complexity in terms of CPU cyles/seconds and requires finer optimization.

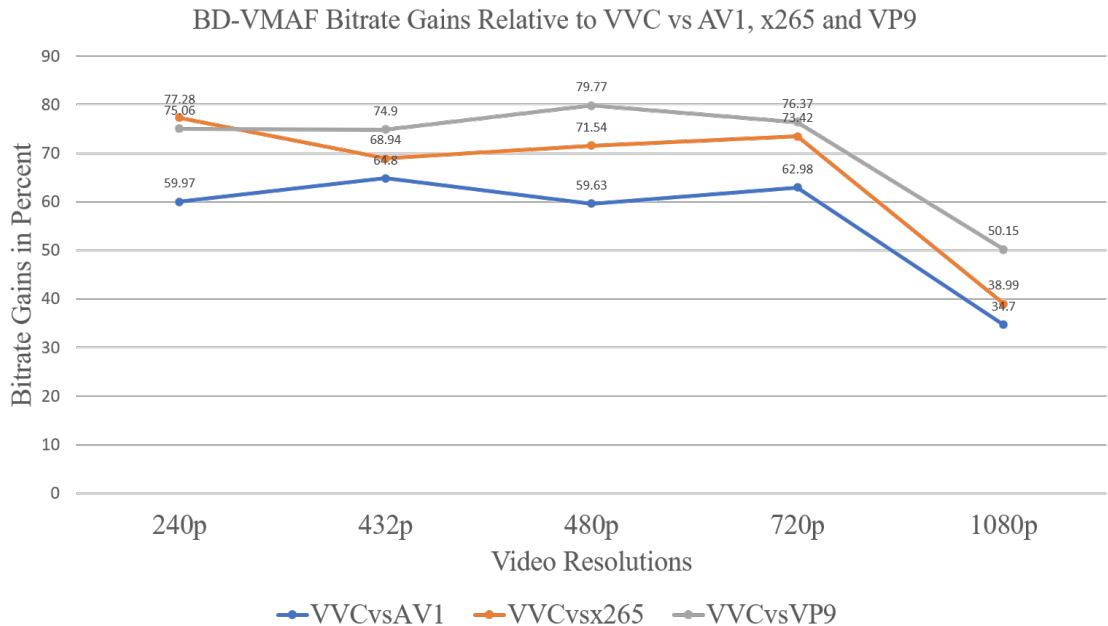

Figure 7.34: BD-VMAF Bitrate Gains of VVC vs All Codecs per Resolution

## 7.4 Subjective & Objective Video Quality Assessments for x265, VP9, AV1 Codecs

### 7.4.1 Video Codec configurations

**Source Sequences** For our subjective assessment, we selected 12 different videos of different resolutions 416x240p, 768x432p, 832x480p, 1280x720p, 1920x1080p and 2500x1600p as we collectively call it *Mixed Video bag* dataset each comprising of natural scenes and different motion content that is from HEVC [2], UT LIVE [20] and Tampere [3], respectively. The goal here was to study the subjective quality assessment with different resolutions and measure the VMAF perceptual metric simulating different video quality conditions. So we came up with three different scenarios.

- High Quality - When user gets a high bandwidth so correspondingly we used

a higher VMAF score of ($\geq 90$)

- Medium Quality - Bitrate recommended by default settings from the encoding ladders with VMAF score (70-85)

- Low Quality - Extremely low bandwidth situations where the video quality is bad and for this we used VMAF score ($\leq 60$)

***Training and Testing Sequences*** We created 36 test sequences from each of the 12 reference video tests with 3 different qualities (low, medium, high) with both objective video quality measured in PSNR and subjective video quality [104] metric in VMAF.

## 7.4.2   Preparing Subjects to View and Assess the Videos

Three subjective experiment sessions were conducted separately on the test sequences in the three codec groups. All three experiment sessions were conducted in a bright lit room with 3H distance meaning the subjects [105] are seated from the screen at an optimal viewing distance, measured in inches. Subjects were briefed about the video quality assessment and were explained different artifacts in videos for example, at low bandwidth the videos might get blocky or pixellated and might be buffered. We then ask the subjects on how would they evaluate the quality of video overall. During the training session, few videos were shown with different video qualities and then the overall process was demonstrated. Then the subjects were given a scoring sheet and asked to score the quality in terms of 1-Bad, 2-Worse, 3-Fair, 4-Good, 5-Excellent.

All video test sequences were randomized and shown on a display, which is a SAMSUNG U28E690D LCD TV, with 4K screen resolution. There are 3 different groups: Group A(SVT-AV1), Group B(VP9) Group C(x265). Each group has been

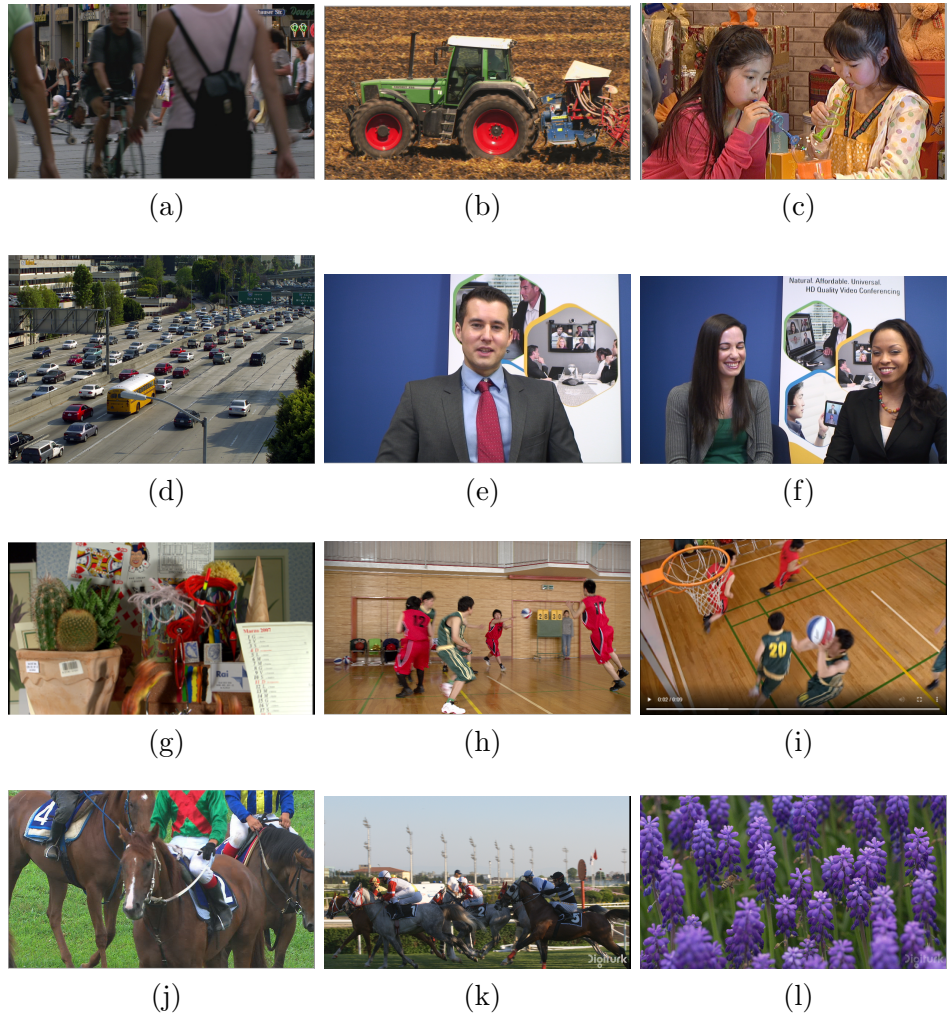

Figure 7.35:  Mixed bag video dataset for Subjective video quality assessment [1] (a), (b) Pedestrian, Tractor, video with resolution 768x432 of 50, 25, 25 fps respectively from UT LIVE Video Quality Database. (c) Blowing Bubbles of 480x240 from Class D with 50 fps, (d) Traffic of 2500x1600 from Class A with 50 fps, (e), (f) Johnny and KristenandSara of 1280x720 from Class E with 60 fps, (g), (h) Cactus, BasketballDrive video with resolution 1920x1080, 50 fps, (i), (j) Racehorse, BasketballDrill video of 832x480, 50 fps, (k), (l) ReadysetGo, HoneyBee videos with resolution 1920x1080 of 60 fps publicly available from Ultra Video group, Tampere University.

put through a training session where test sequences were shown and then explained on how to evaluate the video and then score them. A total of 32 subjects with an

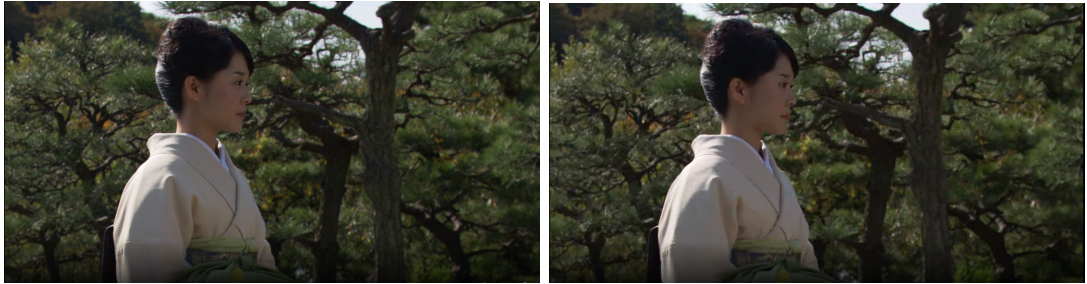

(a) Kimono with High Quality VMAF score 97.

(b) Kimono with Medium Quality VMAF score 85.

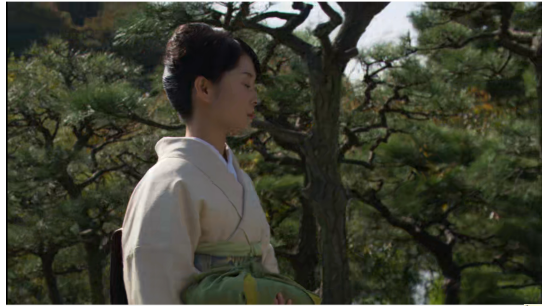

(c) Kimono with Low Quality VMAF score 73.

average age of 27 (age range 21-55) participated for this subjective assessment.

**Correlation Metrics Performance Comparison**

We used two different correlation metrics: 1. Spearman Rank Order Correlation Coefficient (SROCC) 2. Pearson linear Correlation Coefficient (POCC) to measure the performance of the subjective video quality assessment. The correlation performance of two tested objective quality metrics for three codec groups (in terms of SROCC values) as SROCC measures the non-linear relationship between the scores and the original values from the encodings. Earlier, subjective studies shown in [1], show there is a good correlation of above 0 [104] show

The results are summarized in Table 7.22 for all codec groups based on both PSNR and VMAF. Let's consider SROCC metric for VMAF first and from the tested codec versions and configurations, it can be observed that SVT-AV1 achieves the highest correlation of 0.78. On the other hand, VP9 stands at 0.74 and x265 0.63, respectively. In terms of the objective metric PSNR, the SROCC correlations for SVT-AV1 0.64 and VP9 stands at 0.63 and 0.61 for x265.

It can be observed that VMAF outperforms PSNR significantly with the highest SROCC and POCC values, while PSNR results in much lower performance, especially in x265 codec group. It is also noted that, for all test quality metrics, the SROCC values for three codec groups are all below 0.9, which indicates that further enhancement is still needed to achieve more accurate prediction.

| *Subjective VQA with Codecs* | *SVT-AV1* | | *VP9* | | *x265* | |
|---|---|---|---|---|---|---|
| *Correlation Metrics* | *PSNR* | *VMAF* | *PSNR* | *VMAF* | *PSNR* | *VMAF* |
| *SROCC* | 0.64 | *0.75* | 0.63 | *0.74* | 0.61 | *0.635* |
| *POCC* | 0.623 | *0.78* | 0.6 | *0.70* | 0.58 | *0.633* |

Table 7.22: Correlation Metrics for the Subjective Video Quality Assessment for the Mixed bag Dataset

## 7.5 Conclusion

An overview of the emerging VVC encoding tools was described and then BD-rates for both PSNR and VMAF for the codecs x265, VP9, SVT-AV1 against the reference encoder VVC-VTM were compared. Overall, VVC consistently beats the challenging competitor codecs and provides significant coding gains and performance. We also did a subjective video quality assessment with different video qualities encoded by varios codecs of the likes x265, VP9, SVT-AV1 and in this case the latter SVT-AV1 wins the majority both in terms of PSNR and VMAF by measuring their corresponding correlation metrics. VVC was not considered for this study as we do not have a conventional media player that can play .vvc bitstream files yet. Additionally, VVC is still in its early stages and with more tools need to be finalized and fully optimized so as for it to compete with libAOM or SVT-AV1 which is potentially the future in the video streaming industry.

# Chapter 8

# Conclusion and Future Work

## 8.1 Conclusion

We have provided a codec-agnostic dynamic framework that can be used to achieve better compression efficiency without sacrificing quality across different encoding standards. The segments based DRASTIC optimization approach is able to achieve coding gains compared against the YouTube recommended bitrates. We have targeted the constant QP method using single pass (1-Pass) to achieve optimal bitrate/resolution encodings, but this can be easily extended to 2-Pass, 3-Pass or other Multi-pass encoding methods.

The dissertation considered applications in H.265/HEVC, VP9, SVT-AV1, VVC Codecs by introducing new GOP structures for encoding and then applied DRASTIC optimization. The second application was analyzing different activities/camera motions in the video and then using Motion vectors on a frame level to classify the motion of the video content. Using this methodology, we then are able to adaptively encode videos and achieve 35% and 52% bitrate savings for the example videos. Thirdly, we started to analyze the Pareto surface of each video and, after careful observation, we decided to do Segment-based encoding or breaking the video into 3-second segments and then build models by fitting the Pareto surface and by this we can simply capture the entire video content with few encoding parameters. Using these models, we can predict encoding parameters for the next segment using a fast approach that also satisfies dynamic constraints. We were able to conserve approximately 9% and 13% bitrate savings at 1080p videos using this approach.

Fourthly, we introduced new GOP structures in the libVPx encoding standard with VP9 encoder and demonstrated that Segment-based encoding for different video content and can provide 8% bitrate savings. Fifthly, we studied the SVT-AV1 codec tools and employed new GOPs for the segment based encoding, and provided results for 1080p and 480p videos. The final chapter was divided into three sections where we provided an overview of VVC encoding standard and different components of the VTM encoder, performed a codec comparison and also performed a subjective video quality assessment for a mixed video dataset of different resolutions and reported the correlation metrics.

Currently, there are two pending publications that are derived from the dissertation. First, the DRASTIC Segment-based encoding with constrained video delivery for video compression standards x265, VP9, AV1 will be submitted as a full journal paper to the *IEEE Open Access*. Second, the comparison of the VP9, x265, SVT-AV1, VVC Codecs with VMAF as a leverage metric has been accepted to *SPIE, Applications of Digital Image Processing XLIII, 2020.*

## 8.2 Future Work

Segment-based encoding has been studied at the GOP level with selected encoding parameters for all the encoders. I recommend new set of directions for the future work which we looked upon but did not have the time and resources to complete it within the scope of this dissertation.

*Simulated Annealing approach with 5% & 10% Encoding samples:* We built the initial regression models using 250 encoding configurations for 3-second segments. It would be interesting to explore methods to reduce the number of required encodings. For example, using a quite sophisticated sampling approach that uses simulated annealing with 5% or 10% samples from the original Pareto, we may be able to build a model and test it with different constraints for the videos.

*Global Modeling approach with Scalable resolution:* This modeling idea is to build the forward models for one resolution (e.g., 720p or 1080p), and then use the 1080p models for predicting the optimal encoding parameters for a different video segment. To make it work, we would choose a dataset of the same resolution and segment the videos into 3s chunks and then build forward models on each of them.

In-order to effectively capture the entire video content, we have to use different encoding parameter sets as inputs to the encoding. Parameters like GOPs, QPs, Filters for reconstruction, Motion Vectors (MVs) and Motion Vector Prediction (MVP) at frame level, Residual Transform Unit (TUs) size and RDO decisions at the frame level. Using Leave-One-Out (LOOCV) Cross Validation, we can choose the optimal model for a particular video segment and then use it for a different video segment. By scalable we mean the video can be scaled to different resolutions from the original resolution. For example, a 1080p model trained on a 3s video segment can be applied to predict the encoding parameters for a 720p video segment.

*SVT-AV1 and libAOM Comparison for webRTC application:* With ever growing VoD and streaming applications, we need more efficient and reliable systems that are ubiquitous and can deliver high quality video at extremely low bandwidth scenarios. Since AV1 is the promised future codec for video delivery and streaming for real-time video communications, we can study it with the webRTC framework and deploy it in different networking conditions.

*DASH based Codec-agnostic Video Delivery system:* With DASH based delivery gaining popularity, a client-server model with different encoders built in and switching based on video content on available constraints will be a very interesting study.